\documentclass[a4paper,11pt]{article}
\usepackage{jcappub}
\usepackage{amsmath, amssymb, amsfonts, amsthm}
\usepackage[usenames,dvipsnames,svgnames,table]{xcolor}
\usepackage{graphicx}
\usepackage{subfigure}
\newcommand{\at}[2][]{#1|_{#2}}
\makeatletter
\newcommand{\Rmnum}[1]{\expandafter\@slowromancap\romannumeral #1@}
\makeatother

\newcommand{\beq}{\begin{equation}}
\newcommand{\eeq}{\end{equation}}

\newcommand{\ber}{\begin{eqnarray}}
\newcommand{\eer}{\end{eqnarray}}

\def\mpl{M_{p}}
\def\nn{\nonumber}

\def\l{\left}
\def\r{\right}

\def\ie{{\rm i.e.}}

\begin{document}

\newcolumntype{L}[1]{>{\raggedright\arraybackslash}p{#1}}
\newcolumntype{C}[1]{>{\centering\arraybackslash}p{#1}}
\newcolumntype{R}[1]{>{\raggedleft\arraybackslash}p{#1}}

\title{Emergent Cosmology Revisited}

\author[a]{Satadru Bag,}
\author[a]{Varun Sahni,}
\author[b,c]{Yuri Shtanov,}
\author[d]{Sanil Unnikrishnan}

\affiliation[a]{Inter-University Centre for Astronomy and Astrophysics, Pune 411007, India} \affiliation[b]{Bogolyubov Institute for Theoretical Physics, Kiev 03680, Ukraine} \affiliation[c]{Department of Physics, Taras Shevchenko Kiev National University, Kiev, Ukraine}\affiliation[d]{Department of Physics, The LNM Institute of Information Technology, Jaipur 302031,
India}

\emailAdd{satadru@iucaa.ernet.in}
\emailAdd{varun@iucaa.ernet.in}
\emailAdd{shtanov@bitp.kiev.ua}
\emailAdd{sanil@lnmiit.ac.in}

\abstract{We explore the possibility of emergent cosmology using the {\em effective potential}
 formalism.
We discover new models of emergent cosmology which satisfy the constraints posed
by the cosmic microwave background (CMB). We demonstrate that, within the framework of modified gravity, the emergent scenario
can arise in a universe which is spatially open/closed.
By contrast, in general relativity (GR) emergent cosmology
arises from a spatially closed
 past-eternal Einstein Static Universe (ESU). In GR the
ESU is unstable, which creates fine tuning
problems for emergent cosmology.
However, modified gravity models including Braneworld models, Loop Quantum Cosmology (LQC)
and Asymptotically Free Gravity
result in a stable ESU. Consequently, in these models
emergent cosmology arises from a larger class
of initial conditions
including those in which the universe eternally oscillates
about the ESU fixed point. We demonstrate that such an oscillating universe is
necessarily accompanied by {\em graviton production}. For a large region in parameter space
graviton production is enhanced through a parametric resonance, casting serious doubts
as to whether this emergent scenario can be past-eternal.
}
\keywords{Effective potential, Einstein Static Universe, Inflation, Emergent Cosmology, CMB constraints on inflation, Particle production}
\arxivnumber{1403.4243}
\maketitle

\section{Introduction}\label{sec: intro}

The inflationary scenario has proved to be successful in describing a universe which is
remarkably similar to the one which we inhabit. Indeed, one of the central aims of the
ongoing effort in the study of cosmic microwave background (CMB) observations is to
converge on the correct model describing inflation \cite{Ade:2013uln}.

However, despite its very impressive achievements, the inflationary paradigm leaves some questions
unanswered. These pertain both
to the nature of the inflaton field and to the state of the
universe prior to the commencement of inflation.
Indeed, as originally pointed out in \cite{Borde:2001nh}, inflation
(within a general relativistic setting)
 could not have been past eternal.
This might be seen to imply one of several alternative  possibilities including the following:

\begin{enumerate}

\item The universe quantum mechanically tunnelled into an inflationary phase.

\item The universe was dominated by radiation (or some other form of matter)
 prior to inflation and might therefore have encountered
a singularity in its past.

\item The universe underwent a non-singular bounce prior to inflation.
Before the bounce the universe was contracting.

\item The universe existed `eternally' in a quasi-static state, out of which inflationary
expansion emerged.

\end{enumerate}

One should point out that, at the time of writing, none of the above possibilities
is entirely problem free. Nevertheless, our focus in this paper
will be on the last option, namely that of an {\em Emergent Cosmology}.



The idea of an emergent universe is not new and an early semblance of this
concept can be traced back to the seminal work of Eddington \cite{Eddington:1930zz}
and Lema\^{\i}tre \cite{Lemaitre:1927zz}, which was based on the Einstein Static Universe
\cite{Einstein:1917ce}. Indeed, in 1917, Einstein introduced the idea of a closed
and static universe sourced by a cosmological constant and matter. Subsequently it was
found that: (a) the observed universe was expanding \cite{Hubble:1929ig}, (b) the Einstein Static Universe (ESU) was unstable. It therefore became unlikely that ESU could
describe the present universe but allowed for our universe to have emerged from a static
ESU-phase in the past.

With the discovery of cosmic expansion Einstein distanced himself from his own early
ideas referring to them, years later, as his biggest blunder \cite{Pais:1982up}.

Interest in the ESU subsequently waned, although models in which the ESU featured as an
intermediate stage --- called {\em loitering\/} --- received a short-lived burst of
attention in the late 1960's, when it was felt that a universe which loitered at $z
\simeq 2$ might account for the abundance of QSO's at that redshift; an observation that
inspired Zeldovich to write his famous review on the cosmological constant \cite{Zel'dovich:1968zz}.

The present resurgence of interest in ESU and emergent cosmology owes much to the CMB
observations favouring an early inflationary stage, supplemented by the fact that an
inflationary universe is geodesically incomplete \cite{Borde:2001nh} and might therefore have
had a beginning.

This paper commences with a discussion of emergent cosmology in the context of general
relativity in section \ref{sec:emergent_GR}. Since GR-based ESU is unstable, this
scenario suffers from severe fine-tuning problems, as originally pointed out in
\cite{Ellis:2002we,Ellis:2003qz}. One can construct stable ESUs in the context of the Braneworld
scenario \cite{Shtanov:2002mb}, Loop Quantum Cosmology (LQC) \cite{Ashtekar:2006uz}
and Asymptotically Free Gravity. This is
the focus of section \ref{sec:emergent_grav}.
 When viewed in the classical context,
a stable ESU allows the universe to oscillate `eternally' about the ESU fixed point
\cite{Mulryne:2005ef,Parisi:2007kv}. If the universe is filled with a scalar field, then these
oscillations can end, giving rise to inflation.
However, this scenario is feasible only for an appropriate choice of the inflaton
potential. Equally important is the fact that an oscillating universe generically gives
rise to graviton production, which forms the focus of section \ref{sec:graviton}. For a
large region in parameter space, the production of gravitons proceeds
through a parametric resonance, which seems to question the possibility of whether a universe
could have oscillated `eternally' about the ESU fixed point. Our conclusions are drawn in section
\ref{sec:conclusions}.

Our main results seem to suggest that, while emergent cosmology (EC) can be constructed
on the basis of both GR and modified gravity, the restrictions faced by working EC models
are many. Consequently, realistic EC is possible to construct only in a small region of
parameter space, and that too for a rather restrictive class of inflationary potentials.

\section{Emergent cosmology and the effective potential formalism}\label{sec:emergent_GR}


\begin{figure}[t]
\begin{center}
\includegraphics[width=0.482\textwidth]{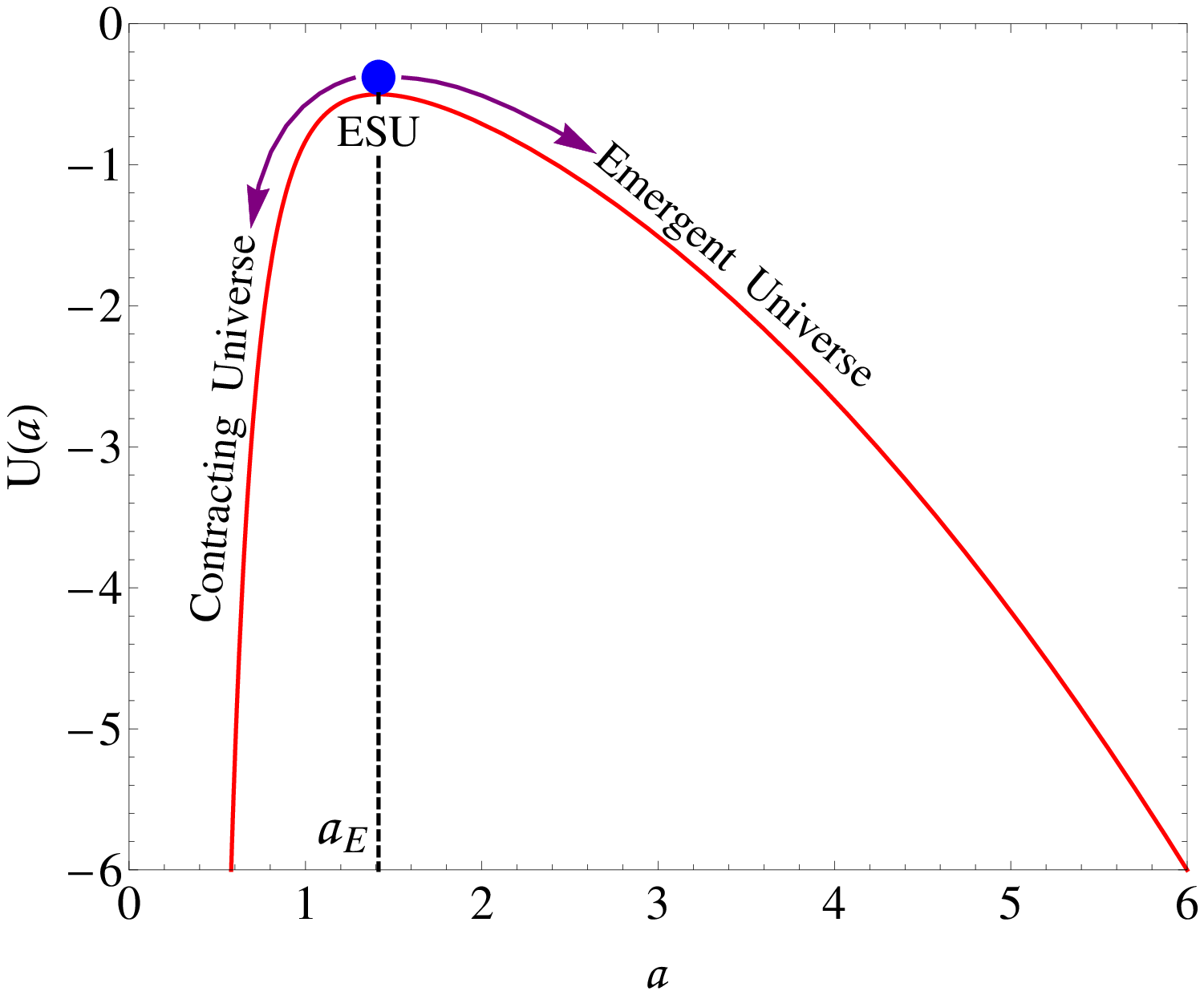}
\includegraphics[width=0.468\textwidth,height=6.06 cm]{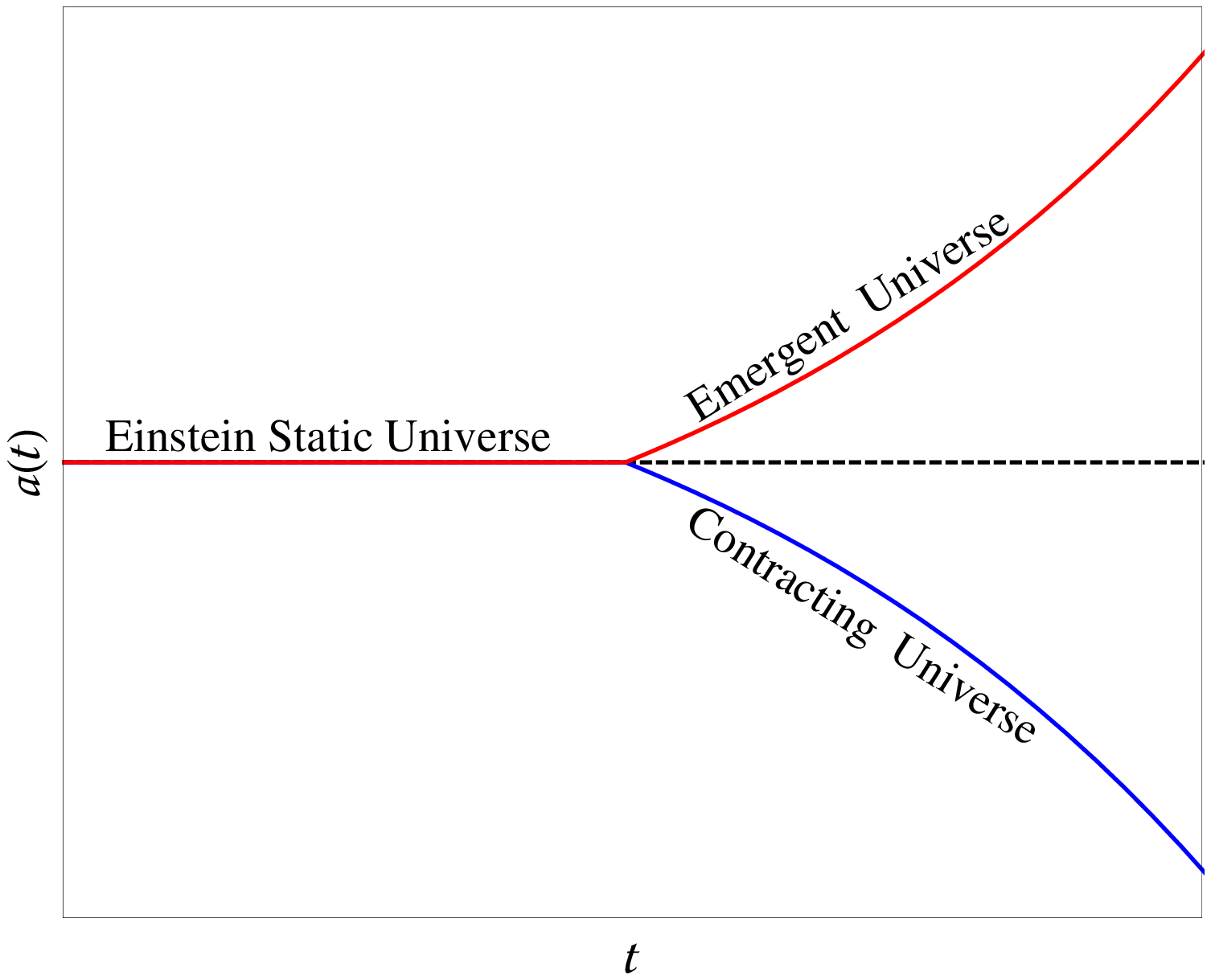}
\caption{The effective potential (left panel) is schematically shown for a universe consisting of
two components, one of which satisfies the strong energy condition $\rho+3P \geq 0$, while
the other violates it. The maxima of the effective potential corresponds to the
Einstein Static Universe (ESU) for which the expansion factor is a constant.
However ESU is unstable to small perturbations and can therefore be perturbed
either into an accelerating emergent cosmology, or into a contracting singular
universe (right panel).
}
\label{fig:ESU}
\end{center}
\end{figure}

An Einstein Static Universe (ESU) is possible to construct provided the universe is closed
and is filled with at least two components of matter: one of which satisfies the strong
energy condition (SEC) $\rho + 3P \geq 0$, whereas the other violates it. The
cosmological constant with $P = -\rho$ presents us with an example of the latter, as does
a massive scalar field which couples minimally to gravity.

In order to appreciate the existence of ESU, and therefore of emergent cosmology (EC),
consider the following set of equations which describe the dynamics of a FRW universe:
\ber
\left (\frac{\dot a}{a}\right )^2 &=& \frac{\kappa}{3}\sum_{i=1}^2\rho_i - \frac{\Bbbk}{a^2}~,
\quad ~~\Bbbk = 0, \pm 1 \label{eq:einstein} \\
\left (\frac{\ddot a}{a}\right ) &=& - \frac{\kappa}{6}\sum_{i=1}^2\left (\rho_i + 3P_i\right ),
\eer
where $\kappa = 8\pi G=\mpl ^{-2}$.
We shall assume that the pressure of the first component satisfies the SEC ($w_1 \geq
-1/3$), whereas that of the second component violates it ($w_2 < -1/3$), where $w =
P/\rho$ is the parameter of equation of state.

Equation (\ref{eq:einstein}) can be recast in terms of the effective potential $U(a)$ as
follows \cite{Sahni:1999gb}:
\beq
\frac{1}{2}{\dot a}^2 + U(a) = E \equiv -\frac{\Bbbk}{2}\;,
\label{eq:energy}
\eeq
where
\beq
U(a) = - \frac{\kappa}{6}a^2\sum_{i=1}^2\rho_i~.
\eeq

The ESU arises when the following two conditions are simultaneously satisfied:
\beq \label{eq:ESU}
{\dot a} = 0~, \qquad {\ddot a} = 0 \;.
\eeq
Substituting ${\dot a} = 0$ into (\ref{eq:energy}) gives
\begin{equation} \label{eq:crit_U}
U (a_E) = - \dfrac{\Bbbk}{2}\;,
\end{equation}
where $a_E$ is the scale factor for the ESU\@. The second condition implies that the ESU corresponds to an extremum of $U(a)$ (see FIG.~\ref{fig:ESU}):
\beq \label{eq:ESU2}
{\ddot a} = 0 ~\Rightarrow~ U'(a_E) = 0\;.
\eeq
In this case, one finds the following relationship between the curvature of the ESU, $a_E^{-2}$, and the densities
$\rho_i$\, for a closed universe ($\Bbbk=1$):
\beq
\frac{\rho_1}{\rho_2} = -\frac{1+3w_2}{1+3w_1}~, \qquad \kappa\rho_2 = \left
(\frac{1+3w_1}{w_1-w_2}\right ) \frac{1}{a_E^2}~. \label{eq:GR_ESU}
\eeq
For a closed $\Lambda$CDM ESU, we have $\rho_1 \equiv \rho_m$, $\kappa\rho_2 = \Lambda$,
$w_1 = 0$, $w_2 = -1$, and equation (\ref{eq:GR_ESU}) yields the familiar results
\beq
\Lambda = \frac{\kappa}{2}\rho_m = \frac{1}{a_E^2}~.
\eeq
On the other hand, for a closed $\mbox{\em radiation} + \Lambda$ ESU, we have $\rho_1
\equiv \rho_r$, $\kappa\rho_2 = \Lambda$, $w_1 = 1/3$, $w_2 = -1$, and one finds
\beq
\Lambda = \kappa\rho_r = \frac{3}{2a_E^2}~.
\eeq

The form of $U(a)$ in Fig.~\ref{fig:ESU} immediately suggests that the ESU with $U'(a_E)
= 0$, $U''(a_E) < 0$ is unstable, and infinitesimally small homogeneous perturbations
will either (i)~cause the ESU to contract towards a singularity, or (ii)~lead to an
accelerating expansion at late times, in other words, to {\em Emergent Cosmology\/}. In
this latter case, an exact solution describing the emergence of a $\Lambda$-dominated
accelerating universe from a $\mbox{\em radiation} + \Lambda$ based ESU is given by
\cite{Harrison:1967zz}
\beq
a(t) = a_i\left[ 1+ \exp{\left ( \frac{\sqrt{2} t}{a_i}\right )}\right]^{1/2} ,
\eeq
where ${a_i}^2 = 3/(2\Lambda) = 3/(2\kappa\rho_r)$.

The above example provided us with a toy model for the emergent scenario in which `inflation'
takes place eternally. A more realistic scenario can be constructed if one replaces $\Lambda$ by
the inflaton \cite{Ellis:2002we} thereby allowing inflation to end and the universe to reheat.

\subsection{Inflationary Emergent Cosmology}
\label{sec:inflation_can}

A spatially homogeneous scalar field minimally coupled to gravity
is described by the Lagrangian density
\beq
{\cal L} = \dfrac{1}{2}\dot \phi ^2 -\; V(\phi)~,
\label{eqn: Lagrangian can}
\eeq
and satisfies the evolution
equation
\begin{equation} \label{eq:scalar_evolution}
\ddot \phi+3H\dot \phi+V'\left(\phi\right)=0~.
\end{equation}
The energy density and pressure of such a field are given, respectively, by
\begin{subequations}
\label{eq:scalar}
 \begin{align}
  \rho_{\phi}=\dfrac{1}{2}\dot \phi ^2+V\left(\phi\right)\;, \label{eq:scalar_rho} \\
P_{\phi}=\dfrac{1}{2}\dot \phi ^2-V\left(\phi\right)\;, \label{eq:scalar_P}
 \end{align}
\end{subequations}
and, by virtue of (\ref{eq:scalar_evolution}), satisfy the usual energy conservation
equation
\begin{equation} \label{eq:energy_conservation}
\dot \rho_\phi =-3H \left(\rho_\phi + P_\phi \right)\;.
\end{equation}

\begin{figure}[htb]
\begin{center}
\includegraphics[width=.45\textwidth]{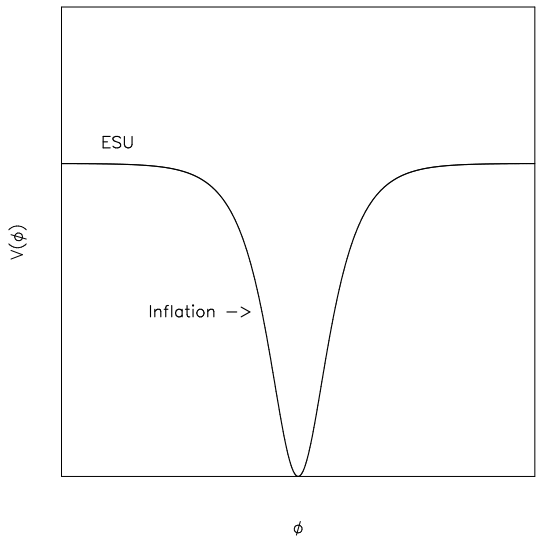}
\caption{The potential (\ref{eq:tanh2})
can give rise to
an emergent cosmology commencing in a {\em Einstein Static Universe} (ESU).
ESU is followed
by inflation after which $\phi$ oscillates and the universe reheats.
Since the potential is symmetric, emergent cosmology can be realized by the inflaton
rolling either towards the right, or the left.}
\label{fig:cosh}
\end{center}
\end{figure}

An emergent universe can be constructed if the inflaton potential $V (\phi)$ has one (or
more) flat wings.
\begin{itemize}
\item Consider first the potential
\beq
V (\phi) = V_0\tanh^{2p}{(\lambda\phi/\mpl)}\;;~~~p=1,2,3.... \label{eq:tanh2}
\eeq
Large absolute values of $\vert\lambda\phi\vert \gg \mpl$ lead to a flat potential with
$V (\phi) \simeq V_0$; see Fig.~\ref{fig:cosh}. 
In this case, the equation of motion (\ref{eq:scalar_evolution}) becomes ${\ddot\phi}
+ 3H{\dot\phi} \simeq 0~.$ Since $H = 0$ in an ESU, one immediately finds
${\dot\phi}^2 = {\rm const}$. Consequently, a scalar field sustaining an ESU behaves
exactly like a two-component fluid, with one component being stiff matter with
equation of state $P=\rho = {\dot\phi}^2/2$, while the other is the cosmological
constant $\Lambda \equiv \kappa V_0$.

Substituting $\rho_1 = {\dot\phi}^2/2$, $\rho_2 = V_0$, $w_1 = 1$, $w_2 = -1$ into
(\ref{eq:GR_ESU}), one gets
\beq
{\dot\phi}^2 = V_0 = \frac{2}{\kappa a_E^2}~, \label{eq:fine_tune}
\eeq
which demonstrates that the kinetic term must be {\em precisely matched\/} to the
asymptotic value of the potential term in an ESU scenario; see also \cite{Ellis:2002we}. Note that the post-ESU inflationary phase can last for a sufficiently long duration, as illustrated in Fig.~\ref{fig:gr_inf_em}.

\begin{figure}[ht]
\begin{center}
\includegraphics[width=0.65\textwidth]{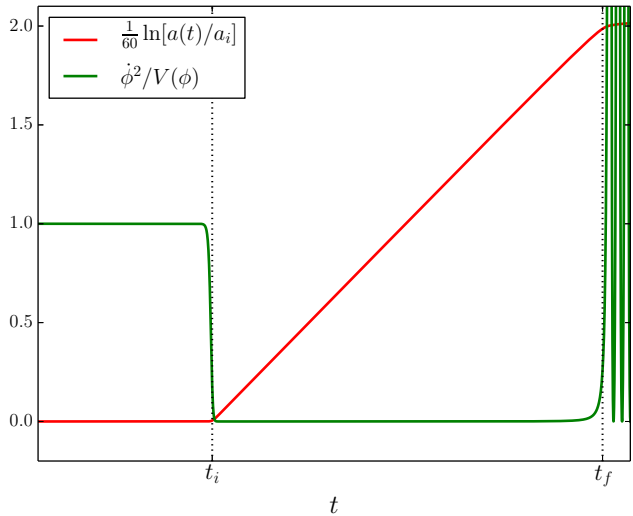}
\caption{The emergent scenario is illustrated by the scale factor (red curve) and slow roll 
parameter (green curve). Inflation commences when $ t \sim t_i$ and lasts for 120 e-folds. 
Note that the (quasi-flat) left branch of the potential in 
(\ref{eq:tanh2}) needs to have a small
positive slope in order for the ESU to end in inflation. If this 
branch has a negative slope, then the ESU will end in a contracting universe -- see Fig.~\ref{fig:ESU}.}
\label{fig:gr_inf_em}
\end{center}
\end{figure}

\subsection{CMB Constraints}
\label{subsec:GR_CMB}

Next, we shall demonstrate that emergent cosmology based on (\ref{eq:tanh2}) is consistent with recent
measurements of the Cosmic Microwave Background (CMB) made by the Planck satellite \cite{Ade:2013uln}; see also \cite{Labrana:2013oca}.

As shown in Fig.~\ref{fig:cosh}, ESU ends and inflation commences once the inflaton
field $\phi$
begins to roll down its potential. 
In the slow-roll approximation, the scalar ($n_{_S}$) and tensor ($n_{_T}$) spectral
indices and the tensor-to-scalar ratio ($r$) are given by
\ber
n_{_S}- 1 &=& -\frac{16\,p\,\lambda^{2}\l(p + 8 N p \lambda^{2} + \sqrt{1 + 8 p^{2} \lambda^{2}}\r)}
{\l(\,8 N p \lambda^{2} + \sqrt{1 + 8 p^{2} \lambda^{2}}\,\r)^{2}\, -\, 1}~, \label{eq: ns cosh pot} \\
&& \nn\\
n_{_T} &=& -\frac{16 p^{2}\lambda^{2}}{\l(\,8 N p \lambda^{2} + \sqrt{1 + 8 p^{2} \lambda^{2}}\,\r)^{2}\, -\, 1}~,
\label{eq: nt cosh pot} \\
{}&& \nonumber \\
r &=& \frac{128 p^{2} \lambda^{2}}{\l(\,8 N p \lambda^{2} + \sqrt{1 + 8 p^{2}
\lambda^{2}}\,\r)^{2}\, -\, 1}~, \label{eq: r cosh pot}
\eer
respectively, where $N$ is the number of $e$-folds counted from the end of inflation.
The values of $n_{_S}$ and $r$ are plotted as a function of $\lambda$ in
Fig.~\ref{fig: ns and r}. We find that CMB constraints are easily satisfied if
$\lambda > 0.1$.

\begin{figure}[t]
\begin{center}
\includegraphics[width=0.49\textwidth,height=5.23 cm]{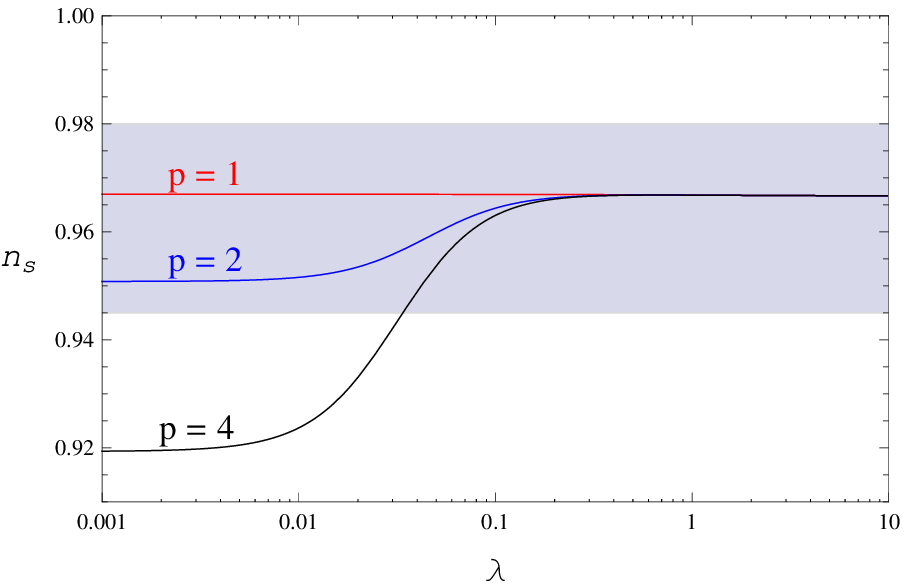}
\includegraphics[width=0.49\textwidth]{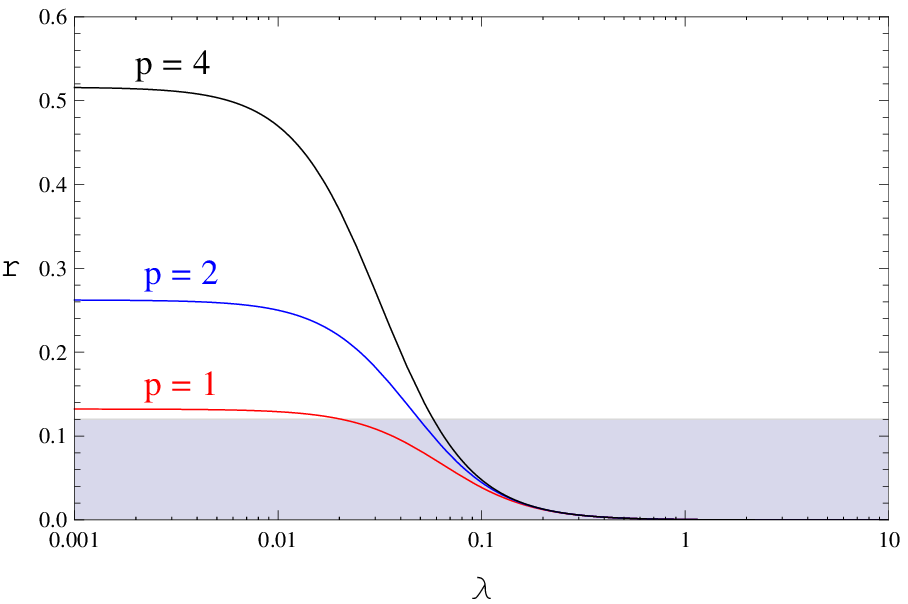}
\caption{In the left panel, the scalar spectral index $n_{_S}$ is plotted as a
function of $\lambda$ for three different values of the parameter $p$ in the
potential~(\ref{eq:tanh2}) while, in the right panel, the value of tensor-to-scalar is
plotted as a function of $\lambda$  for the same set of values for $p$. The number of
$e$-folds is $N = 60$. Note that when $\lambda > 0.1$, the scalar spectral index
approaches a constant value of $0.967$, whereas $r$ decreases as $\lambda^{-2}$.
The shaded region refers to 95\% confidence limits on $n_{_S}$ and $r$ determined by Planck \cite{Ade:2013uln}.}
\label{fig: ns and r}
\end{center}
\end{figure}

In this case, equations (\ref{eq: ns cosh pot})--(\ref{eq: r cosh pot}) are
simplified, respectively, to
\ber
n_{_S}- 1 &\simeq& -\frac{2}{N}~, \\
n_{_T} &\simeq& -\frac{1}{4 N^{2}\,\lambda^{2}}~,\\
r &\simeq& \frac{2}{N^{2}\,\lambda^{2}}~,
\eer
which are {\em independent\/} of the value of $p$ in (\ref{eq:tanh2}).

Interestingly, the above expression for $n_{_S}$ is {\em exactly the same\/} as in
Starobinsky's  $R + R^{2}$ model \cite{Starobinsky:1980te}.  In fact, for $\lambda = 1/\sqrt{6}$,
the expressions for $n_{_T}$ and $r$ match those in the Starobinsky
inflation\,! 

\begin{figure}[t]
\begin{center}
\includegraphics[width=0.5\textwidth]{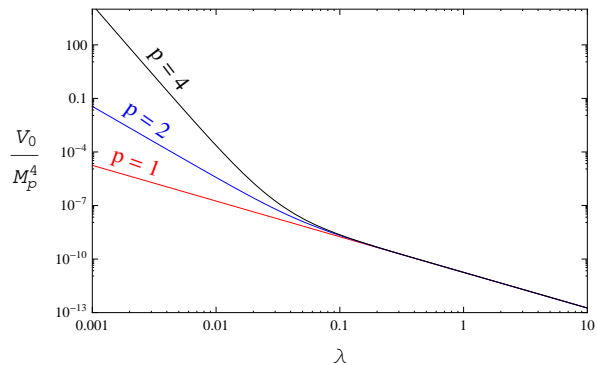}
\caption{The value of $V_{0}$ is plotted as a function of $\lambda$ for three
different values of the parameter $p$ in the potential~(\ref{eq:tanh2}). The number of
$e$-folds is taken to be $N = 60$.} \label{fig: v0}
\end{center}
\end{figure}

The value of the parameter $V_{0}$ in the potential~(\ref{eq:tanh2}) can be fixed
using CMB normalization, which gives $P_{_{S}}(k_{\ast}) = 2.2\times10^{-9}$ at the
pivot scale $k_{\ast} = 0.05\,\mathrm{Mpc}^{-1}$~\cite{Ade:2013uln}. One finds
\beq
\frac{V_{_0}}{\mpl^{4}}\,=\, \frac{192 \pi^{2} p^{2} \lambda^{2}\l(8 N p \lambda^{2}
+ \sqrt{1 + 8 p^{2} \lambda^{2}} + 1\r)^{2p}P_{_{S}}(k_{\ast})} {\l[\l(8 N p
\lambda^{2} + \sqrt{1 + 8 p^{2} \lambda^{2}}\r)^{2} - 1\r]^{p + 1}}~,
\eeq
which is plotted as a function of $\lambda$ in Fig.~\ref{fig: v0}. Since $\lambda >
0.1$ is preferred from the CMB bounds on $n_{_S}$ and $r$, it follows that $V_0 <
10^{-8}\mpl^{4}$.

\item Consider next the potential
\beq
V (\phi) = V_0\left[\exp\l(\frac{\beta\,\phi}{\mpl}\r)\,-\, 1\right]^2~.
\label{eq:pot_exp}
\eeq
shown in Fig.~\ref{fig:exp} and earlier discussed in \cite{Ellis:2003qz}.

For suitable values of $V_0$ and $\beta$, inflation can occur both from the (flat)
left branch and from the (steep) right branch of this potential.

In this section, we shall focus on the left branch since this is the branch which
leads to an ESU, and hence to emergent cosmology in the GR context. For $\beta >
0.1$, one finds
\ber
n_{_S}- 1 &\simeq& -\frac{2}{N}\,-\,\frac{3}{\beta^{2}\,N^{2}}~, \\
n_{_T} &\simeq& -\frac{1}{\beta^{2}\,N^{2}}~, \\
r &\simeq& \frac{8}{\beta^{2}\,N^{2}}~.
\label{eq: ns r exp-pot}
\eer
If $N \geq 60$, then $\beta > 0.14$ is required in order to satisfy the CMB bound $r
< 0.11$.
The value of $V_{0}$ is determined from the relation
\beq
\frac{V_0}{\mpl^{4}}\,=\,12\pi^{2}\l(\frac{P_{_{S}}(k_{\ast})}{N^{2}\,\beta^{2}}\r)~.
\eeq
Substituting here $\beta = 0.5$ gives $V_{_0}\,=\,2.9\times 10^{-10}\,\mpl^{4}$,
~$n_{_S} \simeq 0.96$ and $r \simeq 0.009$, which are in good agreement with the
\emph{Planck\/} results \cite{Ade:2013uln}.
Note that Starobinsky's  $R + R^{2}$ model of inflation~\cite{Starobinsky:1980te} can equivalently be described by a scalar field with potential~(\ref{eq:pot_exp}) but with $\beta = \sqrt{2/3}$; see \cite{Maeda:1987xf}.

\begin{figure}[htb]
\begin{center}
\includegraphics[width=.45\textwidth]{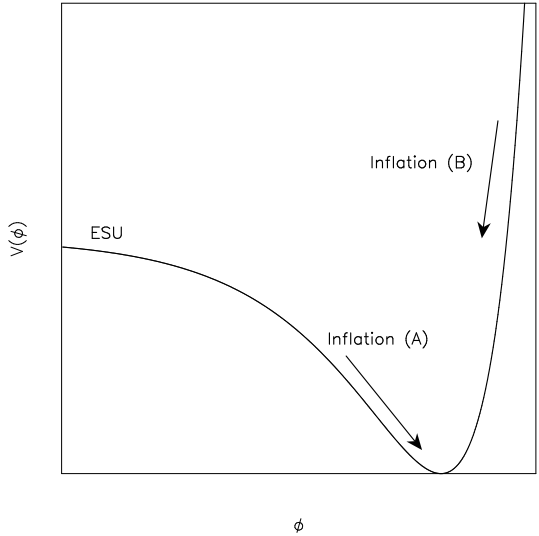}
\caption{\small Potentials (\ref{eq:pot_exp}) and (\ref{eq:bw_inf_phi2}) can give rise to
an emergent cosmology from a {\em Einstein Static Universe} (ESU)\@.
If the ESU is GR-based, as in section \ref{sec:emergent_GR}, then inflation will occur from the left (A) branch.
For modified gravity models explored in section \ref{sec:emergent_grav}, inflation can occur from both A and B branches.
}
\label{fig:exp}
\end{center}
\end{figure}
\end{itemize}

To summarize, we have demonstrated that viable models of emergent cosmology are possible
to construct in a general relativistic setting. Unfortunately, although our models easily
satisfy CMB constraints, they run into severe fine-tuning issues --- since the initial
value of ${\dot\phi_i}^2$ must precisely match the asymptotic form of the inflaton
potential --- a consequence of the ESU being unstable in GR; see (\ref{eq:fine_tune}). (Note that emergent cosmology can also be sustained by a spinor field, as demonstrated in \cite{Cai:2013rna}.)

As we shall show in the next section, emergent cosmology based on modified gravity can
circumvent this difficulty since an ESU can be associated with a {\em stable\/} fixed point in this case.

\section{Emergent cosmology from modified gravity}\label{sec:emergent_grav}

In this section we focus on emergent cosmology based on modified gravity. We shall
confine our discussion to three distinct models of modified gravity, namely:
(i)~Braneworld cosmology and its generalizations, (ii)~Loop Quantum Cosmology (LQC), and
(iii)~a phenomenological model of {\em asymptotically free\/} gravity.

\subsection{Emergent scenario in Braneworld Cosmology}
\label{sec:bouncing_bw}
Consider first the following generalization of the Einstein equations which is known
to give rise to a non-singular bouncing cosmology \cite{Sahni:2012er}:
\ber
H^2 &=& \frac{\kappa}{3}\rho\left\lbrace 1 - \left(\frac{\rho}{\rho_c}\right )^m\right\rbrace
-\frac{\Bbbk}{a^2}~, \quad \kappa = 8\pi G=\mpl^{-2}~, \label{eq:bounce01} \\
\frac{\ddot a}{a} &=& -\frac{\kappa}{6}\left\lbrack (\rho+3P) - \bigg\lbrace(3m+1)\rho +
3(m+1)P\bigg\rbrace\left(\frac{\rho}{\rho_c}\right )^m\right\rbrack~. \label{eq:bounce02}
\eer
The bounce occurs near the critical density $\rho_c$.
When $m=1$, these equations reduce to the following set of equations describing the
dynamics of a Friedmann--Robertson--Walker (FRW) metric on a brane \cite{Shtanov:2002mb}
\ber
H^2 &=& \frac{\kappa}{3}\rho\left\lbrace 1 - \frac{\rho}{\rho_c}\right\rbrace
-\frac{\Bbbk}{a^2}~, \label{eq:bounce1}\\
\frac{\ddot a}{a} &=& -\frac{\kappa}{6}\left\lbrace (\rho+3P) -
\frac{2\rho}{\rho_c}(2\rho+3P)\right\rbrace~, \label{eq:bounce2}
\eer
where $\rho_c$ is related to the five-dimensional Planck mass \cite{Shtanov:2002mb}.
Both (\ref{eq:bounce01}), (\ref{eq:bounce02}) \& (\ref{eq:bounce1}), (\ref{eq:bounce2})
reduce to the FRW limit when $\rho_c \to \infty$, while
 (\ref{eq:bounce1}), (\ref{eq:bounce2}) are valid in LQC when $\Bbbk=0$ \cite{Ashtekar:2006uz}.
In the presence of several components one replaces $\rho = \sum_i\rho_i$ and $P = \sum_i
P_i$ in all these equations.

The emergent scenario which we discuss in this paper is based on `normal' matter
that satisfies the strong energy condition, along with a scalar field that violates
it. As pointed out in the previous section, emergent cosmology arises when the scalar
field potential has a flat wing along which $V(\phi)$ is effectively a constant, which we
denote by $\Lambda$. Equation (\ref{eq:bounce1}) in such a two-component universe becomes
\begin{equation} \label{eq:frw_bwL}
H^2=\left(\dfrac{\kappa}{3}\rho+\dfrac{\Lambda}{3}\right)\left(1-\dfrac{\rho}{\rho_c}-\dfrac{\Lambda}{\kappa
\rho_c}\right)-\dfrac{\Bbbk}{a^2}\;.
\end{equation}
For a spatially closed universe, this equation can be recast as (\ref{eq:energy})
with the effective potential
\begin{equation} \label{eq:effective_potential}
U (a) = -\dfrac{1}{2}a^2\left(\dfrac{\kappa}{3}\rho+\dfrac{\Lambda}{3}\right)\left(1-\dfrac{\rho}{\rho_c}-\dfrac{\Lambda}{\kappa
\rho_c}\right)\;,
\end{equation}
where
\begin{equation} \label{eq:gen_matter}
\dfrac{\kappa}{3}\rho=\dfrac{A}{a^l}~, \quad l = 3\left(1+w\right)\;, \quad w > -1/3~.
\end{equation}
The effective potential $U (a)$ in \eqref{eq:effective_potential} exhibits a pair of
extremes (a maximum and a minimum). The scale factor
and the energy density corresponding to these extremes are given, respectively, by
\begin{equation} \label{eq:gen_a^l}
a^l_{\pm}=\dfrac{3A\left(l-2\right)\left(\kappa \rho_c-2\Lambda\right)}{4\Lambda \left(\kappa \rho_c-\Lambda\right)}\left[1\pm\sqrt{1-\dfrac{16\left(l-1\right)\Lambda \left(\kappa \rho_c-\Lambda\right)}{\left(l-2\right)^2 \left(\kappa \rho_c-2\Lambda\right)^2}}~\right]\;,
\end{equation}
\begin{equation} \label{eq:gen_rho}
\rho_{\pm}=\dfrac{\left(l-2\right)}{4\kappa\left(l-1\right)}\left(\kappa \rho_c-2\Lambda\right)\left[1\pm\sqrt{1-\dfrac{16\left(l-1\right)\Lambda \left(\kappa \rho_c-\Lambda\right)}{\left(l-2\right)^2 \left(\kappa \rho_c-2\Lambda\right)^2}}~\right]\;.
\end{equation}
Note that $a_- < a_+$ and $\rho_- < \rho_+$. The extremes can exist as long as the term under the square root in
\eqref{eq:gen_a^l} and \eqref{eq:gen_rho} remains positive, which imposes the following
conditions on $\Lambda$\,:
\begin{equation} \label{eq:crit_L}
\Lambda<\Lambda_{\rm crit} = \dfrac{\kappa \rho_c}{2}\left[1-\sqrt{\dfrac{q}{q+4}}~\right]~,
\quad \mbox{where} \quad q=\dfrac{16\left(l-1\right)}{\left(l-2\right)^2}\;,
\end{equation}
or
\begin{equation} \label{eq:crit_highL}
\Lambda>\kappa \rho_c\;.
\end{equation}
For the first condition in \eqref{eq:crit_L}, $\left(a_-,\rho_+\right)$ is associated with the minimum of $U(a)$ while
$\left(a_+,\rho_-\right)$ is associated with the maximum of $U(a)$. On the other hand, for the second condition in \eqref{eq:crit_highL} ( i.e. $\Lambda>\kappa \rho_c$), only $a_+$ in \eqref{eq:gen_a^l} and $\rho_-$ in \eqref{eq:gen_rho} survive, and account for a minima in $U(a)$. However, according to \eqref{eq:frw_bwL},
a spatially closed universe ($\Bbbk=1$) does not support a solution with $\Lambda\geq\kappa \rho_c$. Hence this minima is shown, in Appendix \ref{app:bw_open}, to correspond to an ESU in a {\em spatially open} universe ($\Bbbk=-1$).

In this section our focus has been on the spatially closed universe ($\Bbbk=1$) for which
we can obtain static (or oscillating) solutions in the regime $\Lambda<\Lambda_{\rm
crit}<\kappa \rho_c$; i.e. according to \eqref{eq:crit_L}. As noted in the previous
section, a scalar field rolling along a flat potential with $V'(\phi) \equiv 0$, behaves
just like stiff matter plus a cosmological constant. Our analysis in the following
subsections will therefore focus on this important case.

\subsubsection{The effective potential in the presence of stiff matter only}
\label{subsec:stiffL0}

From \eqref{eq:gen_a^l} and \eqref{eq:gen_rho}, one can see that, as $\Lambda$ tends to
zero, $a_+$ approaches infinity and $\rho_-$ declines to zero, but both $a_-$ and
$\rho_+$ remain finite. In other words, in the absence of the cosmological constant, the
(unstable) maximum of the effective potential $U (a)$ disappears, while the stable
minimum remains in place. This new feature of brane cosmology distinguishes it from GR,
in which the effective potential has no extreme value when $\Lambda=0$. As we shall see
later, the persistence of a stable minimum of $U (a)$ in the absence of $\Lambda$ carries
over to other modified gravity models as well.

\begin{figure}[t]
\begin{center}
\scalebox{0.75}[0.75]{\includegraphics{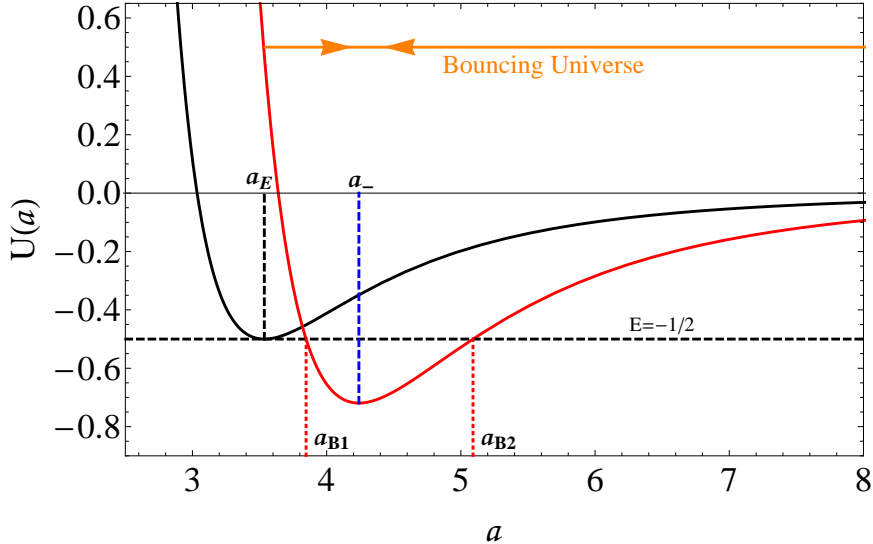}}
\caption{The effective potential is shown for a universe consisting of stiff matter with
$P = \rho$. Two values of the parameter $A$ are chosen.
For $A=A_E$ (black curve), the minimum of $U(a)$ lies precisely on
 the $E = -1/2$ line and the solution is static at $(a_E,\rho_E)$ which corresponds to the
Einstein Static Universe (ESU)\@. For $A > A_E$, the minimum of $U(a)$ shifts towards
higher $a$ and lies below the $E = -1/2$ line. In this case the ESU is absent and the universe
oscillates around $(a_-,\rho_E)$, i.e. around the minimum of the effective potential (red curve).
Note that the value of the energy density at the minimum of $U(a)$ is a fixed quantity
and is given by (\ref{eq:stiff_rhoE}). From the figure it is clear that the motion of a
closed universe is always bounded, and the turning points of its scale factor are
$a_{B1}$ and $a_{B2}$. By contrast, a spatially flat/open universe bounces at small $a$
but need not turn around and collapse. It is important to note that the
above form of the potential $U(a)$ is robust and remains qualitatively unchanged if stiff matter is
replaced by any other form of matter satisfying the SEC, \ie, for $\rho \propto A/a^l$ with $l
\geq 2$ in (\ref{eq:gen_matter}). In our illustration we assume $\kappa=1$, $\rho_c=1$.} \label{fig:stiff_esu}
\end{center}
\end{figure}

In the absence of the cosmological constant, and with a universe consisting only of stiff matter ($P = \rho$), the minimum of $U (a)$ is determined from \eqref{eq:gen_a^l} to be at
\begin{equation} \label{eq:stiff_a-}
a^6_-=\dfrac{15}{2}\dfrac{A}{\kappa \rho_c}\;.
\end{equation}
As mentioned earlier in \eqref{eq:ESU} -- \eqref{eq:ESU2}, the Einstein Static 
Universe (ESU) arises when ${\dot a} = 0$ and ${\ddot a} = 0$. 
As a result, the ESU is associated with an extremum of $U(a)$, so that
 $U'(a_E) = 0$ 
  and $U(a_E)=-\Bbbk/2$ are jointly satisfied. In other words, $a_E$ (the scale factor at the ESU) is the critical
value of $a_-$ for which the minimum of $U (a)$ lies on the $E=-\Bbbk/2$ line (see Fig.~\ref{fig:stiff_esu}). 

The stiff matter density at the minimum of $U (a)$ is determined from \eqref{eq:gen_rho} to be
\begin{equation} \label{eq:stiff_rhoE}
\rho_+=\rho_E=\dfrac{2}{5}\rho_c\;,
\end{equation}
where $\rho_c$ is the braneworld constant defined in (\ref{eq:bounce1}), (\ref{eq:bounce2}). The fact that $\rho_+$ is independent of
$A$ implies that, for a universe which is oscillating about $a_-$,  the density of matter
at $U (a_-)$ {\em is the same\/} as in ESU\@. (This interesting result is {\em
independent\/} of the equation of state of matter and holds for $l \geq 2$ in
\eqref{eq:gen_matter}). Hence we have denoted $\rho_+$ by $\rho_E$, the matter density at ESU. Note that for a closed
universe ($\Bbbk=1$) if $U (a_-) <-\dfrac{1}{2}$, then ESU will no longer be a solution of the field equations. Instead,
the universe will oscillate around $U (a_-)$. With this result taken into account, the minimum in the effective
potential $U (a)$ becomes
\begin{equation} \label{eq:min_U_pri}
U (a_-) = -\dfrac{\kappa}{6}a_-^2\rho_E \left(1-\dfrac{\rho_E}{\rho_c}\right)\;,
\end{equation}
while the ESU condition \eqref{eq:crit_U} for a closed universe ($\Bbbk = 1$) takes the form
\begin{equation} \label{eq:U_esu}
U (a_E) = -\dfrac{1}{2}=-\dfrac{\kappa}{6}a_E^2\rho_E \left(1-\dfrac{\rho_E}{\rho_c}\right)\;.
\end{equation}
Comparing \eqref{eq:min_U_pri} and \eqref{eq:U_esu} we easily find
\begin{equation} \label{eq:min_U}
U (a_-) = -\dfrac{1}{2}\dfrac{a_-^2}{a_E^2}\;.
\end{equation}
By using \eqref{eq:stiff_rhoE} and \eqref{eq:U_esu}, $a_E$ is determined to be
\begin{equation} \label{eq:stiff_aE}
a_E^2=\dfrac{25}{2\kappa \rho_c}\;.
\end{equation}
By using \eqref{eq:energy}, with $\Bbbk = 1$, the condition to have a physical solution becomes
\begin{equation} \label{eq:solution_condition}
U (a_-) \leq -\dfrac{1}{2} ~~\Rightarrow~~ a_-\geq a_E~~\Rightarrow~~ A \geq A_E\;,
\end{equation}
where $A_E$ is the value of the parameter $A$ in \eqref{eq:gen_matter} for which the ESU
conditions are met. In the last inequality of \eqref{eq:solution_condition}, we have used
\eqref{eq:stiff_a-} and the fact that $a_E$ is simply the value of $a_-$ at ESU.

The effective potential is shown in Fig.~\ref{fig:stiff_esu} for a universe containing
stiff matter. As $a \to 0$, the potential $U (a)$ diverges causing the universe
to bounce and avoid the big bang singularity. For large $a$, the
braneworld equations \eqref{eq:bounce1}, \eqref{eq:bounce2} reduce to the GR limit
($\rho/\rho_c \ll 1$), and $U (a)$ asymptotically approaches zero. The motion of a
spatially closed universe is therefore always bounded. The minima in $U (a)$ represent
stable universes. Two values of the parameter $A$ in (\ref{eq:gen_matter}) are
considered. For $A=A_E$, the universe is static (ESU) at $a = a_E$ (black curve), while,
for $A > A_E$, the universe oscillates around $a = a_-$ (red curve).

\subsubsection{The effective potential in the presence of stiff matter and a cosmological constant }
\label{subsec:stiffL}

For stiff matter and $\Lambda$, the scale factor and stiff matter density at the extremes
of $U(a)$ are determined by setting $l = 6$ in \eqref{eq:gen_a^l} and \eqref{eq:gen_rho}.
Consequently, one obtains
\begin{equation} \label{eq:stiffL_a^6}
a^6_{\pm}=\dfrac{3A\left(\kappa \rho_c-2\Lambda\right)}{\Lambda \left(\kappa \rho_c-\Lambda\right)}\left[1\pm\sqrt{1-\dfrac{5\Lambda \left(\kappa \rho_c-\Lambda\right)}{\left(\kappa \rho_c-2\Lambda\right)^2}}~\right]
\end{equation}
and
\begin{equation} \label{eq:stiffL_rho}
\rho_{\pm}=\dfrac{1}{5\kappa}\left(\kappa \rho_c-2\Lambda\right)\left[1\pm\sqrt{1-\dfrac{5\Lambda \left(\kappa \rho_c-\Lambda\right)}{\left(\kappa \rho_c-2\Lambda\right)^2}}~\right]\;.
\end{equation}
As before,
$\left(a_-,\rho_+\right)$ is associated with the minimum of $U(a)$
while $\left(a_+,\rho_-\right)$ is associated with the maximum of $U(a)$, and
$a_- < a_+$, $\rho_- < \rho_+$. The unstable ESU associated with the maximum is GR-like and was
previously encountered in section \ref{sec:emergent_GR}. Hereafter we shall focus on the stable ESU that is associated with the minimum of $U(a)$.

It follows from \eqref{eq:stiffL_a^6} that $a_{\pm}$ increases as the parameter $A$
increases, while $\rho_{\pm}$ is constant for a given $\Lambda$. This
 empowers \eqref{eq:min_U} and \eqref{eq:solution_condition} to hold true even in the completely general case  when $\Lambda \neq 0$; see \eqref{eq:gen_a^l} and \eqref{eq:gen_rho}.
For a certain value of the parameter $A=A_E$, ESU conditions are satisfied and the corresponding scale factor can be determined using \eqref{eq:crit_U} and \eqref{eq:stiffL_rho} to be
\begin{equation} \label{eq:stiffL_aE}
a_E^2=\dfrac{3}{\left(\kappa \rho_+ + \Lambda \right)\left(1-\dfrac{\rho_+}{\rho_c}-\dfrac{\Lambda}{\kappa \rho_c}\right)}\;.
\end{equation}

Note that the condition for the existence of two extreme values in
$U(a)$, namely \eqref{eq:crit_L}, is simplified to
\begin{equation} \label{eq:crit_stiffL}
\Lambda<\Lambda_{\rm crit}=\kappa \rho_c\left[\dfrac{1}{2}-\dfrac{\sqrt{5}}{6}\right]\;.
\end{equation}

\begin{figure}[t]
\centering
\subfigure[]{
\includegraphics[width=0.488\textwidth]{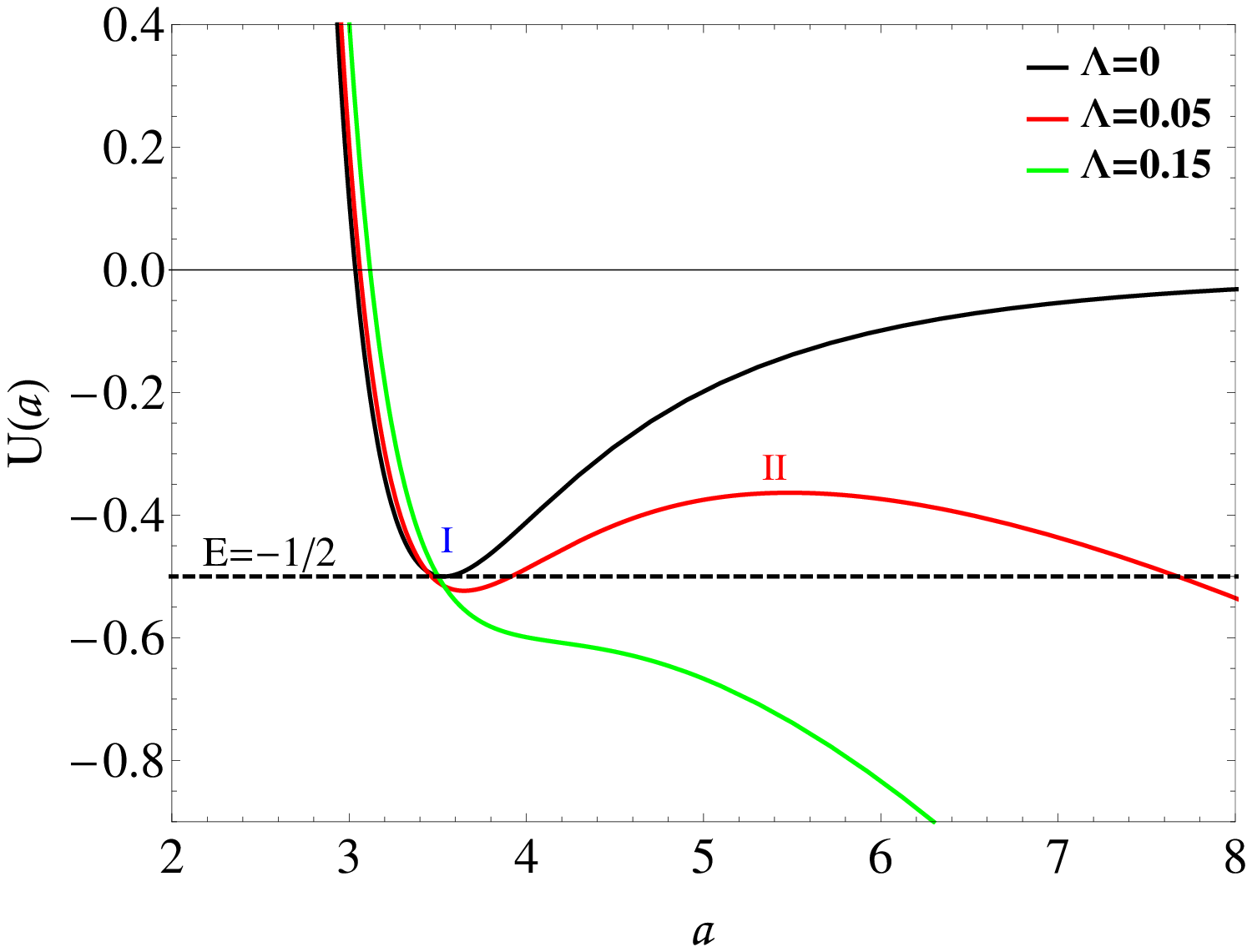}\label{fig:stiff_lam}}
\subfigure[]{
\includegraphics[width=0.488\textwidth,height=5.817 cm]{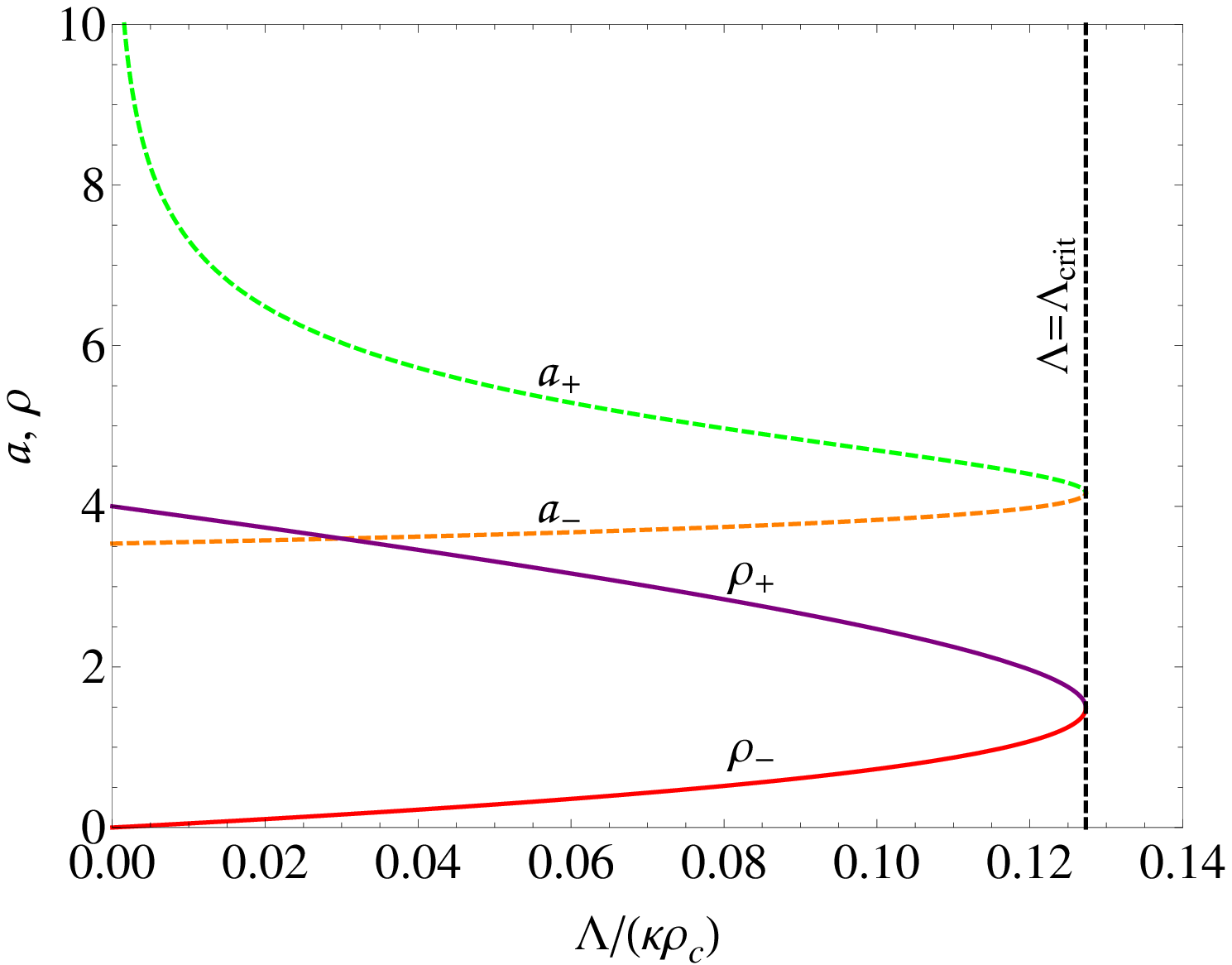}\label{fig:stiffL_move}}
\caption{\label{fig:stiffL_lam_inc}{\bf (a)} The effective potential is plotted for different values of $\Lambda$ for a two-component universe consisting of stiff matter and a cosmological constant. (i)~For $\Lambda=0$, there
is only a single minimum representing a stable fixed point denoted by I (black); see
section \ref{subsec:stiffL0}. (ii)~For $\Lambda>0$, there appears a maximum associated
with the unstable fixed point, denoted by II (red). (iii)~As $\Lambda$ further increases,
these two fixed points approach each other as illustrated in Fig.~\ref{fig:stiffL_move}.
At $\Lambda=\Lambda_{\rm crit}$, they merge to produce an inflection point in $U (a)$,
and, for  $\Lambda>\Lambda_{\rm crit}$, no fixed points are present in the system
(green line in left panel).
{\bf (b)} The scale factor corresponding to the minimum (maximum) of $U(a)$, denoted by I (II) in
Fig.~\ref{fig:stiff_lam} and by $a_-$ ($a_+$) in (\ref{eq:gen_a^l}), is plotted as a
function of $\Lambda$ in the right panel.
The right panel also shows the value of the stiff matter density at the minimum
(maximum) of $U(a)$, and denoted by $\rho_+$ ($\rho_-$) in (\ref{eq:gen_rho}). As
$\Lambda$ increases, the two fixed points, stable and unstable, move towards each other
and merge at $\Lambda=\Lambda_{\rm crit}$, beyond which no fixed point exists.
Units of $\kappa=1$ and $\rho_c=1$ are assumed together with a suitable choice of $A$,
for purposes of illustration.
The unit along the y-axis is arbitrary.
}
\end{figure}

The form of the effective potential is very sensitive to the value of $\Lambda$, as shown
in Fig.~\ref{fig:stiff_lam}. From this figure, we see that the potential $U(a)$ diverges
as $a \to 0$, which allows the universe to bounce at small $a$ and avoid the big bang
singularity. For $\Lambda=0$, there exists only a single minimum in $U (a)$, as discussed
in section \ref{subsec:stiffL0}. The maximum,  representing an unstable fixed point,
appears when $\Lambda > 0$. As $\Lambda$ increases, the maxima and the minima in
$U\left(a\right)$ approach each other. They merge when $\Lambda=\Lambda_{\rm crit}$,
which results in an inflection point in the effective potential, see
Fig.~\ref{fig:stiffL_move}. For $\Lambda>\Lambda_{\rm crit}$, the potential has no extreme
value at finite $a$,  which is indicative of the absence of fixed points in the
corresponding dynamical system.\footnote{ Note that $U(a)$ exhibits a minimum also for
$\Lambda>\kappa \rho_c$. However in this case $U(a)>0$ which
 does not allow an ESU in a spatially closed universe, but permits an ESU in an open universe,
as shown in Appendix \ref{app:bw_open}.}

While figures \ref{fig:stiff_lam} and \ref{fig:stiffL_move} have been constructed for
matter with $P=\rho$ (in view of possible links to the kinetic regime during
pre-inflation), the main results of our analysis, based on \eqref{eq:gen_a^l},
\eqref{eq:gen_rho} and \eqref{eq:crit_L}, should remain qualitatively true for any matter
component with $w > -1/3$, with $\Lambda_{\rm crit}$ depending upon $w$.

\subsubsection{Phase space analysis}

The energy conservation equation \eqref{eq:energy_conservation} for stiff matter reads
\begin{equation}\label{eq:rhos_dot}
 \dot \rho=-6H\rho\;.
\end{equation}
Differentiating \eqref{eq:frw_bwL} with respect to time
leads to the Raychaudhuri equation:
\begin{equation}\label{eq:stiff_Hdot}
\dot H=-\dfrac{\kappa}{3}\left[\left(2\rho-\dfrac{\Lambda}{\kappa}\right)-\dfrac{\left(\kappa\rho+\Lambda\right)}{\kappa\rho_c}\left(5\rho-\dfrac{\Lambda}{\kappa}\right)\right]-H^2\;.
\end{equation}
The fixed points of the dynamical system described above are characterized by
\begin{equation}\label{eq:fixed_point_condition}
\dot \rho=0 ~~\Rightarrow~~ H=0 \quad \mbox{and} \quad \dot H=0 ~~\Rightarrow~~ \ddot a=0\;,
\end{equation}
precisely the same conditions as for ESU\@. For $\Lambda>0$, the two fixed points in the
$\left(\rho,H\right)$ plane are $\left(\rho_+,0\right)$ and $\left(\rho_-,0\right)$,
where $\rho_{\pm}$ are given by \eqref{eq:stiffL_rho}. The condition for the existence of
these fixed points was described earlier in \eqref{eq:crit_stiffL}. For $\Lambda=0$, only
one fixed point $\left(\rho_+,0\right)$ exists. To analyse stability, the nonlinear
dynamical system should be linearized using the {\em linearization theorem} (for instance, as
described in \cite{Arrowsmith:1992}) near the two fixed points. The eigenvalues of the
Jacobian matrix of the linearized system (as described in Appendix \ref{app:linearization}) at the two fixed points are
\begin{equation}\label{eq:stiff_eigen}
 \lambda_{\Rmnum{1},\Rmnum{2}}^2=4\kappa \rho_{\pm}\left[1-\dfrac{1}{\rho_c}\left(5\rho_{\pm}+\dfrac{2\Lambda}{\kappa}\right)\right]\;,
\end{equation}
where the indices \Rmnum{1} and \Rmnum{2} stand for the two fixed points
$\left(\rho_+,0\right)$ and $\left(\rho_-,0\right)$, respectively. By using the values of
$\rho_{\pm}$ from \eqref{eq:stiffL_rho}, the properties of the eigenvalues at the two
fixed points are listed in Table~\ref{table:stiff_eigen} for $\Lambda<\Lambda_{\rm crit}$
in \eqref{eq:crit_stiffL}.

\begin{table}[htb]
\centering
 \begin{tabular}{C{2.5cm}C{2.5cm}C{2.5cm}C{5cm} }
 \\[-1.0em]
 \hline \hline
 \\[-0.5em]
  Fixed point & $\left(\rho,H\right)$ & Eigenvalues & Stability \\ [0.6ex]
  \hline
  \\[-0.6em]
  \Rmnum{1} & $\left(\rho_+,0\right)$ & $\lambda_{\Rmnum{1}}^2<0$ & Centre type stable point \\[2.7ex]
  \Rmnum{2} & $\left(\rho_-,0\right)$ & $\lambda_{\Rmnum{2}}^2>0$ & Saddle type unstable point\\[1.0 ex]
  \hline
 \end{tabular}
 \caption{Eigenvalues of the Jacobian matrix and stability of the fixed points for stiff matter
and a cosmological constant.}
 \label{table:stiff_eigen}
\end{table}

For the fixed point \Rmnum{2} $\left(\rho_-,0\right)$, real eigenvalues with opposite sign
imply that the fixed point is a saddle and therefore unstable. The eigenvalues for the
fixed point \Rmnum{1} $\left(\rho_+,0\right)$ are imaginary and complex conjugates of
each other, suggesting that the fixed point in this case is of centre type (stable), as
expected. But the linearization theorem does not guarantee this for linearized systems of
centre type \cite{Arrowsmith:1992}, hence this result needs to be confirmed numerically. The
phase portrait for general matter with $w>-1/3$ can be studied in a similar fashion. In
general, the energy conservation equation becomes
\begin{equation}\label{eq:gen_rho_dot}
 \dot \rho=-3H\rho\left(1+w\right)\;,
\end{equation}
while the Raychaudhuri equation is given by
\begin{equation}\label{eq:gen_Hdot}
 \dot H=-\dfrac{\kappa}{3}\left[\dfrac{\rho}{2} \left(1+3w\right)
 -\dfrac{\Lambda}{\kappa}-\dfrac{\left(\kappa\rho+\Lambda\right)}{\kappa\rho_c}
 \left\lbrace \left(2+3w\right)\rho-\dfrac{\Lambda}{\kappa}\right\rbrace\right]-H^2\;.
\end{equation}

Again, for $0<\Lambda<\Lambda_{\rm crit}$ the system possesses two fixed points
\Rmnum{1}, \Rmnum{2} in the $\left(\rho,H\right)$ plane, given by $\left(\rho_+,0\right)$
and $\left(\rho_-,0\right)$, respectively, where $\rho_{\pm}$ and $\Lambda_{\rm crit}$
were determined in \eqref{eq:gen_rho} and \eqref{eq:crit_L}. The eigenvalues of the
linearized system at the two fixed points are found to be
\begin{equation}\label{eq:gen_eigen}
\lambda_{\Rmnum{1},\Rmnum{2}}^2=\left(1+3w\right)\left(1+w\right)
\dfrac{\kappa }{2}\rho_{\pm}\left[1-\dfrac{1}{\rho_c}\left\lbrace
\dfrac{4\left(2+3w\right)}{\left(1+3w\right)}\rho_{\pm}+\dfrac{2\Lambda}{\kappa}\right\rbrace\right]~.
\end{equation}
Using the values of $\rho_{\pm}$ from \eqref{eq:gen_rho}, it can be shown that, for
$\left(\rho_-,0\right)$, we have $\lambda_{\Rmnum{2}} ^2>0$, and real eigenvalues of
opposite sign confirm the fixed point \Rmnum{2} to be a saddle. For \Rmnum{1}, the
eigenvalues are again imaginary and complex conjugates of each other, given by
$\lambda_{\Rmnum{1}}^2<0$. Although this suggests that the fixed point \Rmnum{1} is a
centre, the linearization theorem does not assure this, hence this result needs to be
confirmed numerically.

\begin{figure}[t]
\centering
\subfigure[\ For $\Lambda=0$]{
\includegraphics[width=0.487\textwidth]{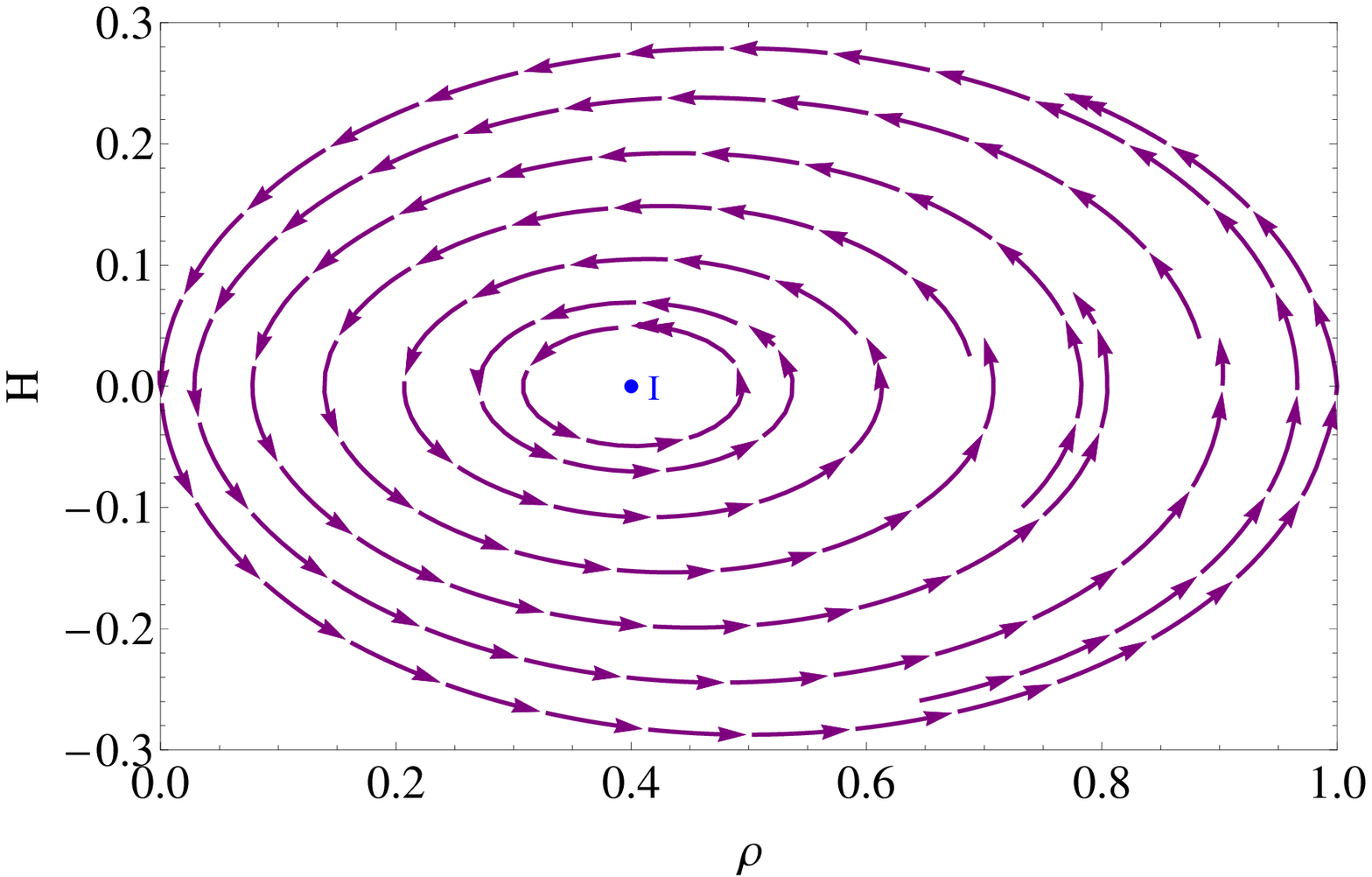}\label{fig:phase_spaceL0}}
\subfigure[\ For $\Lambda=0.1~\kappa \rho_c$]{
\includegraphics[width=0.487\textwidth, height=4.865 cm]{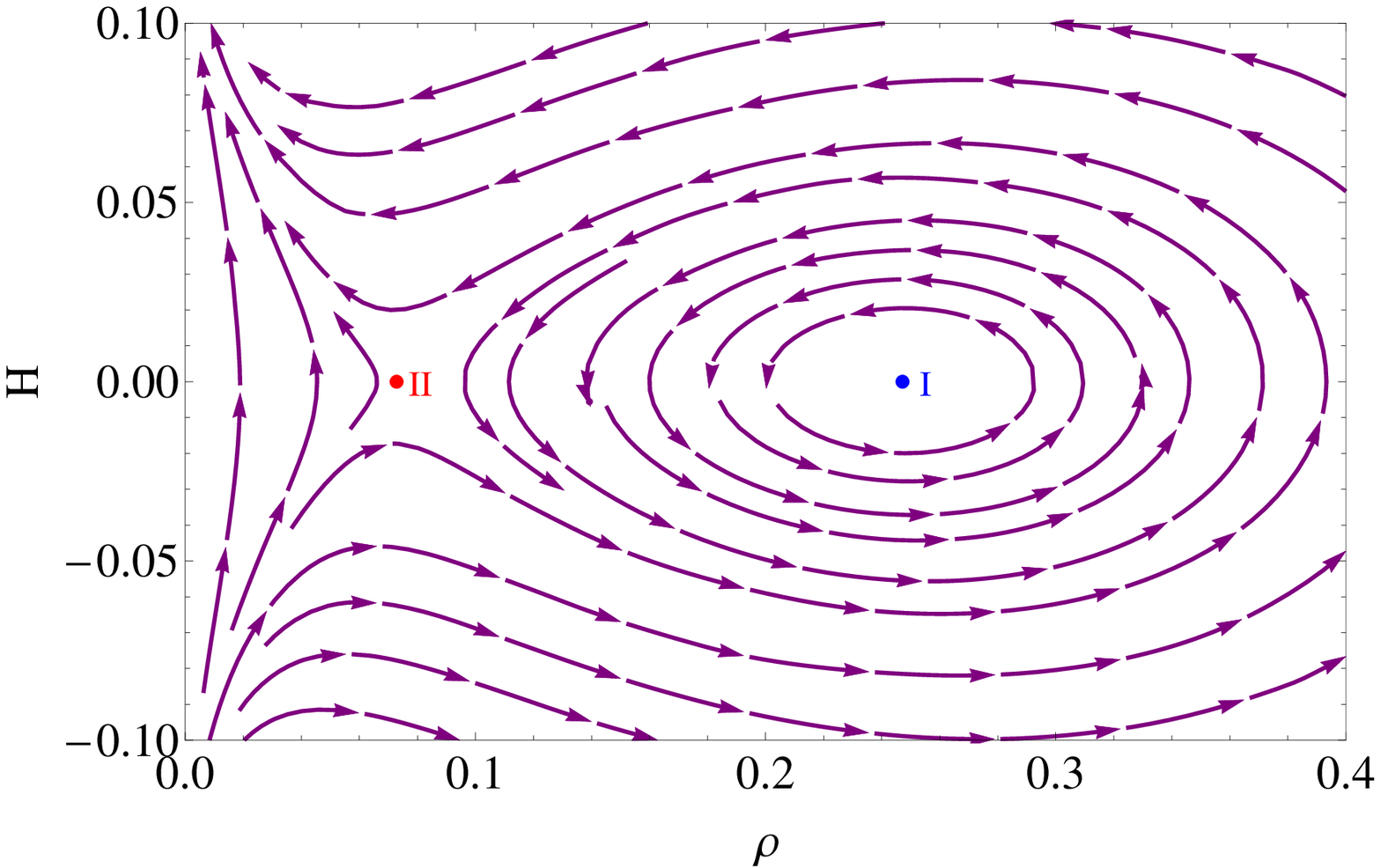}\label{fig:phase_spaceL1}}

\vspace{5.00mm}

\subfigure[\ For $\Lambda=0.15~\kappa \rho_c> \Lambda_{\rm crit}$]{
\includegraphics[width=0.487\textwidth]{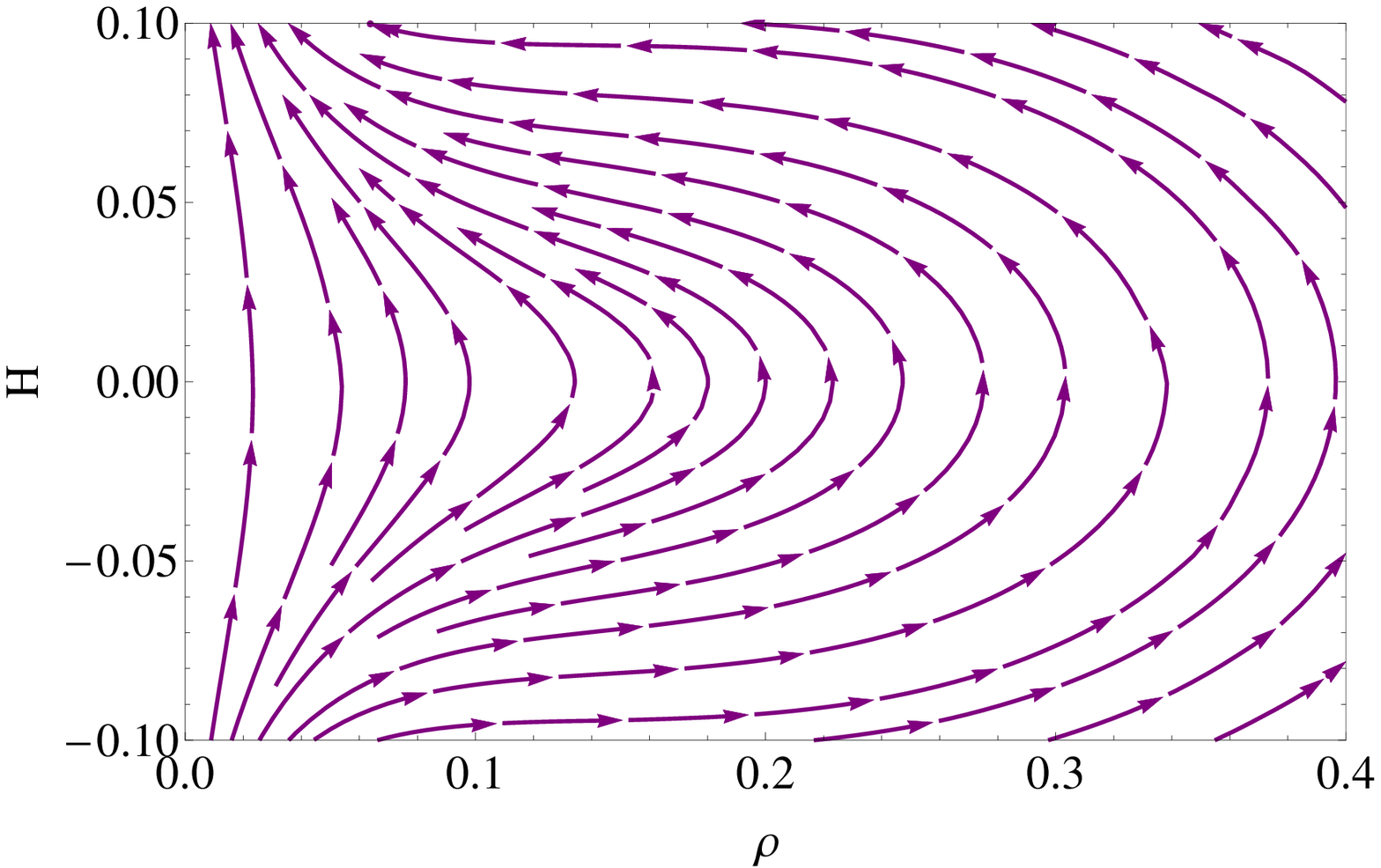}\label{fig:phase_spaceL15}}
\caption{\label{fig:phase_space}{\bf (a)} Phase portrait of a universe consisting only of stiff matter.
In this case, only the centre type fixed point \Rmnum{1} is present.
The motion of the spatially closed braneworld is oscillatory and bounded.
{\bf (b)} Phase portrait for a closed braneworld
 consisting of stiff matter and a cosmological constant:
$0<\Lambda<\Lambda_{\rm crit}$. For $\Lambda > 0$ the unstable saddle type fixed point
\Rmnum{2} appears. As $\Lambda$ increases the two fixed points, \Rmnum{1} and \Rmnum{2},
 move towards each other and coincide at $\Lambda=\Lambda_{\rm crit}$, giving
 rise to an inflection point in $U\left(a\right)$.
 {\bf (c)} Phase portrait for stiff matter with $\Lambda>\Lambda_{\rm crit}$. There is no fixed point in the dynamical system. In the numerical calculation we assume, for simplicity, that $\kappa=1$ with $\rho_c=1$.
}
\end{figure}

We have plotted the phase space for stiff matter and $\Lambda$ using \eqref{eq:rhos_dot}
and \eqref{eq:stiff_Hdot}. Since we focus on the spatially closed case, we need to add
the additional constraint
\begin{equation}\label{eq:sp_closed_condition}
H^2<\left(\dfrac{\kappa}{3}\rho+\dfrac{\Lambda}{3}\right)\left(1-\dfrac{\rho}{\rho_c}-\dfrac{\Lambda}{\kappa \rho_c}\right)\;,
\end{equation}
which follows from \eqref{eq:frw_bwL}.
The resulting phase portrait is shown in Fig.~\ref{fig:phase_spaceL0} for $\Lambda=0$.
Note that only the centre type stable fixed point \Rmnum{1} exists in this case, as
expected.

For $\Lambda>0$, the unstable saddle point \Rmnum{2} appears along with the centre
\Rmnum{1}\@. As $\Lambda$ increases, these two fixed points move towards each other.
The phase portrait for a typical value of $\Lambda$ is illustrated in
Fig.~\ref{fig:phase_spaceL1}. When $\Lambda$ reaches its critical value in
\eqref{eq:crit_stiffL}, i.e., at $\Lambda=\Lambda_{\rm crit}$, the stable and unstable
points merge giving rise to an inflection point in the effective potential $U (a)$, as
discussed earlier. For $\Lambda>\Lambda_{\rm crit}$, fixed points are absent, and flows
in the phase portrait are depicted in Fig.~\ref{fig:phase_spaceL15}, suggesting that,
after the bounce, the matter density declines monotonically as the
universe expands.\footnote{Note that for $\Lambda>\kappa \rho_c$, the dynamical system admits a centre type stable fixed point given by $(\rho_-,0)$ which has no physical significance for a
spatially closed universe, in view of \eqref{eq:sp_closed_condition}. However this fixed point is of great importance in constructing the emergent scenario in a spatially open universe as shown in Appendix \ref{app:bw_open}.}

\subsubsection{Emergent scenario }
\label{subsec:emergent_bw}
The above discussion showed that: (i)~the instability associated with ESU in the GR
context can be avoided by studying such a model in the braneworld context; (ii)~in this
case, in addition to the unstable critical point II reminiscent of GR-based ESU, there
appears a stable critical point I around which the universe can oscillate.

The construction of a realistic emergent cosmology requires that the universe be able to
exit its oscillatory phase. In order to do this, the inflationary potential must be
chosen judiciously, so that:
\begin{itemize}
\item[{\rm [A]}] $V(\phi)$ have an asymptotically flat branch where $V(\phi) \simeq
    V_0$. This permits the appearance of the stable minimum I in the effective
    potential $U(a)$. At this minimum, which corresponds to the ESU, the field's
    kinetic energy and the scale factor of the universe are given, respectively, by
\begin{equation} \label{eq:bw_scalar_phidot2}
\dot \phi^2=\dfrac{2}{5}\left(\rho_c-2V_0\right)\left[1+\sqrt{1-\dfrac{5V_0 \left(\rho_c-V_0\right)}{\left(\rho_c-2V_0\right)^2}}~\right]
\end{equation}
and
\begin{equation}\label{eq:bw_aE_scalar}
 a_E^2=\dfrac{3\mpl^2}{\rho \left(1-\rho/\rho_c\right)}\;.
\end{equation}
The corresponding value of $\rho$ can be determined from \eqref{eq:scalar_rho} using the value of $\dot \phi^2$ in \eqref{eq:bw_scalar_phidot2}.
Note that \eqref{eq:bw_scalar_phidot2} is identical to \eqref{eq:stiffL_rho} for $\rho_+$,
while  \eqref{eq:bw_aE_scalar} is the same as \eqref{eq:stiffL_aE}.

\item [{\rm [B]}] $V (\phi)$ should increase monotonically beyond some value of
    $\phi$ so that the effective cosmological constant, mimicked by $V$, increases
with time. This allows the stable and unstable fixed points to merge and the ESU
phase to end. Thereafter the universe inflates in the usual fashion. The potential
described by (\ref{eq:pot_exp}), which is shown in Fig.~\ref{fig:exp} and was earlier
discussed in \cite{Ellis:2003qz} and \cite{Mulryne:2005ef}, clearly satisfies this purpose.
\end{itemize}

It was earlier shown, in section \ref{sec:emergent_GR},
 that the (flat) left wing of $V (\phi)$ can give rise
to emergent cosmology in GR, provided inflation occurs from the (A) branch in Fig~\ref{fig:exp}. For braneworld-based emergent cosmology one requires inflation
to take place from the much steeper (B) branch of Fig.~\ref{fig:exp}. This leads to a
problem since along this branch the potential in (\ref{eq:pot_exp}) becomes an
exponential, which is ruled out by recent CMB constraints \cite{Ade:2013uln}.
Therefore, to examine this scenario further, we replace (\ref{eq:pot_exp}) by the
following potential which adequately serves our purpose:
\begin{equation} \label{eq:bw_inf_phi2}
V\left(\phi\right)=V_0\left[1-\theta(\phi)\left (\frac{\phi}{M}\right)^\gamma\right]^2\;,
\quad \gamma >0 \;,
\end{equation}
where $\theta(\phi)$ is a step function: $\theta(\phi) = 0$ for $\phi < 0$, and
$\theta(\phi) = 1$ for $\phi \geq 0$, which ensures that
\beq
V (\phi) = \left\{
\begin{array}{cll}
V_0 &\mbox{for} &\phi < 0 \, , \smallskip \\
0 &\mbox{for} &\phi = M \, , \smallskip \\
\displaystyle V_0\left(\frac{\phi}{M}\right)^{2\gamma} &\mbox{for} &\phi \gg M \, .
\end{array}
\right.
\eeq
Note that this potential qualitatively resembles the one shown in Fig.~\ref{fig:exp}.

\begin{figure}[t]
\begin{center}
\scalebox{0.45}[0.45]{\includegraphics{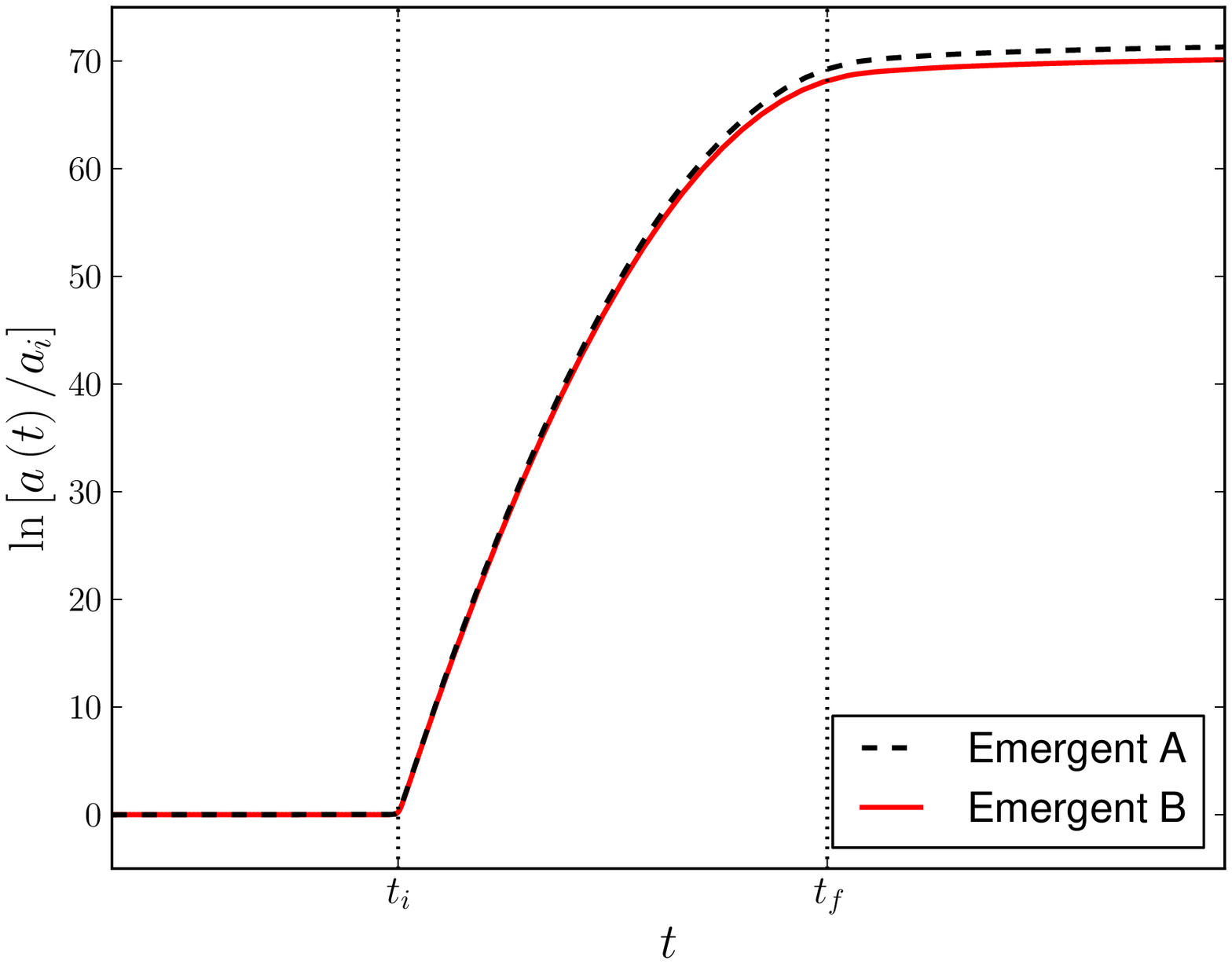}}
\scalebox{0.45}[0.45]{\includegraphics{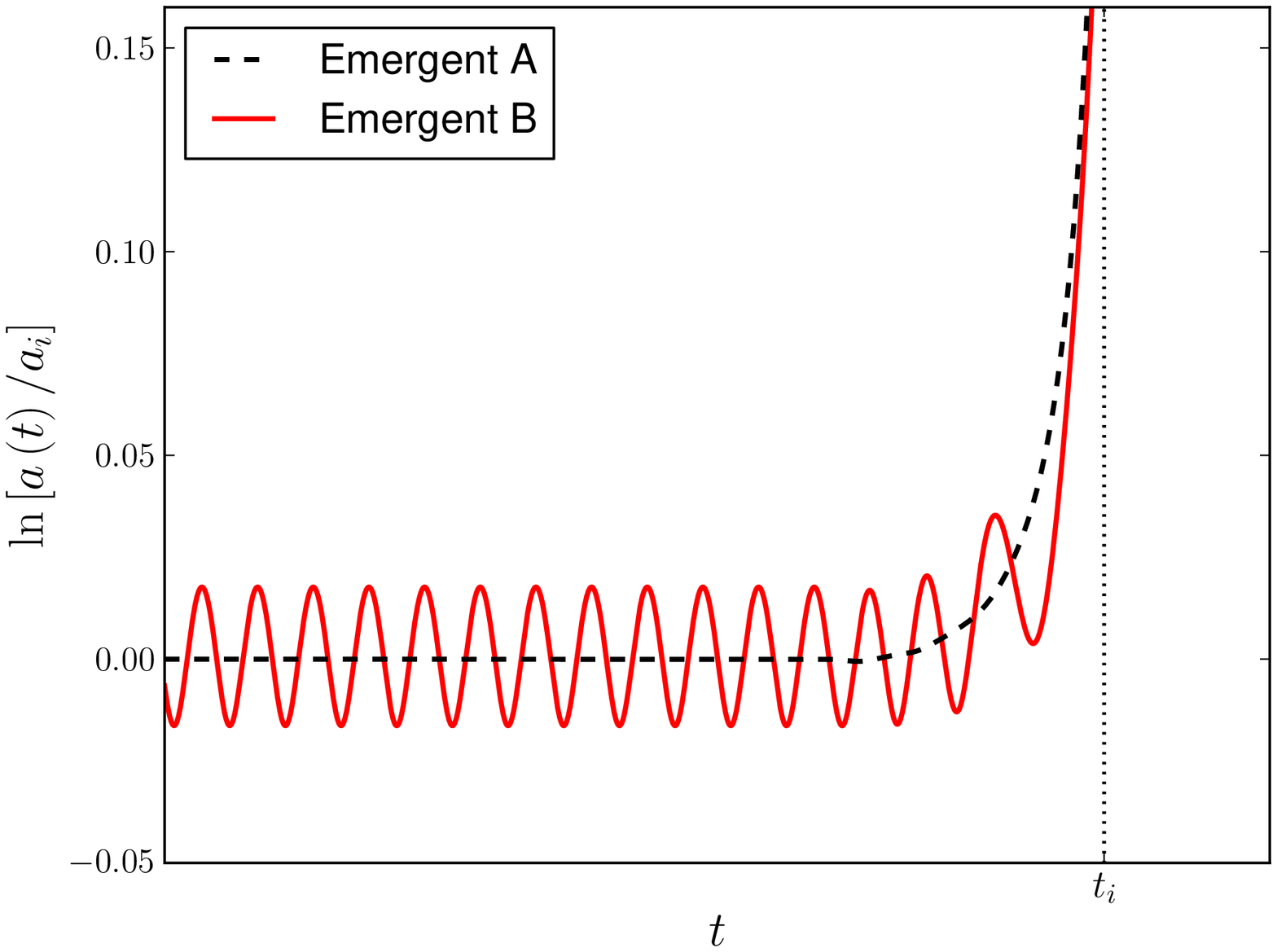}}
\caption{{\bf Top panel:}
Time evolution of the scale factor $a (t)$: $t_i$ and $t_f$ stand for the
beginning and end of inflation, respectively.
\textit{Emergent A:\/} the universe was at ESU prior to inflation.
\textit{Emergent B:\/} the universe was perturbed slightly away from ESU before inflation.
{\bf Bottom panel} shows a magnified view of the top panel prior to inflation
and demonstrates that universe B oscillates about the minimum of the effective potential.
$\kappa=1$ and $\rho_c=1$ are assumed.}
\label{fig:bw_inf_esu}
\end{center}
\end{figure}

\begin{figure}[t]
\begin{center}
\scalebox{0.45}[0.45]{\includegraphics{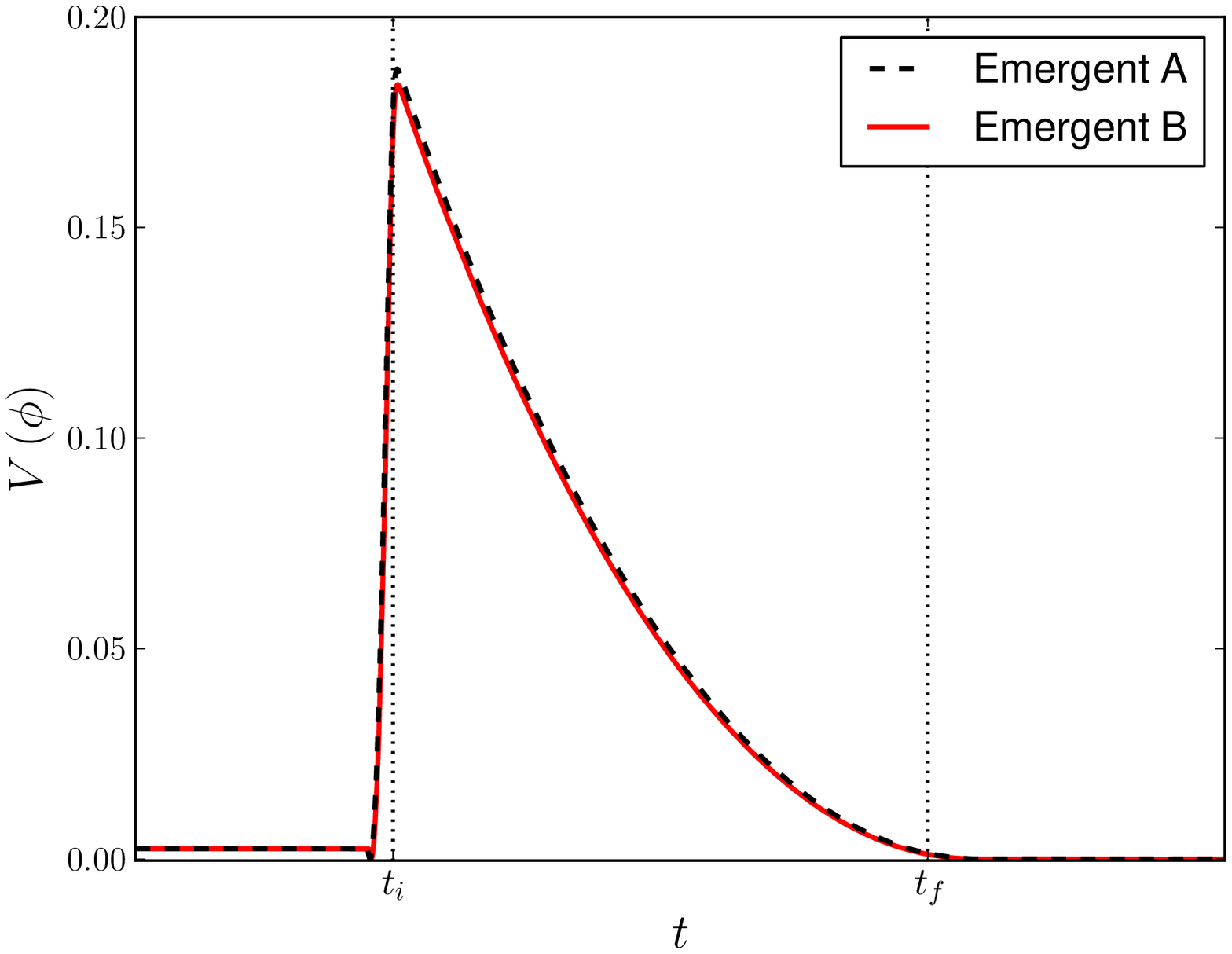}}
\caption{
Time evolution of $V\left(\phi\right)$ in the emergent scenario described by (\ref{eq:bw_inf_phi2}).
$\kappa=1$, $\rho_c=1$ are assumed.
}
\label{fig:bw_inf_V}
\end{center}
\end{figure}

Once the scalar field begins to roll up its potential, $V (\phi)$ (mimicked by $\Lambda$
in the previous section) increases, and the stable and unstable fixed points in
Fig.~\ref{fig:phase_spaceL1} move towards each other (see Fig.~\ref{fig:stiffL_move}),
merging when $V = V_{\rm crit}$, where
\begin{equation} \label{eq:critical_V}
V_{\rm crit} = \rho_c\left[\dfrac{1}{2}-\dfrac{\sqrt{5}}{6}\right] \,.
\end{equation}
For $V > V_{\rm crit}$, there is no fixed point, and the universe escapes from attractor
I.

For positive $H$ in \eqref{eq:scalar_evolution}, the scalar field experiences a very
large damping which causes it to stop climbing the potential $V (\phi)$. The inflationary
regime commences once the scalar field begins to roll slowly down its potential.

The system of equations (\ref{eq:scalar_evolution}), (\ref{eq:scalar}) and
(\ref{eq:bounce1}), (\ref{eq:bounce2}) has been integrated numerically in the context of spatially closed universe ($\Bbbk=1$) using the potential given in \eqref{eq:bw_inf_phi2}. The $\theta(\phi)$ function is realized numerically by $(1+\tanh(\tau \phi))/2$ with a very large $\tau$. Our results
are shown in Fig.~\ref{fig:bw_inf_esu} for the following two cases: \textit{Emergent
A:\/} the universe was at ESU prior to the inflation; \textit{Emergent B:\/} the universe
was perturbed away from ESU before inflation. The beginning and end of inflation are
denoted by $t_i$ and $t_f$, respectively, on the time axis. The lower panel to
Fig.~\ref{fig:bw_inf_esu} shows a zoomed-in view of the scale factor prior to inflation.
This panel demonstrates that universe A stays at ESU ($a_E$) eternally in the past,
whereas universe B exhibits oscillations around the minimum ($a_-$) of the effective
potential $U (a)$. (Note that $a_-$ is slightly displaced from $a_E$, as illustrated in
Fig.~\ref{fig:stiff_esu} for $\Lambda=0$.) The discerning reader may realize at this point that in
Emergent A, the fine-tuning requirement on the field's kinetic energy, given
 by  (\ref{eq:bw_scalar_phidot2}), is similar in spirit to
 that imposed in GR-based Emergent cosmology, namely (\ref{eq:fine_tune}). In other words, given an effective potential with a
minimum, its much more `likely' for the universe to oscillate about the ESU (Emergent B)
than to sit exactly at the ESU fixed point (Emergent A). The time evolution of the
inflaton potential $V(\phi)$ is illustrated in Fig.~\ref{fig:bw_inf_V}. The CMB
constraints for this scenario can easily be satisfied, as shown in Appendix \ref{app:bw_CMB}.

\subsection{Emergent scenario in Asymptotically Free Gravity}
\label{sec:emergent_AFG}

In this section we discuss the emergent scenario within the context of a theory of gravity which
becomes {\em asymptotically free} at high energies. In this phenomenological model
the gravitational constant depends upon the matter density as follows:
$G(\rho) = G_0\exp{(-\rho/\rho_c)}$. As a consequence, the FRW equations become
\ber
 H^2 &=& \frac{\kappa}{3}\rho e^{-\rho/\rho_c} - \frac{\Bbbk}{a^2}~, \label{eq:AFGH} \\
{\dot H} &=& \frac{\kappa}{2}(1+w)\rho e^{-\rho/\rho_c}\left [\frac{\rho}{\rho_c} - 1\right ]
+ \frac{\Bbbk}{a^2}~,
\label{eq:AFG}
\eer
where $\kappa = 8\pi G_0$ and $G_0$ is the asymptotic value of the gravitational constant:
$G(\rho \to 0) = G_0$. It is easy to see that the low-energy limit of this theory
is GR, while at intermediate energies $0 \ll \rho < \rho_c$ the field equations
(\ref{eq:AFGH}), (\ref{eq:AFG}) resemble those for the braneworld (\ref{eq:bounce1}),
(\ref{eq:bounce2}). From (\ref{eq:AFGH}), (\ref{eq:AFG}) we find that at large densities gravity becomes {\em asymptotically free:\/} $G(\rho) \to 0$ for $\rho \gg \rho_c$. In this case:

\begin{itemize}

\item
The universe will bounce if $\Bbbk=1$.

\item
The universe will `emerge' from Minkowski space (${\dot a} = 0, {\ddot a} = 0$ as $a \to 0$) if $\Bbbk = 0$.

\item
The universe will `emerge' from the Milne metric ($a \propto t$ as $a \to 0$) if $\Bbbk=-1$.

\end{itemize}

In all three cases the big bang singularity is absent.

The effective potential for this theory is
\beq
U(a) = -\frac{\kappa}{6}a^2\rho e^{-\rho/\rho_c}\;,
\label{eq:eff_pot}
\eeq
where the evolution of $\rho$ is given is \eqref{eq:gen_matter}.
In order to link the emergent scenario with inflation we shall consider $\rho$ to represent the
density of the inflaton field. Recall that the inflaton
moving along a flat direction ($V' = 0$) behaves like a fluid consisting of two
non-interacting components, namely stiff matter and the cosmological constant.
Hence $\rho$ in (\ref{eq:AFGH}), (\ref{eq:AFG}) \& (\ref{eq:eff_pot}) is effectively replaced by
$\rho_\phi \equiv \rho_{\rm stiff} + \rho_\Lambda$ where $\rho_{\rm stiff} \propto a^{-6}$
and $\rho_\Lambda = \Lambda/\kappa$.

\begin{figure}[t]
\begin{center}
\scalebox{0.7}[0.7]{\includegraphics{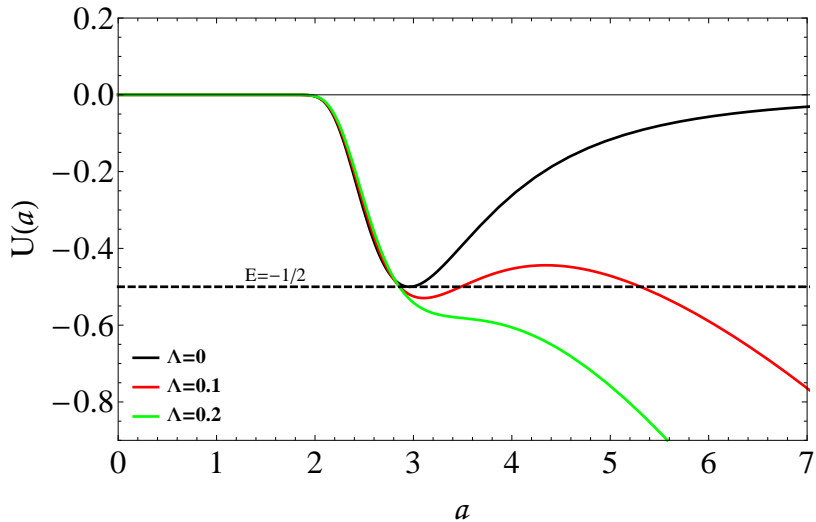}}
\caption{The effective potential, $U(a)$, is shown for {\em Asymptotically free gravity}
 for various values of $\Lambda$ (in units $\kappa=1$ and $\rho_c=1$). For $\Lambda=0$ there is only a single minima in $U(a)$, shown by the black line. A maxima appears when $\Lambda>0$ (red line).
As $\Lambda$ increases the two extrema move towards each other and merge when $\Lambda = \Lambda_{\rm crit}$
which results in
 an inflection point in $U(a)$. For $\Lambda>\Lambda_{\rm crit}$ there is no extrema in $U(a)$ (green line).
}
\label{fig:AFG_eff_pot}
\end{center}
\end{figure}

If $\Lambda = 0$ then the effective potential in \eqref{eq:eff_pot} exhibits a single
minimum shown by the black line in Fig.~\ref{fig:AFG_eff_pot}. As noted in section \ref{sec:emergent_GR}, the existence of an ESU implies the simultaneous implementation of
the following conditions:

\begin{enumerate}

\item ${\ddot a} = 0 \Rightarrow U'(a_E) = 0$,

\item ${\dot a} = 0 \Rightarrow U(a_E) = -\Bbbk/2$.

\end{enumerate}

Focussing on a closed universe ($\Bbbk=1$) dominated by stiff matter ($\Lambda = 0$) one finds
that the scale factor and density associated with ESU are given by
\begin{equation} \label{eq:AFG_rhoE}
\rho_E=\dfrac{2}{3}\rho_c\,, ~~~
a_E ^2=\dfrac{9}{2\kappa \rho_c} e^{2/3}~.
\end{equation}

Including $\Lambda > 0$, one finds that the form of $U(a)$ changes to accommodate {\em both}
a maxima and a minima.
The scale factor and density associated with these extrema are given by
\begin{eqnarray}
a_{\pm}^6 &=& \dfrac{3A}{2\Lambda \kappa \rho_c}\left(2\kappa \rho_c -3\Lambda\right)\left[1\pm \sqrt{1-\dfrac{12\Lambda\kappa\rho_c}{\left(2\kappa \rho_c -3\Lambda\right)^2}}~\right]\;, \label{eq:AFGL_a^6} \\
\rho_{\pm}^6 &=& \dfrac{\left(2\kappa \rho_c -3\Lambda\right)}{6 \kappa}\left[1\pm \sqrt{1-\dfrac{12\Lambda\kappa\rho_c}{\left(2\kappa \rho_c -3\Lambda\right)^2}}~\right]\;. \label{eq:AFGL_rho}
\end{eqnarray}
Again $\left(a_-,\rho_+\right)$ corresponds to the minimum, while $\left(a_+,\rho_-\right)$ is associated with the maximum in $U\left(a\right)$. Note that $\rho_{\pm}$ are
independent of the value of the parameter $A$ defined in \eqref{eq:gen_matter}. This implies that for a universe which oscillates about $a_-$, the matter density at $U(a_-)$ is the same as in the ESU. Recall that this situation was earlier encountered for the Braneworld in section \ref{sec:bouncing_bw} and implies that equations \eqref{eq:min_U} and \eqref{eq:solution_condition} hold in the present case too. Once more we shall
focus on the stable ESU corresponding to the minima.
The ESU is given by the scale factor
\begin{equation} \label{eq:AFGL_aE}
a_E ^2=\dfrac{3}{\left(\kappa \rho_+ +\Lambda\right)}\exp\left[\dfrac{\kappa \rho_++\Lambda}{\kappa \rho_c}\right]\,.
\end{equation}

As $\Lambda$ increases the minima and maxima in (\ref{eq:AFGL_a^6}) approach each other, merging when
\begin{equation} \label{eq:AFG_crit_L}
\Lambda = \Lambda_{\rm crit}=\dfrac{4}{3}\kappa \rho_c \left(1-\dfrac{\sqrt{3}}{2}\right) \,,
\end{equation}
which results in an inflection point in $U\left(a\right)$. 
The effective potential in {\em Asymptotically free gravity\/} shown in
Fig.~\ref{fig:AFG_eff_pot} resembles that in the braneworld model\footnote{Note that
unlike the braneworld case in section \ref{sec:bouncing_bw}, this asymptotically free gravity
model admits no minima in $U\left(a\right)$ for large $\Lambda$, i.e. for $\Lambda>\kappa
\rho_c$.} in Fig.~\ref{fig:stiffL_lam_inc}. We therefore find that, as in the braneworld,
an emergent scenario can be supported by the inflaton potential in
 \eqref{eq:bw_inf_phi2}.

\subsection{Emergent scenario in Loop Quantum Cosmology}
\label{sec:LQC}

In Loop Quantum Cosmology (LQC), the FRW equation for a
 spatially closed universe ($\Bbbk=1$) consisting of matter which satisfies the SEC and
$\Lambda$ is \cite{Ashtekar:2006uz,Parisi:2007kv}
\begin{equation} \label{eq:LQC_frw}
H^2 = \left(\dfrac{\kappa}{3}\rho + \dfrac{\Lambda}{3}-\dfrac{1}{a^2} \right)\left(1-\dfrac{\rho}{\rho_c}-\dfrac{\Lambda}{\kappa \rho_c}+\dfrac{3}{\kappa \rho_c a^2}\right)\;,
\end{equation}
where $\rho_c\sim \mpl^4$. In this case the effective potential depends on the curvature, and for
$\Bbbk=1$ one determines it from \eqref{eq:energy} to be
\begin{equation} \label{eq:LQC_eff_pot}
U\left(a\right)=-\left(\dfrac{\kappa}{6}\rho a^2 + \dfrac{\Lambda}{6} a^2-\dfrac{1}{2} \right)\left(1-\dfrac{\rho}{\rho_c}-\dfrac{\Lambda}{\kappa \rho_c}+\dfrac{3}{\kappa \rho_c a^2}\right)-\dfrac{1}{2}\;,
\end{equation}
where $\rho$ is given by \eqref{eq:gen_matter}. The ESU conditions \eqref{eq:ESU} provide
two possibilities for an {\em Einstein Static Universe} in this scenario, which have been
 listed in Table~\ref{table:LQC_ESU}. An unstable ESU appears for $\Lambda>0$ which
is identical to the ESU in GR, earlier
 discussed in section \ref{sec:emergent_GR}. The stable ESU appears due to
 LQC modifications and only exists for $\Lambda>\kappa \rho_c$ \cite{Parisi:2007kv}.

\begin{table}[htb]
\centering
 \begin{tabular}{C{3.15cm} C{1.65cm} C{4.45cm} C{4.5cm}}
 \\[-1.0em]
 \hline \hline
 \\[-0.6em]
  Fixed point (ESU) & $\Lambda$ & $\rho$ & $a_E$ \\[0.5ex]
  \hline \\[-1.0ex]
  Unstable & $\Lambda>0$ & $\rho_{GR}=\dfrac{2\Lambda}{\kappa (1+3w)}$ & $a_{GR}^2=\dfrac{1+3w}{\Lambda(1+w)}$  \\[1.75em]

  Stable  & $\Lambda>\kappa \rho_c$ & $\rho_{LQC}=\dfrac{2(\Lambda-\kappa \rho_c)}{\kappa (1+3w)}$ & $a_{LQC}^2=\dfrac{1+3w}{(\Lambda-\kappa \rho_c)(1+w)}$  \\[2.0ex]
  \hline
 \end{tabular}
 \caption{Density and scale factor for the
{\em Einstein Static Universe} (ESU) in LQC. The stable ESU is denoted by
 `LQC' while the unstable fixed point resembles the `GR' case.}
 \label{table:LQC_ESU}
\end{table}

\begin{figure}[t]
\centering
\subfigure[]{\includegraphics[scale=0.5]{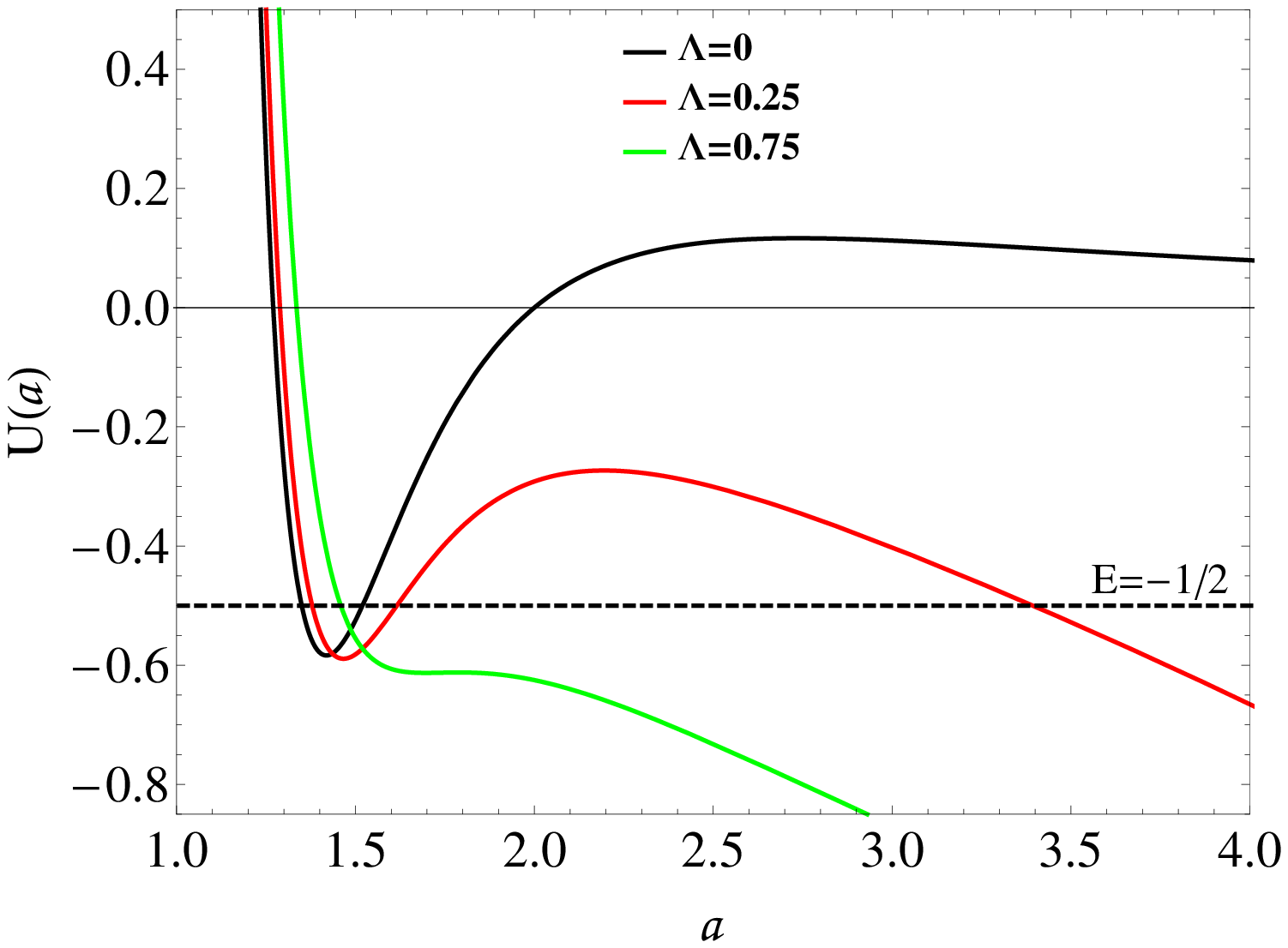}\label{fig:LQC_pot_a}}
\subfigure[]{\includegraphics[scale=0.5]{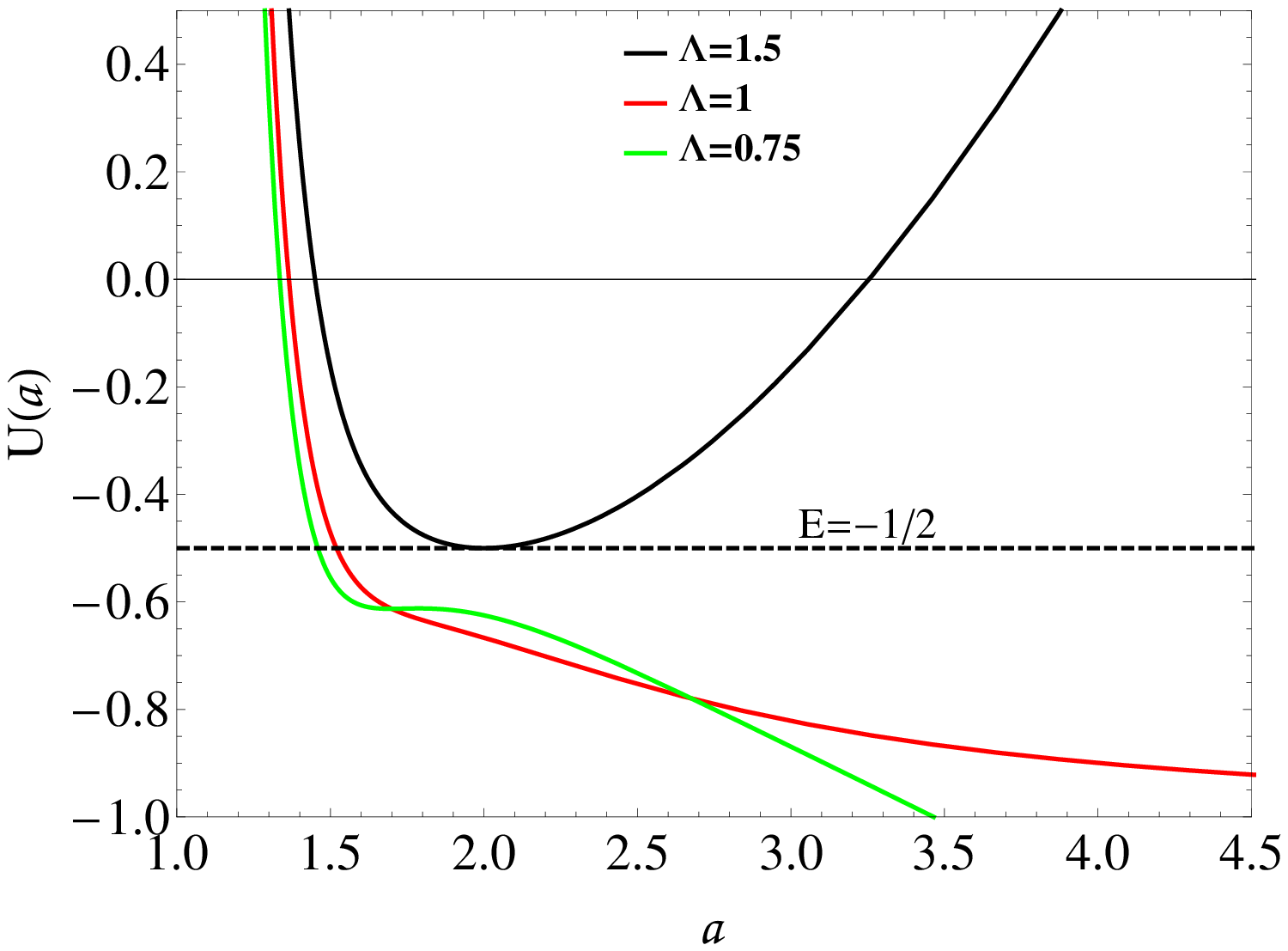}\label{fig:LQC_pot_b}}
\caption{\label{fig:LQC_pot} The effective potential for LQC is plotted for {\em small} values of
$\Lambda$ in the left panel {\bf(a)}, and for {\em large} values of $\Lambda$ in the right panel {\bf(b)}.
In both cases one assumes that the universe consists of stiff matter in addition to the cosmological
constant. This combination mimics the behavior of a scalar field rolling along a {\em flat}
direction in the potential ($V' = 0$) as described by (\ref{eq:bw_inf_phi2}) or (\ref{eq:tanh2}).
We choose a typical value of the parameter $A$ in \eqref{eq:gen_matter} while assuming $\mpl=1$ with
 $\rho_c \sim \mpl^4$.
The left panel indicates that inflation could proceed via the potential (\ref{eq:bw_inf_phi2})
illustrated in Fig.~\ref{fig:exp}.
Whereas the right panel supports inflation described by (\ref{eq:tanh2})
and illustrated in Fig.~\ref{fig:cosh}.
}
\end{figure}

As discussed earlier, a scalar field driven scenario, in which the scalar rolls along a flat potential
 ($V' = 0$), is equivalent to stiff matter together with a cosmological constant $\Lambda$. The effective potential in this case is shown for various values $\Lambda$ in Fig.~\ref{fig:LQC_pot}.
It is worth noting that every extremum (maxima or minima) in $U(a)$ does not necessarily result in an
ESU since every solution of $\ddot a=0$ may not support $\dot a=0$. 

Figures \ref{fig:LQC_pot_a} and \ref{fig:LQC_pot_b} demonstrate that the
emergent scenario in LQC can arise in two distinct ways:

\begin{enumerate}

\item As shown in Fig.~\ref{fig:LQC_pot_a} a minimum in $U(a)$ can exist for small values of
$\Lambda$. As $\Lambda$ increases this minimum gets destabilized. This indicates that an
inflaton potential such as (\ref{eq:bw_inf_phi2}) discussed earlier in the braneworld context, could also
give rise to an emergent scenario in LQC.
Note that unlike the braneworld case, a stable ESU does not exist in LQC for small $\Lambda$
-- see Table \ref{table:LQC_ESU}.
However this does not prevent the universe from {\em oscillating about} the minimum of $U(a)$,
thereby giving rise to a stable emergent scenario.
It is easy to see that since $\rho \ll \rho_c$ during inflation, the CMB constraints on
the parameters of the inflaton potential in (\ref{eq:bw_inf_phi2}) will be similar to those discussed in
Appendix \ref{app:bw_CMB} in the braneworld context.

\item The presence of a minimum in $U(a)$ is illustrated in Fig.~\ref{fig:LQC_pot_b}
for {\em large} values of $\Lambda$: $\Lambda>\kappa \rho_c$. This minimum is associated with a stable ESU as demonstrated in Table \ref{table:LQC_ESU}. As $\Lambda$ {\em decreases} this minimum gets destabilized. This suggests that a potential such as (\ref{eq:tanh2}), earlier discussed in the GR context,
could give rise to an emergent scenario in LQC.

Possibility 2 might however be problematic in two respects:

As noted in Table \ref{table:LQC_ESU}, the value of $V_{_0} (\equiv \Lambda/\kappa)$ in the flat wing of the emergent cosmology potential must be larger than $\rho_c \sim \mpl^4$ in order for LQC effects to successfully drive emergent cosmology.

\noindent
(i) $V_{_0} > \mpl^4$ might question the semi-classical treatment pursued by us in
this section. 

\noindent
(ii)  While the potential (\ref{eq:tanh2}) can successfully
drive an emergent scenario in LQC, CMB bounds derived in section \ref{subsec:GR_CMB}
suggest $V_{_0} < 10^{-8}\mpl^4$, which conflicts with the LQC requirement
\footnote{While the CMB bounds in section \ref{subsec:GR_CMB} were derived using GR, this would also be a good approximation
to the LQC case in which $\rho \ll \rho_c$ once inflation commences, and LQC effects can be ignored.}
$V_{_0} \gtrsim \mpl^4$.

While (i) lies outside the scope of the present paper, we demonstrate in Appendix \ref{app:LQC_CMB}
 that CMB constraints
can be satisfied even with $V_{_0} > \mpl^4$ provided the scalar field Lagrangian possesses
{\em non-canonical} kinetic terms \cite{Unnikrishnan:2012zu}.

\end{enumerate}

All of the emergent scenarios discussed in this section passed through
a prolonged (formally infinite) duration quasi-static stage during which the universe was
located either at the Einstein Static fixed point (ESU) or oscillated around it.
In the next section we examine the semi-classical properties of the oscillatory
universe, focusing especially on its impact on graviton production.

\section{Graviton production in an oscillatory universe}\label{sec:graviton}

In a flat FRW universe, each of the two polarization states of the graviton behaves as a
massless minimally coupled scalar field \cite{Grishchuk:1974ny}. This is also true for
the massless graviton modes in the higher-dimensional theories (see, e.g., \cite{SV} for
the case of braneworld model). While the conformal flatness of the FRW space-time ensures
that the creation of conformally coupled fields (including photons) does not happen, no
such suppression mechanism exists for fields that couple non-conformally to gravity
\cite{Birrell:1982,Grib:1994bk}. Indeed, it is well known that gravitons are generically
created in a FRW universe, and this effect has been very well studied in the context of
inflation \cite{Starobinsky:1979ty,Allen:1987bk,Sahni:1990tx}. One, therefore, naturally
expects gravity waves to be created in an oscillating universe such as the one examined
in the previous section, in the context of emergent cosmology. That this is indeed the
case will
be demonstrated below.

In the emergent scenario, the universe oscillates around a fixed value of the scale
factor for an indefinite amount of time before (gradually changing value of) the potential
ends the oscillatory regime and leads to inflation. While oscillating, the universe
produces gravitons by a quantum-mechanical process. A large graviton density
(compared with that of the existing matter), can disrupt
the oscillatory regime making it difficult for emergent cosmology to be past-eternal. 
Here we investigate this issue in the context of the
 braneworld scenario described in section \ref{sec:bouncing_bw}.

\subsection{Resonant particle production}

Tensor metric perturbations (or gravity waves) in general relativity are described by the
quantities $h_{ij}$ defined as
\beq
\delta g_{ij} = a^2 h_{ij} \, , \quad i, j = 1, 2, 3 \, .
\eeq
The tensor $h_{ij}$ is transverse and traceless and obeys the following equations
\cite{Mukhanov:2005sc}:
\begin{equation} \label{gen_grav}
h''_{ij}+2\mathcal{H} h'_{ij}-(\bigtriangledown^2-2\Bbbk)h_{ij} =0\;,
\end{equation}
where ${\cal H} = a' / a$, and $\Bbbk~(=0,\pm1)$ is the spatial curvature of the FRW
universe. Here, the prime denotes differentiation with respect to the conformal time
$\eta$. For a spatially closed universe ($\Bbbk=1$), we make the rescaling $h_{ij} =
\chi_{ij}/ a$ and pass to the generalized Fourier transform on the three-sphere
\beq
\chi_{ij} (\eta, x)  = \sum_{n \ell} \chi_{n \ell} (\eta) Y_{ij}^{n \ell} (x) \, ,
\eeq
where $Y_{ij}^{n\ell} (x)$ are the normalized transverse and traceless tensor
eigenfunctions of the Laplacian operator on a unit three-sphere. They are labeled by the
main quantum number $n$ and by the collective quantum numbers $\ell = \{p, l, m\}$, which
have the following meaning \cite{Allen:1994yb}\,:
\beq \label{qnumbers} \begin{array}{ll}
n = 3, 4, 5, \ldots &\quad \mbox{(main quantum number)} \, ,  \medskip \\
p = 1, 2 &\quad \mbox{(polarization)}  \, , \medskip \\
l = 2, 3, \ldots, n - 1 &\quad \mbox{(angular momentum)} \, ,  \medskip \\
m = - l , - l + 1 , \ldots, l &\quad \mbox{(angular momentum projection)} \, .
\end{array}
\eeq
The eigenvalues $- k_n^2$ of the Laplacian operator $\nabla^2$ for transverse traceless
tensor modes on a unit three-sphere depend only on $n$ and are given by \cite{Allen:1994yb}
\beq \label{kn2}
k_n^2 = n^2 - 3\;, \quad n = 3, 4, 5, \ldots \, .
\eeq
Equation \eqref{gen_grav} then leads to the following equations for the Fourier
coefficients $\chi_{n \ell} (\eta)$\,:
\begin{equation} \label{closed_grav}
\chi''_{n \ell} + \left( k_n ^2 + 2 - \dfrac{a''}{a} \right) \chi_{n \ell} =0\;.
\end{equation}

Sufficiently close to the minimum of the effective potential, the universe can exhibit oscillatory
motion. Subject to a small perturbation, the oscillatory motion with frequency $\omega$
satisfies
\begin{equation}
a (t) = a_- + \delta a \cos \omega t \, ,
\end{equation}
where the frequency is given by
\beq
\omega^2 = \dfrac{d^2 U ( a_- ) }{da^2} \;,
\eeq
and the amplitude is
\begin{equation}
\delta a^2 = - \dfrac{2}{\omega^2} \left[U (a_-) + \dfrac{1}{2}\right]\;.
\end{equation}

To be able to use \eqref{gen_grav} and \eqref{closed_grav}, we express this motion in
terms of the conformal time $\eta$\,:
\begin{equation} \label{eta(t)}
\eta = \int \dfrac{dt}{a (t)}=\int \dfrac{dt}{a_-+\delta a \cos \omega t}\approx \dfrac{t}{a_-}
\quad \mbox{for} \quad \dfrac{\delta a}{a_-} \ll 1\;.
\end{equation}
Now, the functions $a$ and $a''/a$ can be calculated as
\begin{subequations}
 \begin{align}
  a (\eta)=a_-+\delta a \cos \zeta \eta \;, \label{a(eta)} \\
\dfrac{a''}{a} = - \dfrac{\delta a}{a_-} \zeta^2 \cos \zeta \eta \;, \label{a''/a}
 \end{align}
\end{subequations}
where
\begin{equation} \label{zeta}
\zeta=a_- \omega
\end{equation}
is the frequency in the conformal time $\eta$.

We would like to apply the theory of parametric resonance, as described in
\cite{Shtanov:1994ce}, to equation \eqref{closed_grav}.  This equation has the general
form that was under consideration in \cite{Shtanov:1994ce}\,:
\begin{equation} \label{reso}
\chi''_{n \ell} + \left[\zeta_n^2 + \epsilon g(\zeta \eta) \right] \chi_{n \ell} = 0\;,
\end{equation}
where $g (x)$ is a $2 \pi$-periodic function, and $\epsilon$ is a convenient small
parameter.  In our case, we have
\begin{equation} \label{zeta_n^2}
\zeta_n^2 = k_n^2 + 2 = n^2 - 1\;, \quad n = 3, 4, 5, \ldots \, ,
\end{equation}
\beq \label{g}
g (x) = \zeta^2 \cos x \, ,
\eeq
and
\beq
\epsilon = \frac{\delta a}{a_-} \, .
\eeq

Equation (\ref{reso}) with the function $g (x)$ given by (\ref{g}) is just the Mathieu
equation. It is known that the first resonance band for this equation, which is dominant
for small values of $\epsilon$, lies in the neighbourhood of the frequency
\beq
\zeta_{\rm res} = \frac{\zeta}{2} \, ,
\eeq
and the resonant amplification takes place for eigenfrequencies satisfying the condition
\beq \label{band}
\varDelta_n^2 < g_1^2 \, .
\eeq
Here,
\begin{equation}
\varDelta_n=\dfrac{1}{\epsilon} \left(\zeta_n^2 - \zeta_{\rm res}^2 \right)\;,
\end{equation}
and
\begin{equation} \label{gs}
g_1 = \dfrac{1}{2}\zeta^2
\end{equation}
is the Fourier amplitude of the harmonic function $g (x)$.

Within the resonance band (\ref{band}), particle production proceeds exponentially with
time, so that the number of quanta in the mode grows as
\beq
N_n = \frac{1}{1 - \varDelta_n^2 / g_1^2} \sinh^2  \mu_n \eta \, ,
\eeq
where
\begin{equation} \label{mu}
\mu_n = \dfrac{\epsilon}{\zeta}\sqrt{g_1^2 - \varDelta_n^2}\;.
\end{equation}
The condition (\ref{band}) for instability in the first resonance band can be expressed as %
%
\begin{equation} \label{reso1}
\left| \zeta_n^2 - \left(\frac{\zeta}{2}\right)^2 \right| < \frac{\epsilon \zeta^2}{2} \;.
\end{equation}

\subsubsection{Radiation-dominated universe with \texorpdfstring{$\Lambda=0$}{Lg}}
\label{subsec:reso_rad}

From the viewpoint of the effective potential $U (a)$, the created gravitons behave as
radiation. Therefore, for simplicity of the analysis, we consider a radiation-dominated
universe, in which the produced gravitons just increase the existing radiation energy
density (i.e., their effect will consist in the increase of $A$ in
\eqref{eq:gen_matter}). For further simplification, we first consider the case with no
cosmological constant. The frequency of oscillations (with respect to the cosmic time
$t$) turns out to be constant in this case\,:
\begin{equation} \label{rad_omega}
\omega ^2=\dfrac{d^2U (a_-)}{da^2} = \dfrac{4}{9}\kappa \rho_c=\dfrac{4}{9 \mpl^2} \rho_c\;.
\end{equation}
Therefore, adding more to the existing radiation energy density (increasing the quantity
$A$ in \eqref{eq:gen_matter}) does not affect $\omega$. The energy densities at
the extremes of the effective potential are also constant (independent of $A$ in
\eqref{eq:gen_matter})\,:
\begin{equation} \label{rad_rho+}
\rho_+=\dfrac{1}{3}\rho_c \, , \qquad \rho_-=0\;.
\end{equation}
The second equation implies that there are no maxima in the effective potential (or the
maximum is reached as $a \rightarrow \infty$).

From \eqref{eq:crit_U} and \eqref{rad_rho+}, the scale factor and the quantity $\zeta$ can be evaluated for
the ESU\,:
\begin{equation} \label{rad_aE}
a_E^2=\dfrac{27}{2\kappa \rho_c}=\dfrac{27}{2}\dfrac{\mpl^2}{\rho_c}\;,
\end{equation}
\begin{equation} \label{rad_zetE2}
\zeta_E^2 = \omega^2 a_E^2 = 6 \;.
\end{equation}
Eventually, the terms appearing in \eqref{reso} can be calculated as
\begin{equation} \label{rad_zet2}
\zeta ^2 = \omega^2 a_{-}^2 = 6\left(\dfrac{a_- ^2}{a_E^2}\right) \geq 6
\end{equation}
and
\begin{equation} \label{rad_epsilon2}
\epsilon^2 = \left(\dfrac{\delta a}{a_-}\right)^2 = \dfrac{1}{6} - \dfrac{9\mpl^2}{4\rho_c a_- ^2}
= \dfrac{1}{6} \left(1 - \dfrac{a_E^2}{a_- ^2} \right) \leq \dfrac{1}{6}\;.
\end{equation}
Although $\omega$ is independent of graviton production, we can see from
\eqref{eq:gen_a^l} that $a_-$ increases as more and more gravitons are produced. This
leads to an increase both in $\zeta$ and in $\epsilon$ (although $\epsilon$ has an
asymptotic saturation value $1/\sqrt{6}$). Now, we assume
\begin{equation} \label{x}
x=\left(\dfrac{a_-}{a_E}\right)^2\geq1\;.
\end{equation}
In this case, the resonance condition \eqref{reso1} has the form
\begin{equation} \label{rad_reso2}
\left( \zeta_n^2 - \frac32 x \right)^2 < \frac32 x (x - 1) \;.
\end{equation}


\begin{table}[ht]
\centering
 \begin{tabular}{C{1.0cm} C{2.5cm} C{2.5cm}}
 \hline \hline \\[-0.5em]
  $\zeta_n^2$ & $a_-/a_E$ & $\epsilon$ \\ [0.5ex]
  \hline \\[-1.0em]
  8 & 1.784$-$5.179 & 0.338$-$0.401 \\
  15 & 2.396$-$7.229 & 0.371$-$0.404 \\
  24 & 3.007$-$9.217 & 0.385$-$0.406 \\ [1ex]
  \hline
 \end{tabular}
\caption{Resonance intervals of the quantities $a_-/a_E$ and $\epsilon$ for several lower
modes in a radiation-dominated universe.}
 \label{table:rad_excitation}
\end{table}

In Table~\ref{table:rad_excitation}, the resonance intervals of the quantity $a_-/a_E$
along with the corresponding intervals of $\epsilon$ are shown for several lower modes
($\zeta_n ^2$ is given by \eqref{zeta_n^2} and \eqref{kn2}) by solving the inequality
\eqref{rad_reso2} in terms of $x$. As a graviton mode gets excited, the quantity
$a_-/a_E$ increases due to the growth of $A$ in \eqref{eq:gen_a^l}. The resonant
intervals of $a_-/a_E$ for neighboring values of $n$ overlap, as demonstrated in
Table~\ref{table:rad_excitation} for several lowest values of $n$.  It is clear then
that, if the initial value of $a_-/a_E$ is sufficiently large, so that it falls in any of
the resonance regions, then this will lead to excitation of all modes, one by one,
resulting in a monotonous increase in $a_-/a_E$ due to graviton production. In this case,
the oscillatory regime is unstable with respect to graviton production. However, if the
initial amplitude of oscillations is sufficiently small, so that $\epsilon < 0.338$, or
$a_-/a_E < 1.784$, then no graviton mode is in the resonance initially, and past eternal
oscillation of the universe is stable with respect to resonant production of gravitons.

Although the values of $\epsilon$ in Table~\ref{table:rad_excitation} turn out to be not much
smaller than unity, they are still considerably small, and we believe that our analysis
in this case is qualitatively correct.

\subsubsection{Stiff-matter dominated universe}
\label{subsec:reso_stiff} For a constant scalar field potential $V (\phi)$, the kinetic
energy density $\frac12 \dot \phi^2$ in \eqref{eq:scalar} evolves as $a^{-6}$, similarly
to the energy density of stiff matter. Therefore, when dealing with inflation based on a
scalar field, it is a good idea to consider the case of a stiff-matter dominated
universe. It was observed previously that the value $a_-$ of the scale factor
corresponding to the minimum of the effective potential actually increases as more and
more gravitons are added to the existing stiff-matter energy density. This was also
verified numerically. Below, we analyze both the simple case with $\Lambda = 0$ and the
more realistic case with non-zero $\Lambda$.

\subsubsection*{[A] No cosmological constant, $\Lambda = 0$}

The analysis for stiff-matter dominated universe is carried out in a manner similar to
that of the universe filled with radiation that was under investigation in section \ref{subsec:reso_rad}. Here, the frequency in cosmic time $\omega$ again turns out
to be constant (independent of $A$ in \eqref{eq:gen_matter}) and is given by
\begin{equation} \label{stiff_omega}
\omega ^2=\dfrac{d^2U (a_-)}{da^2} = \dfrac{8}{5}\kappa \rho_c=\dfrac{8}{5 \mpl^2} \rho_c\;.
\end{equation}
The value $\rho_E$ of the stiff matter energy density at the minimum of the effective
potential was already calculated in \eqref{eq:stiff_rhoE}. Using the value of $a_E$ from
\eqref{eq:stiff_aE}, we figure out the value of $\zeta^2$ at the ESU\,:
\begin{equation} \label{stiff_zetE2}
\zeta_E ^2=\omega^2a_E^2=20\;.
\end{equation}
Again, the terms appearing in \eqref{reso1} can be calculated as
\begin{equation} \label{stiff_zet2}
\zeta ^2=\omega^2a_{-}^2=20 \left(\dfrac{a_- ^2}{a_E^2}\right) \geq 20
\end{equation}
and
\begin{equation} \label{stiff_epsilon2}
\epsilon^2=\dfrac{1}{\omega^2}\left(\dfrac{2}{25}\kappa \rho_c
-\dfrac{1}{a_-^2}\right)=\dfrac{1}{20} \left( 1 - \dfrac{a_E^2}{a_- ^2} \right)   \leq
\dfrac{1}{20}\;.
\end{equation}

Introducing the quantity $x$ as in \eqref{x}, and using \eqref{stiff_zet2} and
\eqref{stiff_epsilon2}, we present the resonance condition \eqref{reso1} in the form
\begin{equation} \label{stiff_reso2}
\left( \zeta_n^2 - 5 x \right)^2 < 5 x (x - 1) \;.
\end{equation}


\begin{table}[ht]
\centering
 \begin{tabular}{C{1.0cm} C{2.5cm} C{2.5cm}}
 \hline \hline \\[-0.5em]
  $\zeta_n^2$ & $a_-/a_E$ & $\epsilon$ \\ [0.5ex]
  \hline \\[-1.0em]
  8 & 1.146$-$1.561 & 0.109$-$0.172 \\
  15 & 1.5$-$2.236 & 0.167$-$0.200 \\
  24 & 1.867$-$2.875 & 0.189$-$0.210 \\ [1ex]
  \hline
 \end{tabular}
 \caption{Resonance intervals of the quantities $a_-/a_E$ and $\epsilon$ for several lower
modes in a stiff-matter dominated universe.}
 \label{table:stiff}
\end{table}

The resonance intervals of $a_-/a_E$ and $\epsilon$, given by inequality
\eqref{stiff_reso2}, are listed for several lowest modes in Table~\ref{table:stiff}.
Again, the overlapping of the resonance bands indicates that, once a graviton mode is
excited, the monotonous growth in $a_-/a_E$ will lead to resonant excitation of the next
modes. However, for $a_-/a_E < 1.146$, or $\epsilon <0.109$, the past eternal oscillation
of the universe is stable with respect to resonant production of gravitons.

\subsubsection*{[B] Stiff matter with cosmological constant $\Lambda$}

If the value of the flat wing of the scalar-field potential is non-negligible, then, in
our model of stiff matter, we must also introduce the cosmological constant $\Lambda$. As
before, the frequency $\omega$ is independent of the quantity $A$ in
\eqref{eq:gen_matter},
\begin{equation} \label{stiffL_omega}
\omega^2 = \dfrac{d^2U (a_-)}{da^2} = -\dfrac{10}{3}\kappa
\rho_+\left(1-\dfrac{2\Lambda}{\kappa \rho_c}\right)+\dfrac{55}{3} \frac{\kappa
\rho_+^2}{\rho_c}-\dfrac{\Lambda}{3}\left(1-\dfrac{\Lambda}{\kappa
\rho_c}\right)\;,
\end{equation}
while $\rho_+$ and $a_E$ are already given in \eqref{eq:stiffL_rho} and
\eqref{eq:stiffL_aE}, respectively. The terms appearing in \eqref{reso1} can be expressed
more generally as
\begin{equation} \label{stiffL_zet2}
\zeta ^2=\omega^2 a_{-}^2=\zeta_E ^2 \left(\dfrac{a_- ^2}{a_E^2}\right) \geq \zeta_E^2
\end{equation}
and
\begin{equation} \label{stiffL_eps2}
\epsilon ^2=\dfrac{1}{\omega ^2}\left( \dfrac{1}{a_E ^2}-\dfrac{1}{a_-^2} \right) =
\dfrac{1}{\zeta_E ^2} \left(1-\dfrac{a_E^2}{a_-^2}\right) \le \dfrac{1}{\zeta_E^2}\;.
\end{equation}
In deriving \eqref{stiffL_eps2}, we used equation \eqref{eq:min_U}.

\begin{figure}[htb]
\begin{center}
\includegraphics[width=0.5\textwidth]{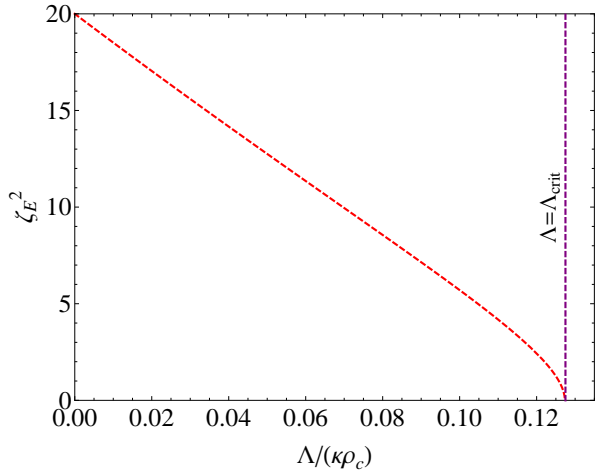}
\caption{
$\zeta_E^2$ vs $\Lambda$ in the natural units $\kappa=1$ and $\rho_c=1$.
}
\label{fig:stiffL_zeta2}
\end{center}
\end{figure}

For a general value of $\Lambda$, the value of $\zeta_E ^2$ cannot be evaluated
analytically. For $\Lambda = 0$, the calculation is presented earlier in this subsection~\ref{subsec:reso_stiff}. In the opposite limit $\Lambda\rightarrow \kappa \rho_c
(1/2-\sqrt{5}/6)$, we have
\beq \label{stiffL_limitL}
a_E^2=\dfrac{15}{\kappa \rho_c} \, , \qquad \omega^2=0 \, , \qquad \zeta_E ^2=0 \, .
\eeq
It is impossible to satisfy the resonance condition \eqref{reso1} in this limit of
$\Lambda$, hence no resonant graviton production takes place. Now, as $\Lambda$ increases
$a_-$ also increases while $\omega$ decreases. Fig.~\ref{fig:stiffL_zeta2} shows that
$\zeta_E ^2$ actually decreases with increasing $\Lambda$, and $\zeta_E ^2$ goes to zero
as $\Lambda$ approaches the limiting value specified in \eqref{eq:crit_stiffL}.

For convenience, along with the variable $x$ defined in \eqref{x}, we also introduce the
variable $y=\zeta_E^2/4$. Then
\begin{equation} \label{stiffL_gen_zet2/4}
\frac{\zeta^2}{4} = \frac{\zeta_E^2}{4} \left(\dfrac{a_- ^2}{a_E^2}\right)=xy\;,
\end{equation}
and, for a general $\zeta_E$, the resonance condition \eqref{reso1} becomes
\begin{equation} \label{stiffL_reso1}
\left(\zeta_n^2 - y x \right)^2 < y x \left(x - 1 \right)\;.
\end{equation}
Solving the inequality with respect to $x$ for a specified value $\zeta_n^2$, we get the
allowed range of $x$ to excite the respective mode.  The point $y = 1$ is critical, so
one needs to distinguish between two cases.

(i)~For $y > 1$, the range of allowed $x$ lies in the interval $\left(x_-, x_+ \right)$,
where the end points are given by
\begin{equation} \label{stiffL_xrange}
x_\pm=\left(\dfrac{a_-}{a_E}\right)^2=\dfrac{2\zeta_n^2-1}{2\left(y-1\right)}\left[1\pm\sqrt{1-\dfrac{4\zeta_n^4\left(y-1\right)}{\left(2\zeta_n
^2 -1\right)^2 y}}~\right]\;.
\end{equation}

(ii)~For $0 < y < 1$, also taking into account that $x > 1$, we find that the resonant
production takes place in the domain
\beq \label{stiffL_xrange1}
x > x_- = \frac{2\zeta_n^2 - 1}{2 (1 - y)} \left[\sqrt{1 + \frac{4 \zeta_n^4 (1 - y)}{(2
\zeta_n^2 - 1)^2 y}} -1\right] \, .
\eeq

It is worth noting that, for any mode, $x_+$ in (\ref{stiffL_xrange}) diverges as
$y\rightarrow1$ while the lower boundary $x_-$ in (\ref{stiffL_xrange1}) blows up as
$y\rightarrow0$ (i.e., as $\Lambda\rightarrow\Lambda_{\rm crit}$, as expected). Note that
the lower boundary in \eqref{stiffL_xrange} and \eqref{stiffL_xrange1} is, by expression,
the same function of $y$ in different domains, so is denoted by the same symbol $x_-$.

The lowest mode ($\zeta_n^2=8$) is of particular interest because it determines the
condition that no gravitons are resonantly produced. This mode is excited for the range
of $x$ determined by
\ber \label{stiffL_xrange_8}
x_\pm &=& \dfrac{15}{2 (y - 1 )}\left[1 \pm \sqrt{1 - \dfrac{256 (y - 1)}{225
y}}\, \right]\;, \quad y > 1 \, , \\
x_- &=& \dfrac{15}{2 (1 - y)}\left[\sqrt{1 + \dfrac{256 (1 - y)}{225 y}} - 1\right]\;,
\quad 0 < y < 1 \, .
\eer

\begin{figure}[t]
\begin{center}
\scalebox{0.65}[0.65]{\includegraphics{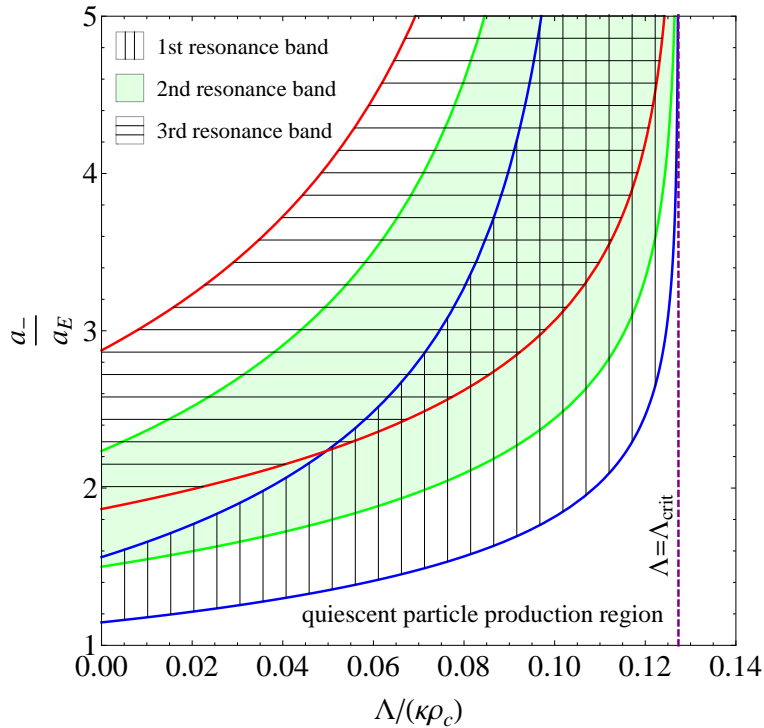}}
\caption{The resonance bands of the quantity $a_-/a_E$ for three lowest graviton modes
are shown as a function of the cosmological constant $\Lambda$. Again, the natural unit
of $\kappa=1$ is assumed along with $\rho_c=1$. The ``quiescent particle production
region'' below the first resonance domain corresponds to the region of parameters in
which no graviton mode is in resonance.} \label{fig:stiffL_reso}
\end{center}
\end{figure}

The resonance bands of the quantity $a_-/a_E$ for three lowest modes are plotted as a
function of $\Lambda$ in Fig.~\ref{fig:stiffL_reso}. The values of $y$ in
\eqref{stiffL_xrange} and \eqref{stiffL_xrange1} for different values of $\Lambda$ are
calculated from \eqref{eq:stiffL_rho}, \eqref{eq:stiffL_aE} and \eqref{stiffL_omega}. The
overlapping bands (spanning all values of $\Lambda<\Lambda_{\rm crit}$) again imply that,
once a particular graviton mode is in the resonance, all the subsequent modes eventually
will be resonantly excited.  However, the range of $a_-/a_E$ which does not excite any
resonance mode actually expands with increasing $\Lambda$. This region, below the first
resonance band, is indicated as the ``quiescent particle production region''.

\subsection{Non-resonant production of gravitons}

In this section, we investigate more thoroughly the case where no resonant production of
gravitons takes place, i.e., where no frequency $\zeta_n$ lies in the resonance band
determined by condition (\ref{reso1}).  Although the graviton modes are not excited {\em
resonantly\/}, their excitation still takes place.  Our aim is to show that this effect
is rather small and does not destroy the stability of the oscillating universe.

Since all graviton modes are assumed to be out of the resonance band, we can apply
perturbation theory for the calculation of their average occupation numbers.  From the
general theory of excitation of a harmonic oscillator with time-dependent frequency
$\Omega(\eta)$, it is known that these occupation numbers are expressed through the
complex Bogolyubov coefficients $\alpha$ and $\beta$.  These coefficients, for boson
fields, obey the normalization condition
\beq
| \alpha|^2 - | \beta|^2 = 1 \, ,
\eeq
and satisfy the following system of equations (see, e.g., \cite{Shtanov:1994ce}):
\beq \label{bogolyubov}
\alpha'=\dfrac{\Omega'}{2\Omega} e^{+2i\int\Omega d\eta}\beta\;, \qquad
\beta'=\dfrac{\Omega'}{2\Omega} e^{-2i\int\Omega d\eta}\alpha\;.
\eeq
If initially (for convenience, we set $\eta_0 = 0$) the oscillator is not excited, then
one can set
\begin{equation} \label{initial_alpha_beta}
\alpha (0) = 1 \, , \qquad \beta (0) = 0\;.
\end{equation}
The average excitation number (average number of particles) at the time $\eta$ is then
given by
\begin{equation} \label{eta_avg_no}
N (\eta)=|\beta (\eta)|^2\;.
\end{equation}

In the case of graviton production, from \eqref{closed_grav}, \eqref{a''/a} and
\eqref{reso}, the time-dependent frequency of the corresponding harmonic oscillator can
be identified as
\begin{equation} \label{OMEGA_2}
\Omega_n^2 = \zeta_n^2+\epsilon \zeta^2\cos \zeta \eta \;.
\end{equation}
For sufficiently small value of $\epsilon$ and outside the resonance band, we have
\begin{equation} \label{OMEGA'/2OMEGA}
\dfrac{\Omega_n'}{2\Omega_n} \approx -\dfrac{1}{4}\dfrac{\epsilon \zeta^3 \sin \zeta
\eta}{\zeta_n ^2}\;.
\end{equation}
To the first order in $\epsilon$, from \eqref{bogolyubov} and \eqref{initial_alpha_beta}
we then have the following equation for $\beta$\,:
\begin{equation} \label{beta_n'}
\beta_n' (\eta) \approx -\dfrac{1}{4}\dfrac{\epsilon \zeta^3 \sin \zeta \eta}{\zeta_n ^2} e^{-2i\zeta_n \eta}
= \dfrac{i\epsilon \zeta^3}{8\zeta_n ^2} \left[e^{-i\left(2\zeta_n-\zeta\right)\eta}-e^{-i\left(2\zeta_n+\zeta\right)\eta}\right]\;.
\end{equation}

Now integrating \eqref{beta_n'} from $0$ to $\eta$, we have
\begin{equation} \label{beta_n}
\beta_n (\eta) \approx \dfrac{\epsilon \zeta^3}{8\zeta_n ^2} \left[\dfrac{e^{-i\left(2\zeta_n+\zeta\right)\eta}-1}{2\zeta_n+\zeta}-\dfrac{e^{-i\left(2\zeta_n-\zeta\right)\eta}-1}{2\zeta_n-\zeta}\right]\;.
\end{equation}
Hence,
\ber \label{beta_n^2_1}
|\beta_n (\eta)|^2 &\approx& \left(\dfrac{\epsilon \zeta^3}{8\zeta_n ^2}\right)^2
\left[\dfrac{e^{-i\left(2\zeta_n+\zeta\right)\eta}-1}{2\zeta_n+\zeta}-
\dfrac{e^{-i\left(2\zeta_n-\zeta\right)\eta}-1}{2\zeta_n-\zeta}\right] \nonumber \\
&&{} \times \left[\dfrac{e^{i\left(2\zeta_n+\zeta\right)\eta}-1}{2\zeta_n+\zeta}
-\dfrac{e^{i\left(2\zeta_n-\zeta\right)\eta}-1}{2\zeta_n-\zeta}\right]
\eer
or,
\ber \label{beta_n^2_2}
|\beta_n (\eta)|^2 &\approx& \left(\dfrac{\epsilon \zeta^3}{8\zeta_n ^2}\right)^2 \left(
\dfrac{\sin ^2\left(\zeta_n+\zeta/2\right)\eta}{\left(\zeta_n+\zeta/2 \right)^2}
+\dfrac{\sin^2 \left(\zeta_n-\zeta/2\right)\eta}{\left(\zeta_n-\zeta/2 \right)^2} \right. \nonumber \\
&& {} \left. +\dfrac{1}{2\left[\zeta_n^2- (\zeta/2)^2\right]} \left[
\cos\left(2\zeta_n+\zeta\right)\eta + \cos\left(2\zeta_n-\zeta\right)\eta - \cos 2\zeta
\eta  - 1 \right] \right) \;. \, \,
\eer

In the resonance, as $\zeta_n \rightarrow \zeta/2$, the second term in \eqref{beta_n}
dominates. When the resonance condition is not met, we can safely assume that $\zeta_n
\gg \zeta/2$ (i.e., we are far away from the resonance). Equation \eqref{beta_n^2_2} is
then simplified as
\begin{equation}  \label{beta_n^2_out}
|\beta_n (\eta)|^2 \approx \left(\dfrac{\epsilon \zeta^3}{8\zeta_n ^2}\right)^2
\dfrac{1}{\zeta_n^2}\left[2\sin^2 \zeta_n\eta + \cos 2\zeta_n\eta
-\dfrac{1}{2}\left( 1+\cos 2\zeta \eta \right)\right]
=  \left(\dfrac{\epsilon \zeta^3}{8}\right)^2 \dfrac{\sin^2 \zeta \eta}{\zeta_n^6}\;.
\end{equation}

Now, to calculate the graviton energy density $\rho_{\rm grav}$, the quantity $|\beta_n
(\eta)|^2$ should be summed over the graviton modes with the energy of graviton taken
into account, and then divided by the volume of the space, which is $a_- ^3$.  We recall
that the graviton modes on a three-sphere are labeled by the quantum numbers
(\ref{qnumbers}). The frequency depends only on $n$.  Summing over $l$ and $m$, we get
\beq
\sum_{l = 2}^{n - 1} \sum_{m = - l}^l = \sum_{l = 2}^{n - 1} (2 l + 1) = n^2 - 4 \, .
\eeq
Then, taking into account two polarizations and the fact that the energy of a graviton is
$\zeta_n / a_-$, we have
\begin{eqnarray} \label{rho_grav_out}
\rho_{\rm grav} & = & \dfrac{2}{a_- ^4}\left(\dfrac{\epsilon \zeta^3}{8}\right)^2 \sin^2 \zeta \eta
\sum_{n = 3}^\infty \dfrac{n^2 - 4}{\zeta_n^5}\;, \nonumber \\
& =& \dfrac{2}{a_- ^4}\left(\dfrac{\epsilon \zeta^3}{8}\right)^2
\sin^2 \zeta \eta \sum_{n=3}^\infty
\dfrac{n^2 - 4}{\left(n^2 - 1\right)^{5/2}}\;, \nonumber \\
& = & \dfrac{2}{a_- ^4}\left(\dfrac{\epsilon \zeta^3}{8}\right)^2
\sin^2 \omega t \sum_{n=3}^\infty
\dfrac{n^2 - 4}{\left(n^2 - 1\right)^{5/2}} \;.
\end{eqnarray}
The sum in this expression is convergent, and can be estimated as
\beq
\sum_{n=3}^\infty \dfrac{n^2 - 4}{\left(n^2 - 1\right)^{5/2}} \approx 0.06 \, .
\eeq

\begin{figure}[t]
\begin{center}
\includegraphics[width=0.5\textwidth]{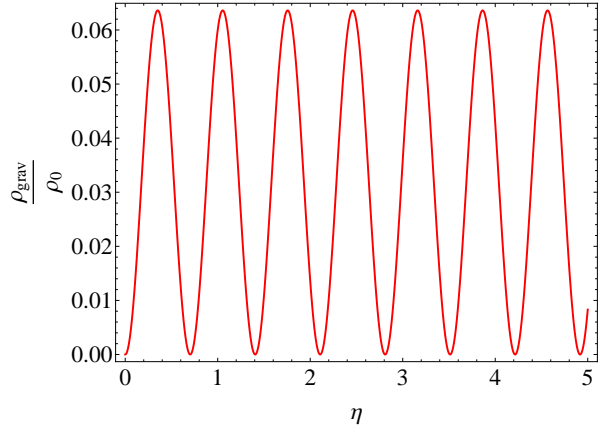}
\caption{The energy density $\rho_{\rm grav}$ as a function of conformal time $\eta$ in
the case of graviton production far away from the resonance, when $\zeta_n \gg \zeta_{\rm
res} = \zeta/2$. Here, we choose $\zeta^2=20$ along with $\rho_0=\dfrac{2}{a_-
^4}\left(\dfrac{\epsilon \zeta^3}{8}\right)^2$.} \label{fig:rho_grav}
\end{center}
\end{figure}

The graviton energy density turns out to be periodic with time as indicated in
\eqref{rho_grav_out}, with period equal to half the period of oscillation of the
universe. The energy density of produced gravitons is shown as a function of conformal
time $\eta$ in Fig.~\ref{fig:rho_grav} for a typical value of $\zeta = \sqrt{20}$, i.e.,
for the case of stiff-matter dominated universe, with no cosmological constant,
oscillating very close to the ESU (see \eqref{stiff_zetE2} and \eqref{stiffL_zet2}).

From Fig.~\ref{fig:stiffL_zeta2} and in the view of \eqref{stiffL_zet2}, it is apparent that, for small values of $\Lambda$, the quantity $\zeta/2$ cannot be regarded as much smaller than the frequency
$\zeta_3 = \sqrt{8}$ of the first mode, which has the leading contribution in
\eqref{rho_grav_out}. Thus, when the `far away from resonance' assumption is not strictly
valid, a more general expression for $\rho_{\rm grav}$ can be calculated directly from
\eqref{beta_n^2_2}\,:
\ber \label{rho_grav}
\rho_{\rm grav} &=& \dfrac{2}{a_- ^4}\left(\dfrac{\epsilon \zeta^3}{8}\right)^2\sum_{n =
3}^\infty \dfrac{n^2 - 4}{\zeta_n ^3} \left( \dfrac{\sin^2
\left(\zeta_n+\zeta/2\right)\eta}{\left(\zeta_n+\zeta/2 \right)^2}
+\dfrac{\sin^2\left(\zeta_n-\zeta/2\right)\eta}{\left(\zeta_n-\zeta/2 \right)^2} \right. \nonumber \\
&&{} \left. +\dfrac{1}{2\left(\zeta_n^2 - \left(\zeta/2\right)^2\right)} \left[
\cos\left(2\zeta_n+\zeta\right)\eta + \cos\left(2\zeta_n - \zeta\right)\eta - \cos 2\zeta
\eta  - 1 \right] \right)\;.
\eer

\begin{figure}[t]
\centering
\subfigure[\ $\zeta^2=6$]{
\includegraphics[width=0.48\textwidth]{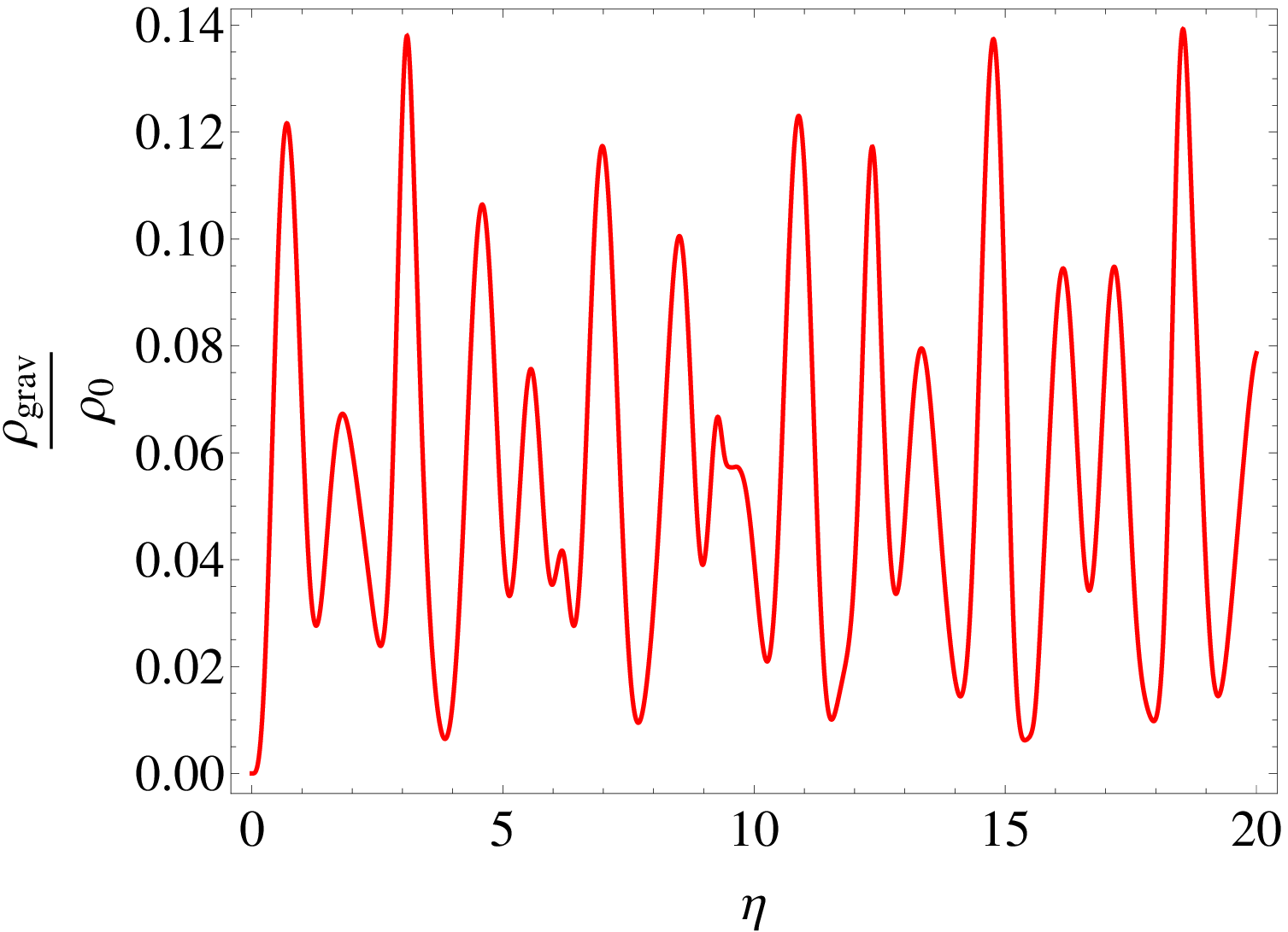}\label{fig:rho_grav_rad}}
\subfigure[\ $\zeta^2=20$]{
\includegraphics[width=0.48\textwidth,height=5.36 cm]{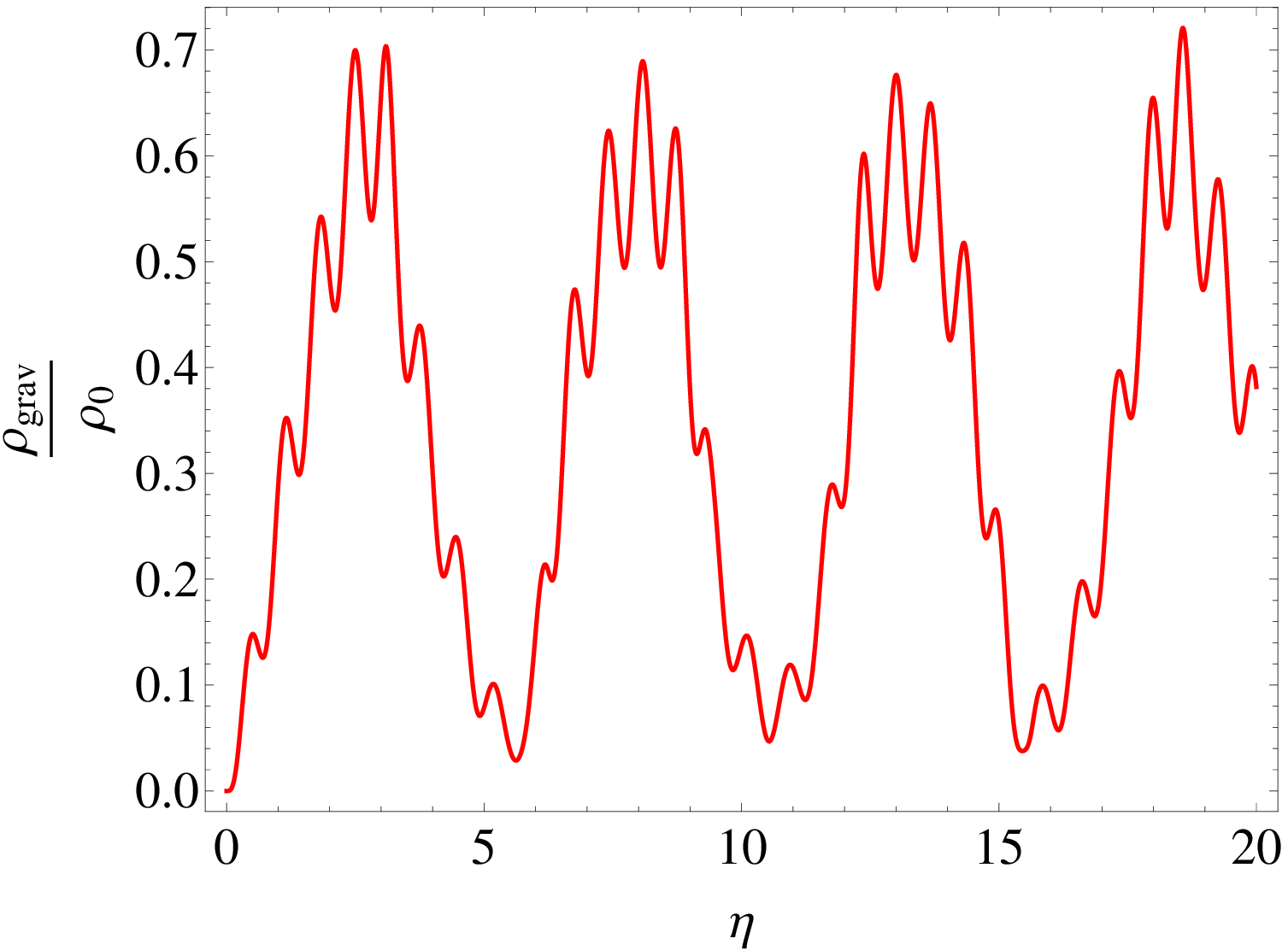}\label{fig:rho_grav_stiff}}
\caption{\label{fig:rho_grav_stiff_rad} The energy density $\rho_{\rm grav}$ as a
function of conformal time $\eta$ with (a) $\zeta^2=6$ and (b)
$\zeta^2=20$ in the case of non-resonant graviton production. These two cases represent
a universe filled with radiation and stiff matter, respectively, with $\Lambda=0$ and
oscillating very closely to the ESU\@. Again, $\rho_0=\dfrac{2}{a_-
^4}\left(\dfrac{\epsilon \zeta^3}{8}\right)^2$.}
\end{figure}

In Figs.~\ref{fig:rho_grav_rad} and \ref{fig:rho_grav_stiff}, the quantity $\rho_{\rm
grav}$ given by (\ref{rho_grav}) is shown as a function of the conformal time $\eta$ for
$\zeta^2=6$ and $\zeta^2=20$, which can represent a radiation and stiff-matter dominated
universe, respectively, with $\Lambda = 0$, while oscillating very close to the ESU.  It
is normalized by the quantity
\beq \label{eq:rho_0}
\rho_0=\dfrac{1}{32} \epsilon^2 \zeta^2 \omega^4=\dfrac{1}{32} \left(\dfrac{\epsilon^2
\zeta_E^2}{1-\epsilon^2 \zeta_E^2}\right) \omega^4=\dfrac{1}{32 a_E ^4}
\left(\dfrac{\epsilon^2 \zeta_E^6}{1-\epsilon^2 \zeta_E^2}\right)\;,
\eeq
where we have used the relation (see \eqref{stiffL_zet2} and \eqref{stiffL_eps2}),
\beq \label{eq:epsilon2zeta2}
 \epsilon^2 \zeta^2 =\dfrac{\epsilon^2 \zeta_E^2}{1-\epsilon^2 \zeta_E^2}\;. \nonumber
\eeq
It is worth noting that, in the limit $\epsilon^2 \rightarrow 1/ \zeta_E^2$, the quantity
$\rho_0$ (hence, also $\rho_{\rm grav}$) becomes infinite.  This is consistent with
\eqref{rho_grav_out} and \eqref{rho_grav} since, in this limit, $a_-$ (hence, also
$\zeta$) is also infinite (see \eqref{stiffL_eps2}).

%
%

Now we have to compare the amount of graviton produced to the existing matter density.
Although Figs.~\ref{fig:rho_grav_rad} and \ref{fig:rho_grav_stiff} show some departure
from Fig.~\ref{fig:rho_grav}, we see that, in all cases, the energy density of the
gravitons is small.  Indeed, let us denote the summation in \eqref{rho_grav} by $s$ (in
\eqref{rho_grav_out}, it is the summation multiplied by the time-dependent part). From
\eqref{rad_rho+}, \eqref{eq:stiff_rhoE} and \eqref{eq:stiffL_rho}, it is clear that
$\rho_c$ is a good estimation for the existing matter density. Hence, the ratio of the
energy density of the produced gravitons to that of existing matter is
\beq \label{comparison}
\dfrac{\rho_{\rm grav}}{\rho_c} \approx 2 s \left(\dfrac{\epsilon \zeta^3}{8}\right)^2
\dfrac{1}{a_- ^4 \rho_c} = 2 s \left(\dfrac{\epsilon \zeta}{8}\right)^2\dfrac{\zeta_E
^4}{a_E^4 \rho_c} =\dfrac{s\omega^4}{32\rho_c} \left(\dfrac{\epsilon^2
\zeta_E^2}{1-\epsilon^2 \zeta_E^2}\right)=\dfrac{s}{32 a_E ^4\rho_c}
\left(\dfrac{\epsilon^2 \zeta_E^6}{1-\epsilon^2 \zeta_E^2}\right)\;.
\eeq
In the last step, we used the general expression for $\zeta^2$ from \eqref{stiffL_zet2}.
Now we can see that $\dfrac{\rho_{\rm grav}}{\rho_c}\ll 1$ due to the presence of the
small parameter $\epsilon^2$ on the right-hand side (all other quantities have finite
values). For a typical case of stiff-matter dominated universe with $\Lambda=0$, we can
estimate the ratio using \eqref{eq:stiff_aE} and \eqref{stiff_zetE2},
\begin{equation} \label{comparison_stiff}
\dfrac{\rho_{\rm grav}}{\rho_c} \approx \dfrac{s (\epsilon \zeta)^2 }{a_E^2
\mpl^2} = \dfrac{8}{5}
\left(\dfrac{s\epsilon^2}{1-20\epsilon^2}\right)\dfrac{\rho_c }{\mpl^4}\ll 1\;.
\end{equation}

Thus, for small deviations from the Einstein Static Universe, the maximum possible energy
density of gravitons produced during the eternal oscillations in the emergent scenario
appears to be tiny compared to the existing matter density, which ensures the stability
of the model.

\section{Conclusions}\label{sec:conclusions}
In this paper we have shown how the effective potential formalism can be used to study the
dynamical properties of the emergent universe scenario. Within the GR setting, the
 effective potential has a single extreme point, a maximum, which corresponds to
the {\em unstable} Einstein Static Universe (ESU). Extending our analysis to modified
gravity theories we find that a new {\em minimum} in the  effective potential appears
corresponding to
a {\em stable} ESU. These results are in broad agreement with earlier studies which also
pointed out the appearance of an ESU in the context of extensions to GR \cite{Mulryne:2005ef,Parisi:2007kv,Boehmer:2007tr};
also see \cite{Mukherjee:2006ds}.

While in GR, the emergent scenario can only occur if the universe is closed, we show that
this restriction does not apply to certain modified gravity models in which the emergent scenario
can occur in spatially closed as well as open cosmologies.

The appearance of a stable minimum in the  effective potential considerably enlarges the
initial data set from which the universe could have `emerged'. In this case,
in addition to  being precisely located at the minimum (ESU) -- which requires
considerable fine tuning of
initial conditions, the universe can
oscillate about it.
Furthermore, we show that the existence of an ESU, while being conducive for emergent cosmology,
is not essential for it. In section \ref{sec:LQC} this is demonstrated for LQC for which a stable ESU
exists only for $\Lambda>\kappa \rho_c$ \cite{Parisi:2007kv}. We demonstrate that
even for $\Lambda\ll \kappa \rho_c$, when a stable ESU no longer exists, the universe
can still oscillate about the minimum of its effective potential allowing an
emergent scenario to be constructed.

However an oscillating universe is always accompanied by
graviton production. While the magnitude of this semi-classical effect depends upon
parameters in the effective potential, 
for a large region in parameter space this effect can be very large, casting doubts
as to whether such an emergent scenario could have been past-eternal.\footnote{ Graviton production
is small, and does not stand in the way of emergent cosmology being past eternal, only
if the universe oscillates very near the minimum of its effective potential.
(Naturally, there is no particle production for a universe located precisely at the minimum of $U(a)$, i.e. for the ESU.)
But this situation might require considerable fine tuning of parameters.}
(The instability of emergent cosmology to quantum effects has also been
recently investigated in  \cite{Mithani:2014jva}.)
Although graviton production has been discussed in detail for an effective potential
derived from the braneworld scenario, the effect itself is semi-classical and generic,
 and would be expected to
accompany any emergent scenario in which the universe emerges from an oscillatory
state.

One might also note that in the emergent scenarios discussed 
in this paper, the post-emergent universe inflates by well over 60 e-folds.
Consequently any feature associated with the transition from an ESU to inflation
is pushed to scales much larger than the present horizon. However it could well be
that in some emergent scenarios this is not the case, and the transition from the
ESU to inflation takes place fewer than $\sim 60$ e-folds
from the end of inflation. In this case the spectrum of inflationary perturbations
would differ from those considered in this paper on large scales, and may contain
a feature in the CMB anisotropy spectrum, $C_\ell$, at low values of $\ell$.

\bigskip
\section*{Acknowledgments}

V.S. and Yu.S. acknowledge support from the India-Ukraine Bilateral Scientific
Cooperation programme.  The work  of Yu.S. was also partially supported by the SFFR of
Ukraine Grant No.~F53.2/028.

\bigskip 
\bigskip
\newpage
\appendix
\section{Emergent scenario in a spatially open Braneworld}\label{app:bw_open}
As mentioned in section \ref{sec:bouncing_bw}, the braneworld admits a minimum in the effective potential $U(a)$ if $\Lambda>\kappa \rho_c$;
 see \eqref{eq:gen_a^l} and \eqref{eq:gen_rho}. Considering stiff matter along with $\Lambda>\kappa \rho_c$, from the equations \eqref{eq:stiffL_a^6} and \eqref{eq:stiffL_rho}, one finds that only $(a_+ , \rho_-)$ survive which now account for the {\em minimum} given by
\begin{equation} \label{eq:stiffL_open_a^6}
a^6_{+}=\dfrac{3A\left(2\Lambda-\kappa \rho_c\right)}{\Lambda \left(\Lambda-\kappa \rho_c\right)}\left[1+\sqrt{1+\dfrac{5\Lambda \left(\Lambda-\kappa \rho_c\right)}{\left(2\Lambda-\kappa \rho_c\right)^2}}~\right]\;,
\end{equation}
\begin{equation} \label{eq:stiffL_open_rho}
\rho_{-}=\rho_E=\dfrac{1}{5\kappa}\left(2\Lambda-\kappa \rho_c\right)\left[\sqrt{1+\dfrac{5\Lambda \left(\Lambda-\kappa \rho_c\right)}{\left(2\Lambda-\kappa \rho_c\right)^2}}-1\right]\;.
\end{equation}

\begin{figure}[h]
\begin{center}
\scalebox{0.7}[0.7]{\includegraphics{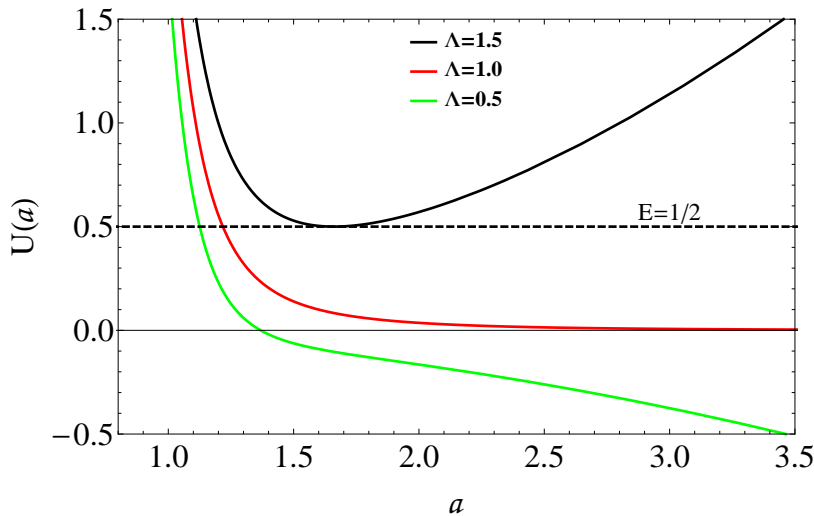}}
\caption{The effective potential for the braneworld in \eqref{eq:effective_potential} is plotted
for large values of $\Lambda$ (in units $\kappa=1$, $\rho_c=1$). For $\Lambda>\kappa \rho_c$, a minima in $U(a)$ appears which can give rise to an ESU provided the universe is open ($\Bbbk=-1$). For $\Lambda \leq \kappa \rho_c$ the minimum disappears which allows the universe to exit the ESU and also to
inflate (for a suitable choice of the inflaton potential).
}
\label{fig:bw_open}
\end{center}
\end{figure}

According to \eqref{eq:frw_bwL}, when $\Lambda>\kappa \rho_c$, a static solution can exist only for a
 spatially open universe ($\Bbbk=-1$). Again $\rho_-$ is the energy density at the ESU, hence denoted by $\rho_E$ in \eqref{eq:stiffL_open_rho}. The scale factor at ESU is calculated for
the spatially open universe using \eqref{eq:frw_bwL} and \eqref{eq:stiffL_open_rho} to be
\begin{equation} \label{eq:stiffL_open_aE}
{a_E}^{-2}=\frac{\left(\kappa\rho_E +\Lambda\right)}{3}\left(\dfrac{\rho_E}{\rho_c}+\dfrac{\Lambda}{\kappa \rho_c}-1\right)\;.
\end{equation}


The effective potential in this case is plotted for various $\Lambda$ in
Fig.~\ref{fig:bw_open} which illustrates the emergence scenario in a spatially open
universe. For $\Lambda>\kappa \rho_c$ the universe can either be an ESU or can oscillate
around the minimum of $U(a)$ at $a_+$. When $\Lambda=\kappa \rho_c$, the minimum
disappears and, for large values of $a$, $U(a)$ asymptotically approaches zero from
above. Therefore $\Lambda<\kappa \rho_c$ destabilizes the ESU and can result in inflation
for the potential $V(\phi)$ in \eqref{eq:tanh2}. After inflation commences the curvature
term, $\Bbbk/a^2$, rapidly declines to zero resulting in a spatially flat universe. (See
\cite{Zhang:2013ykz} for a discussion of an emergent scenario in a spatially flat
braneworld.)

\section{Linearization near a fixed point}\label{app:linearization}
A two dimensional non-linear system given by
\beq
\dot x_i=X_i(x_1,x_2)~~~ {\rm where}~~i=1,2\;,
\eeq
can be linearized in the neighbourhood of its simple fixed point $(\zeta,\eta)$ as \cite{Arrowsmith:1992},
\beq \label{eq:linearization}
\dot x_i\approx {\cal A}_{ij}x_j~~~~~~{\rm where}~~~~~{\cal A}_{ij}=\dfrac{\partial X_i}{\partial x_j}\at[\Big]{(\zeta,\eta)}\;.
\eeq
It should be noted that the {\em linearization theorem} guarantees that the linearized system in the neighbourhood of a fixed point is qualitatively equivalent to the non-linear system as long as the former does not suggest a centre type linearization.

For braneworld consisting of two component fluid: stiff matter with $\Lambda$, the linearized system is given by the co-efficient matrix in \eqref{eq:linearization},
\[
{\cal A}=
  \begin{bmatrix}
    0 & -6\rho_{\pm} \\
    -\dfrac{2}{3}\kappa\left\lbrace 1-\dfrac{1}{\rho_c}\left(5\rho_{\pm}+\dfrac{2\Lambda}{\kappa}\right)\right\rbrace & 0 \\
  \end{bmatrix}
\]

The stability of fixed points of a linear system depends upon the eigenvalues of the Jacobian matrix ${\cal A}$
(see chapter 2 of \cite{Arrowsmith:1992}). For a two dimensional system, if both eigenvalues are real and of opposite sign then the fixed point is a saddle in the phase portrait. On the other hand,
if the eigenvalues are imaginary and complex conjugates of each other then the fixed point is
likely to be a centre, but this must be confirmed numerically.

\section{CMB constraints on the Emergent Scenario in Braneworld Cosmology}\label{app:bw_CMB}

The parameter $\gamma$ in the potential~(\ref{eq:bw_inf_phi2}) can be constrained using the CMB bounds on the scalar spectral index $n_{_S}$ and the tensor-to-scalar ratio $r$ from the recent \emph{Planck} mission~\cite{Ade:2013uln}. As the scalar field rolls down the potential from $\phi >> M$, the potential~(\ref{eq:bw_inf_phi2}) is approximately the same as in the case of chaotic inflation models with $V(\phi) \propto \phi^{2\gamma}$. Therefore, in the slow roll limit one finds that
\ber
n_{_S} - 1 &=& -\frac{2(\gamma + 1)}{\gamma + 2N}~,\nn\\
n_{_T} &=& -\frac{2\gamma}{\gamma + 2N}~,\nn\\
r &=& \frac{16\gamma}{\gamma + 2N}~.\label{eq: ns-nt-r bw-inf}
\eer
At $95\%$ CL \emph{Planck} data allows $n_{_S}$ within the range $[0.945-0.98]$.
For  $n_{_S}$ to lie in this allowed range with $N = 60$, the parameter $\gamma$ must lie in the following range: $[0.202-2.26]$. Furthermore, \emph{Planck} data also indicate that $r < 0.12$ at $95\%$ CL when BAO data is included~\cite{Ade:2013uln}. Using Eq.~(\ref{eq: ns-nt-r bw-inf}) and with $N = 60$, $r < 0.12$ can be realised only if $\gamma \leq 0.9$.
Therefore, the potential~(\ref{eq:bw_inf_phi2}) at $\phi >> M$ can lead to both $n_{_S}$ within the range $[0.945-0.98]$ and $r < 0.12$ if the parameter $\gamma$ is within the following range:
\beq
0.202\,\leq\,\gamma \,\leq\, 0.9\label{eq: alpha range}
\eeq
For example, if $\gamma = 0.8$, one gets $n_{_S} \simeq 0.97$ and $r \simeq 0.1$ and these values are consistent with \emph{Planck} results~\cite{Ade:2013uln}.
Note that in the potential~(\ref{eq:bw_inf_phi2}), we have assumed that $\phi >> M$ during inflation. Inflation ends at $\phi = \sqrt{2}\,\gamma\,\mpl$. Therefore, the approximation $\phi >> M$ is reasonable if $M << \mpl$.

\begin{figure}[hbt]
\begin{center}
\scalebox{1}[1]{\includegraphics{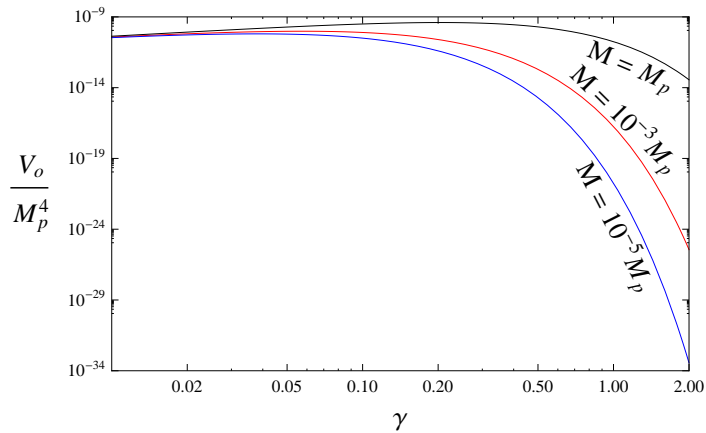}}
\caption{The CMB normalized value of $V_{0}$ in the potential~(\ref{eq:bw_inf_phi2}) is plotted as a function of $\gamma$ for three different values of $M$. In this figure we have taken the number of e-folds $N = 60$.}
\label{fig: bw-inf-V0}
\end{center}
\end{figure}

The constant $V_{0}$ can be fixed using the CMB normalization which indicate that $P_{_{S}}(k_{\ast}) = 2.2\times10^{-9}$ at the pivot scale $k_{\ast} = 0.05\,\mathrm{Mpc}^{-1}$~~\cite{Ade:2013uln}. The expression for $V_{0}$ in terms of $P_{_{S}}(k_{\ast})$ is given by
\beq
\frac{V_{0}}{\mpl^{4}}\,=\, \l(\frac{48\pi^{2} \gamma^{2}P_{_{S}}(k_{\ast}) }{\l[2\gamma(\gamma + 2N)\r]^{\gamma + 1}}\r) \l(\frac{M}{\mpl}\r)^{2\gamma}~.
\label{eqn: V0}
\eeq
Using the above equation, it turns out that $V_{0} = 8\times10^{-16}\mpl^{4}$ for $\gamma = 0.8$ and $M = 10^{-3}\mpl$. In Fig.~\ref{fig: bw-inf-V0}, the CMB normalized value of $V_{0}$  is plotted as a function of $\gamma$. One might note that in deriving~(\ref{eq: ns-nt-r bw-inf}) we assumed $\rho << \rho_c$
during inflation, with $\rho_c$ defined in Eq.~(\ref{eq:bounce1}).
At $60$ e-folds before the end of inflation, one finds $\rho \simeq V(\phi) = 3.45\times10^{-9}\mpl^{4}$ for $\gamma = 0.8$ irrespective of the value of $M$. Therefore, the approximation  $\rho << \rho_c$ during inflation is valid provided $\rho_c >> 10^{-9}\mpl^{4}$.

\section{CMB constraints on the Emergent Scenario in LQC}\label{app:LQC_CMB}
\noindent
As noted in section \ref{sec:LQC}, the emergent scenario in LQC can proceed in two distinct
 ways:

\noindent
(i) If a minimum in the effective potential exists for {\em small} values of the inflaton
potential. In this case the minimum gets destabilized as the inflaton potential {\em increases},
as shown in Fig.~\ref{fig:LQC_pot_a}. A canonical scalar field potential such as (\ref{eq:bw_inf_phi2})
 can accomplish this while
satisfying CMB constraints.

\noindent
(ii) If a minimum in the effective potential exists for {\em large} values of the inflaton
potential. In this case the minimum gets destabilized as the inflaton potential {\em decreases},
as shown in Fig.~\ref{fig:LQC_pot_b}.
 While an inflaton potential such as (\ref{eq:tanh2}) does accomplish this,
it fails to satisfy CMB bounds if the scalar field has {\em canonical kinetic terms}.
One therefore needs to turn to {\em non-canonical scalars} in order to construct
a working example of the emergent scenario in this case, which forms the focus of this appendix.

Consider the following non-canonical scalar field Lagrangian~\cite{Unnikrishnan:2012zu}
\beq
{\cal L}(X,\phi) = X\l(\frac{X}{M^{4}}\r)^{\alpha-1} -\; V(\phi), ~~~ X = \dfrac{1}{2}\dot \phi ^2~,
\label{eqn: Lagrangian NC}
\eeq
where $\alpha$ is a dimensionless parameter
($\alpha \geq 1)$ while $M$ has dimensions of mass. The canonical Lagrangian (\ref{eqn: Lagrangian can}) corresponds to $\alpha = 1$ in (\ref{eqn: Lagrangian NC}).
The energy density and pressure are modified for the non-canonical Lagrangian as follows:
 \begin{subequations}
\label{eq:scalar_NC}
 \begin{align}
  \rho_{\phi}=\left(2\alpha-1\right)X\left(\dfrac{X}{M^4}\right)^{\alpha-1}+V\left(\phi\right)\;, \label{eq:scalar_NC_rho} \\
P_{\phi}=X\left(\dfrac{X}{M^4}\right)^{\alpha-1}-V\left(\phi\right)\;. \label{eq:scalar_NC_P}
 \end{align}
\end{subequations}
The modified scalar field equation of motion is given by
\begin{equation} \label{eq:scalar_NC_evolution}
\ddot \phi+\dfrac{3H\dot \phi}{2\alpha-1}+\left(\dfrac{V'\left(\phi\right)}{\alpha(2\alpha-1)}\right)\left(\dfrac{2M^4}{\dot \phi ^2}\right)^{\alpha-1}=0 \;.
\end{equation}
As long as $V(\phi)$ is constant, the non-canonical scalar is equivalent to
 two non-interacting fluids: $\Lambda$ (which mimics the constant potential $V$) plus
 matter with equation of state
\begin{equation}\label{eq:NC_equ_state}
w=\dfrac{1}{2\alpha-1}\;.
\end{equation}
Thus the non-canonical formalism allows for a wider range of possibilities for the
equation of state: $0 \leq w  < 1$. For instance $\alpha=2 \Rightarrow w = 1/3$, and the
non-canonical scalar plays the role of a radiation+$\Lambda$ filled universe. The
corresponding effective potential $U(a)$ is similar to that shown for the canonical
scalar in Fig.~\ref{fig:LQC_pot_b}, while the scale factor at ESU is determined from
Table~\ref{table:LQC_ESU} to be $a_{LQ}^2=3(\Lambda-\kappa \rho_c)/2$.

We focus on the potential~(\ref{eq:tanh2}) within the non-canonical setting,
assuming that during inflation $\lambda\, \phi << \mpl$, so that (\ref{eq:tanh2}) can be approximated as
\beq
V(\phi) \simeq V_{0}\,\lambda^{2p}\,\l(\frac{\phi}{\mpl}\r)^{2p}~,
\label{eq:pot_cosh aprox}
\eeq
which allows the problem to be tackled analytically.

Following \cite{Unnikrishnan:2012zu} we find that in this case
\ber
n_{_S}- 1 &=& -2\l(\frac{\sigma + p}{2N\sigma + p}\r)~,\nn\\
&& \nn\\
n_{_T} &=& -\frac{2 p}{2N\sigma + p}~,\nn\\
{}&& \nn\\
r &=& \l(\frac{1}{\sqrt{2\alpha -1}}\r)\l(\frac{16p}{2N\sigma + p}\r)~,
\label{eq: ns r cosh pot NC}
\eer
where
\beq
\sigma = \frac{\alpha + p\l(\alpha - 1\r)}{2\alpha -1}~.
\label{eq: sigma NC}
\eeq

\begin{figure}[t]
\begin{center}
\scalebox{0.83}[1]{\includegraphics{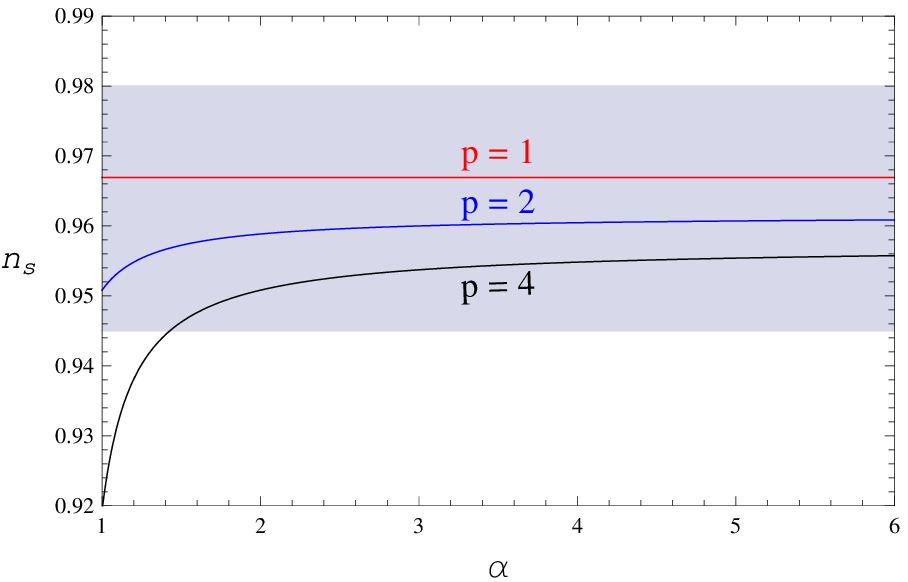}}
\scalebox{0.83}[0.972]{\includegraphics{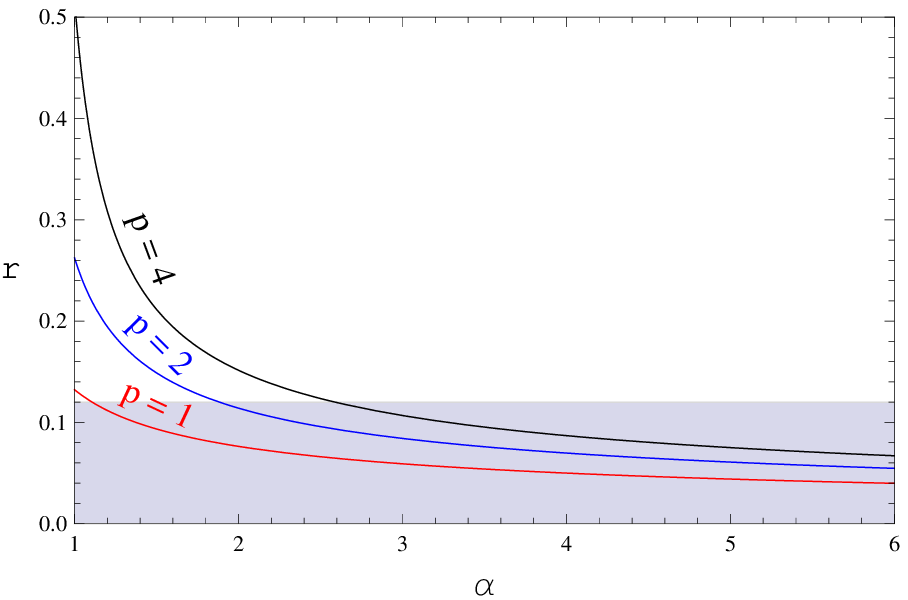}}
\caption{The scalar spectral index $n_{_S}$ (left panel) and the tensor-to-scalar ratio $r$
(right panel) are shown as functions
of $\alpha$, for $p = 1$ and $p = 2$ in (\ref{eq:tanh2}).
The number of e-folds is fixed to $N = 60$.
The shaded region refers to 95\% confidence limits on $n_{_S}$ and $r$ determined by Planck \cite{Ade:2013uln}.}
\label{fig: NC1 ns and r}
\end{center}
\end{figure}

In Fig.~\ref{fig: NC1 ns and r}, the scalar spectral index $n_{_S}$ and the tensor-to-scalar ratio
$r$ are
plotted as functions of $\alpha$. Note that $r$ decreases as $\alpha$ increases.
Substituting $\alpha = 2$ in (\ref{eq: ns r cosh pot NC}), we find
\ber
n_{_S} = 0.97~~&\mathrm{and}&~~~r = 0.076~~~\mathrm{for}~~~p = 1,\nn\\
n_{_S} = 0.96~~&\mathrm{and}&~~~r = 0.11~~~~\mathrm{for}~~~p = 2,
\label{eqn: ns r alpha2}
\eer
which satisfy the \emph{Planck} requirements $n_{_S} \in [0.945-0.98]$ and $r < 0.12$ at $95\%$ CL ~\cite{Ade:2013uln}.

\begin{figure}[t]
\begin{center}
\scalebox{0.83}[1]{\includegraphics{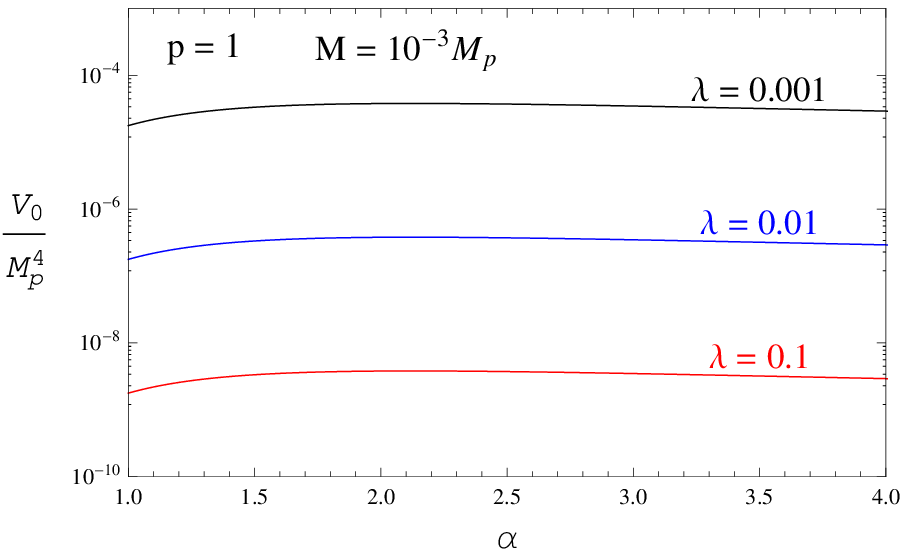}}
\scalebox{0.83}[1]{\includegraphics{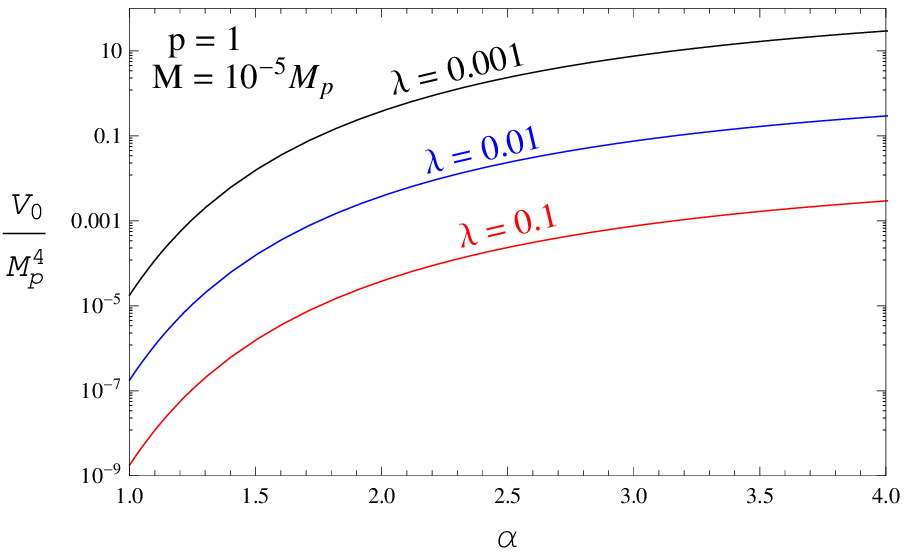}}
\caption{The CMB normalized value of $V_{0}$ in the potential~(\ref{eq:tanh2}) is plotted as a function of $\alpha$. In the left panel, the value of the parameter $M$ in (\ref{eqn: Lagrangian NC}) is set to be $10^{-3}\mpl$,  whereas $M = 10^{-5}\mpl$  in the right panel.}
\label{fig: NC1-V0-vs-alpha-p1}
\end{center}
\end{figure}

\begin{figure}[hbt]
\begin{center}
\scalebox{1}[1]{\includegraphics{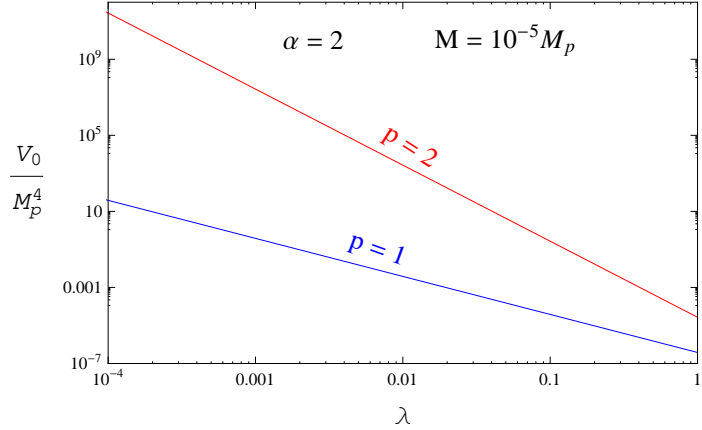}}
\caption{The CMB normalized value of $V_{0}$ in (\ref{eq:tanh2}) is plotted as a function of $\lambda$.
The two parameters $\alpha$ and $M$ in the Lagrangian~(\ref{eqn: Lagrangian NC})
have been fixed to $\alpha = 2$ and $M = 10^{-5}\mpl$.
}
\label{fig: NC1-V0-vs-lambda}
\end{center}
\end{figure}

The value of $V_{0}$ can be fixed using CMB normalization \emph{viz.}~$P_{_{S}}(k_{\ast}) = 2.2\times10^{-9}$ at the pivot scale $k_{\ast} = 0.05\,\mathrm{Mpc}^{-1}$~\cite{Ade:2013uln}.
For (\ref{eqn: Lagrangian NC}) and (\ref{eq:tanh2}), the expression for $V_{0}$ in terms of $P_{_{S}}(k_{\ast})$ is given by
\beq
\frac{V_{0}}{\mpl^{4}}\,=\, \l(\frac{1}{\lambda^{2p}}\r)\l\{\l(\frac{24 \pi^{2} p\, P_{_{S}}(k_{\ast})}{\sqrt{2 \alpha - 1}}\r)
\l[\l(\frac{\alpha}{2 p}\r)\l(\frac{1}{6 \mu^{4}}\r)^{\alpha - 1}\r]^{\frac{p}{\sigma(2\alpha - 1)}}
\l(\frac{1}{2N\sigma + p}\r)^{\frac{\sigma + p}{\sigma}}
\r\}^{\frac{\sigma(2\alpha - 1)}{\alpha}}~,
\label{eqn: V0-cosh pot NC}
\eeq
where $\mu = M/\mpl$.
In Fig.~\ref{fig: NC1-V0-vs-alpha-p1} the CMB normalized value of $V_{0}$ is plotted as a function of $\alpha$ for different values of $\lambda$ and $M$.
One finds that when $M = 10^{-3}\mpl$ the value of $V_{0}$ is nearly independent of $\alpha$,
whereas $V_{0}$ {\em increases dramatically} with increasing $\alpha$,
when $M < 10^{-3}\mpl$.
This result also holds when $p > 1$. Furthermore,
for fixed values of $\alpha$ and $M$, $V_{0}$ increases as $\lambda$ decreases,
 as shown in Fig.~\ref{fig: NC1-V0-vs-lambda}. From these figures its clear
 that one can have $V_{0} \gtrsim \mpl$ by appropriately choosing
 the model parameters $\alpha$, $\lambda$ and $M$. For instance, $\alpha = 2$ and $M = 10^{-5}\mpl$
result in
\ber
V_{0} \gtrsim \mpl^{4}~~&\mathrm{when}&~~~\lambda  \lesssim 6.2\times 10^{-4}~~~\mathrm{for}~~~p = 1,\nn\\
V_{0} \gtrsim \mpl^{4}~~&\mathrm{when}&~~~\lambda  \lesssim 7.2\times 10^{-2}~~~\mathrm{for}~~~p = 2,
\eer
together with the \emph{Planck}-consistent values for $n_{_S}$ and $r$
quoted in (\ref{eqn: ns r alpha2}).

As mentioned earlier, these results are valid only when  $\lambda \phi << \mpl$ during inflation.
In this case,  $N$ e-folds prior to the end of inflation, one finds
\beq
\phi(N) = C_{_1}^{1/(2\sigma)}\l(2N\sigma + p\r)^{1/(2\sigma)}\mpl~,
\eeq
where $\sigma$ was defined in (\ref{eq: sigma NC}) and $C_{_1}$ is given by
\beq
 C_{_1} = \l\{\l(\frac{2 \,p \,\mu^{4(\alpha -1)}}{\alpha}\r)\l(\frac{6\mpl^{4}}{V_{0}\lambda^{2p}}\r)
 \r\}^{\frac{1}{2\alpha - 1}}~.
\eeq
For $\alpha = p = 2$, the above equations give $\phi = 0.1 \mpl$, $60$ e-folds before the end of inflation, making the approximation $\lambda \phi << \mpl$ perfectly reasonable provided $\lambda < 1$.

We therefore conclude that the emergent scenario in LQC based on the
 non-canonical model~(\ref{eqn: Lagrangian NC}) with potential~(\ref{eq:tanh2}) can lead to CMB-consistent values for $n_{_S}$ and $r$ even when $V_{0} \gtrsim \mpl^{4}$, provided that the semi-classical
treatment followed by us is allowed in this regime.



\end{document}